\documentclass[longauth]{aa}  
\usepackage{graphicx}
\usepackage{comment}
\usepackage{txfonts}
\usepackage{caption}
\usepackage{xcolor}
\usepackage{hyperref}
\newcommand{\vrulesep}{\ }
\newcommand{\hrulesep}{\unskip \ \hrule\ \unskip }

\usepackage{graphicx}
\usepackage{lipsum}
\usepackage{orcidlink}
\usepackage{gensymb}
\usepackage{txfonts}
\usepackage{amssymb}
\usepackage{amsmath}
\usepackage{booktabs}
\usepackage{dcolumn}
\usepackage{sidecap}
\usepackage{subfig}
\usepackage{multirow}
\usepackage{threeparttablex}
\usepackage{longtable}
\usepackage{xcolor}
\usepackage{natbib}
\usepackage{pdflscape}
\usepackage{upgreek}
\usepackage{lscape}
\bibpunct{(}{)}{;}{a}{}{,}
\usepackage{array}
\usepackage{placeins}
\newcolumntype{H}{>{\setbox0=\hbox\bgroup}c<{\egroup}@{}}
\newcommand{\ODISEA}[1]{ODISEA\_C4\_#1}
\usepackage{keyval}

\begin{document} 

\title{The Ophiuchus DIsc Survey Employing ALMA (ODISEA). }
\subtitle{Substructures as a function of SED Class and disc mass in 100 systems}
\authorrunning{Bhowmik et al.}
\titlerunning{Substructures as a function of SED Class and disc mass in 100 systems}
\author{Trisha Bhowmik\inst{\ref{instUDP},\ref{instYEMS}}\orcidlink{0000-0002-4314-9070}
\and Lucas Cieza\inst{\ref{instUDP},\ref{instYEMS}}\orcidlink{0000-0002-2828-1153} \and J. M. Miley\inst{\ref{instYEMS},\ref{instESO},\ref{instALMA_JAO}}\orcidlink{0000-0002-1575-680X} 
\and P. H. Nogueira\inst{\ref{instYEMS},\ref{instPedro}}\orcidlink{0000-0001-8450-3606}
\and Camilo González-Ruilova\inst{\ref{instYEMS},\ref{instUSACH},\ref{instCIRAS}}\orcidlink{0000-0003-4907-189X}
\and Prachi Chavan\inst{\ref{instUDP},\ref{instYEMS}}\orcidlink{0000-0003-2406-0684}
\and Anibal Sierra\inst{\ref{instAnibal}, \ref{instUNAM}}\orcidlink{0000-0002-5991-8073}
\and Anuroop Dasgupta\inst{\ref{instUDP},\ref{instYEMS},\ref{instESO}}\orcidlink{0009-0009-8115-8910}
\and Simon Casassus\inst{\ref{instUCH},\ref{instDO}}\orcidlink{0000-0002-0433-9840}
\and Grace Batalla-Falcon\inst{\ref{instUDP}}\orcidlink{0009-0007-4878-0252}
\and Gioele Di Lernia\inst{\ref{instUDP},\ref{instYEMS},\ref{instETH}}
\and Antonio S. Hales\inst{\ref{instYEMS},\ref{instALMA_JAO},\ref{instNRAO}}\orcidlink{0000-0001-5073-2849}
\and  Jeff Jennings \inst{\ref{instJeff}}
\and Santiago Orcajo \inst{\ref{instYEMS},\ref{instSanti1}, \ref{instSanti2}}\orcidlink{0000-0002-7625-1768}
\and Sebasti\'an P\'erez\inst{\ref{instYEMS},\ref{instUSACH},\ref{instCIRAS}}\orcidlink{0000-0003-2953-755X}
\and Dary Ruíz-Rodriguez\inst{\ref{instNRAO}}\orcidlink{0000-0003-3573-8163} 
\and Yangfan Shi\inst{\ref{instYangfin2}}\orcidlink{0000-0001-9277-6495}
\and Jonathan P. Williams\inst{\ref{instJohn}}\orcidlink{0000-0001-5058-695X} 
\and Ke Zhang\inst{\ref{instCoco}\orcidlink{0000-0002-0661-7517}}
\and Alice Zurlo\inst{\ref{instUDP},\ref{instYEMS}} \orcidlink{0000-0002-5903-8316}
       }
\institute{
Instituto de Estudios Astrof\'isicos, 
Facultad de Ingenier\'ia y Ciencias,
Universidad Diego Portales,
Av. Ej\'ercito 441, Santiago, Chile
\email{trisha.bhowmik@mail.udp.cl}
\label{instUDP}
\and
Millennium Nucleus on Young Exoplanets and their Moons (YEMS), Santiago, Chile \label{instYEMS}
\and
European Southern Observatory, 
Alonso de Córdova 3107, 
Vitacura, Santiago, Chile \label{instESO}
\and 
Joint ALMA Observatory, Avenida Alonso de Cordova 3107, Vitacura 7630355, Santiago, Chile\label{instALMA_JAO}
\and
Department of Physics and Astronomy, Texas A\&M University, College Station, TX 77843-4242, USA \label{instPedro} 
\and 
Departamento de Física, Universidad de Santiago de Chile, Av. Víctor Jara 3493, Santiago, Chile \label{instUSACH}
\and 
Center for Interdisciplinary Research in Astrophysics Space Exploration (CIRAS), Universidad de Santiago de Chile, Chile \label{instCIRAS}
\and
Mullard Space Science Laboratory, University College London, Holmbury St Mary, Dorking, Surrey RH5 6NT, UK \label{instAnibal}
\and
Departamento de Astronomía, Universidad de Chile, Camino el Observatorio 1515, Las Condes, Santiago, Chile \label{instUCH}
\and 
{Data Observatory Foundation, Eliodoro Y\'a\~{n}ez 2990, Providencia, Santiago, Chile} \label{instDO}
\and
ETH Zurich, Institute for Particle Physics and Astrophysics, Wolfgang-Pauli-Strasse 27, CH-8093 Zurich, Switzerland \label{instETH}
\and
National Radio Astronomy Observatory, 520 Edgemont Road, Charlottesville, VA 22903-2475, USA \label{instNRAO}
\and
Center for Computational Astrophysics, Flatiron Institute, 162 Fifth Ave., New York, NY 10010, USA \label{instJeff}
\and
Facultad de Ciencias Astronomicas y Geofisicas, Universidad Nacional de La Plata, Paseo del Bosque s/n, 1900 La Plata, Argentina \label{instSanti1}
\and
Instituto de Astrofísica de La Plata (IALP), CCT La Plata-CONICET-UNLP, Paseo del Bosque s/n, La Plata, Argentina \label{instSanti2}
\and
Department of Astronomy, Peking University, Beijing 100871, China  \label{instYangfin2}
\and
Institute for Astronomy, 2680 Woodlawn Dr, Honolulu, HI USA \label{instJohn}
\and
Department of Astronomy, University of Wisconsin-Madison, 475 N Charter St, Madison, WI 53706, USA \label{instCoco}
\and
Universidad Nacional Autónoma de México. Instituto de Astronomía. A.P. 70-264, 04510. Ciudad de México, México \label{instUNAM}
}
\date{Received XXXX; accepted XXXX}
\abstract{
Understanding the origin of substructures in protoplanetary discs is one of the main challenges in the field. However,  their characterization is currently biased towards small samples of the brightest (flux $\gtrsim$ 50 mJy {at} 225 GHz) and largest discs in nearby molecular clouds.}
{We present a complete flux-limited high-resolution study of $\sim$100 discs from the Ophiuchus DIsc Survey Employing ALMA (ODISEA), spanning fluxes of $\sim$4-400 mJy at 225 GHz, to investigate disc substructures as a function of SED Class and disc mass using ALMA Band 8 continuum observations (410 GHz; 0.7 mm).
}
{The survey extends to faint discs containing as little as $\sim$2\,M$_{\oplus}$ of dust. Given the well-established flux–size relationship in discs, sources with flux $\geq 20$\,mJy at 225\,GHz were observed at a nominal resolution of 20\,au, while fainter sources were observed at three times higher resolution. In both samples, we used the \texttt{Frankenstein} code to fit non-parametric models
directly to the visibilities, achieving sub-beam resolution. 
We classified the substructures to place each object within an evolutionary sequence, inspired by our earlier qualitative definitions, linking disc morphology with different stages of giant planet formation. Stages I–V follow a progression
from discs with weak substructures marked by inflection points, through systems
hosting one or more gap–ring pairs and additional weaker structures beyond them,
to discs with central cavities that either preserve multiple substructures or
are reduced to a single dominant ring. We now introduce Stage 0 to describe featureless discs and provide precise and systematic definitions for Stages 0–V.
}
{
{Our results show that, despite higher optical depths, Band~8 is an efficient tracer of disc substructures and can recover, with shorter integration times, the same gaps and cavities observed at longer wavelengths. We find that the most massive discs ($\gtrsim 10$\,M$_{\oplus}$ of dust)
exhibit structures consistent with the
proposed sequence, even when observed at modest resolution. 
In this dust-mass regime, the incidence of evolved substructures (Stages II to V) increases from
{23$\%$ (6/26)} in the Class I/F sources to {$\gtrsim$53$\%$ (17/32)} in the Class II objects.   
In contrast, very few lower-mass discs show the gaps and cavities typically associated with planet
formation, despite being observed at significantly higher resolution. This trend may partly reflect observational limitations, given the very steep flux–size relationship, implying that observations at even higher spatial resolution are required to achieve the same number of {resolution elements across the disc} as in their higher-mass counterparts, which is required to probe the nature of these faint discs robustly.
}
}
{
{Our results support the idea that substructures in discs with $\gtrsim$10\,M$_{\oplus}$ are consistent with the formation of giant planets and highlight the power of Band~8 observations to characterize disc substructures.
Since ALMA can provide $\sim$1\,au resolution in Band~8, where low-mass discs remain relatively bright, this band offers a viable path to probe substructures
in discs with $\lesssim$10\,M$_{\oplus}$, a regime that remains largely unexplored.} 
}
 \maketitle
%

\section{Introduction}

Circumstellar discs are the sites where planets are born and are therefore natural laboratories for planet formation \citep[e.g.,][]{2011ARA&A..49...67W}. ALMA long-baseline observations have shown that substructures, primarily
cavities, rings, and gaps, are ubiquitous in discs observed at high angular resolution \citep{2020ARA&A..58..483A}. However, most long-baseline observations to date have focused on the brightest and largest discs in nearby molecular clouds \citep[e.g.,][]{2018Andrews,2018Long,2021Cieza}. {Recent high-resolution surveys have begun to address this bias by targeting compact and fainter disc populations in Taurus and Lupus, while AGE-PRO provides a complementary evolutionary context by tracing gas and dust properties across Ophiuchus, Lupus, and Upper Sco \citep{2023ApJ...952..108Z,2025A&A...696A.232G,2025ApJ...989....1Z}.} In contrast, although substructures have been reported in some smaller and fainter discs through targeted or method-specific studies \citep[e.g.,][]{2019ApJ...882...49L,2019A&A...626L...2F,2021A&A...645A.139K,2020Gonzalez-Ruilova,2021ApJ...923..121Y}, their overall incidence remains poorly constrained and has not yet been systematically explored across a large, uniform sample. Previous Band 6 (225 GHz/1.3 mm) results of the Ophiuchus DIsc Survey Employing ALMA (ODISEA) project \citep{2019Cieza} illustrate this bias well.  Although ODISEA detected over 200 discs with fluxes spanning 3 full orders of magnitude (from $\sim$0.4 to
$\sim$400 mJy; \cite{2019Williams}), only the 15 brightest objects ($>$ 70 mJy in Band 6) were observed by ALMA with long baselines \citep{2021Cieza}. 

While several alternatives have been proposed to explain the origin of the substructure seen in protoplanetary discs, planet formation remains one of the leading explanations \citep{2018Zhang, 2020ARA&A..58..483A}.  
This scenario is {consistent with} the direct detection of giant planets within the cavities and gaps of discs such as PDS 70, {WISPIT 2 and Elias 2-24b \citep[][Bernadi et al. in press]{2018A&A...617A..44K,2019NatAs...3..749H,2025ApJ...990L...8V,2025ApJ...990L...9C,2021AJ....161..146J}.}   
In this context, and using long-baseline observations of Ophiuchus discs, \cite{2021Cieza} previously proposed an evolutionary sequence in which the diversity of disc substructures can be understood in terms of the formation of giant planets through core accretion and dust evolution. This interpretation has recently been supported by numerical models that reproduce the sequence in detail \citep{2025ApJ...984L..57O}. Given the nature of the sample (the brightest discs in Ophiuchus), the proposed
sequence predominantly applies to discs with dust masses $\gtrsim$50\,M$_{\oplus}$. Such systems provide large solid reservoirs that are favorable for giant planet formation, consistent with core-accretion models and observed correlations between disc mass, stellar mass, and giant planet occurrence \citep{2018haex.bookE.143M,2023ASPC..534..539M,2025ARA&A..63..217I}. However, these objects represent only the high-mass tail of the disc mass distribution.

To measure the sizes of previously unresolved discs and to investigate the incidence of substructures across a broader and more representative range of disc masses among discs, we recently performed a Band~8 (410~GHz / 0.7~mm) continuum survey of the 
$\sim$100 brightest discs in Ophiuchus (Band 6 fluxes $>$ 4 mJy). 
Given the strong correlations between disc flux and size \citep{2017Tripathi, 2018Andrews,2021Tazzari}, objects brighter than 20 mJy in Band 6 were observed at 0.15”  resolution in Band 8. In comparison, objects with Band 6 fluxes between 4 and 20 mJy were observed in Band 8 with a factor of 3 higher resolution (0.05”).  
\cite{2025Dasgupta} have already used the same Band 8 data analyzed here to characterize the dust continuum size distribution of the sample. They find that the distribution is log-normal with a median value of just $\sim$14 au with no significant difference between the younger {(Class I/F)} sources and older (Class II) targets. The lack of disc size evolution suggests that mechanisms capable of slowing radial dust drift must operate across the ODISEA Band~8 sample. While (proto)planet-induced pressure bumps provide a natural explanation, alternative scenarios such as reduced drift efficiency in gas-rich discs may also contribute \citep[e.g.,][]{2024ApJ...976...50W}. If planets drive the observed dust trapping, exoplanet statistics imply they are predominantly low-mass, since massive planets at tens of au are rare \citep[see, e.g.,][]{2021A&A...651A..72V,2021exbi.book....2G}.

In this paper, we reconstruct 1D radial brightness profiles of the discs in our sample by fitting non-parametric models directly to the observed visibilities using the \texttt{Frankenstein} code \citep[hereafter \texttt{Frank}]{2020Jennings}  to our Band 8 data. We then classify each disc based on its substructures (or lack thereof),  placing each object in the evolutionary sequence connected to different stages of giant planet formation as proposed by \cite{2021Cieza}.  We use our results to provide a first-order approximation of the population of planets that are expected to emerge in the Ophiuchus molecular cloud. The paper is organized as follows. In Section 2, we present the sample selection, the observations, and the data reduction. In Section 3, we describe the data analysis, including the precise definitions and procedures we adopt to classify substructures and the evolutionary stages.  In Section 4, we present our results for the entire sample of $\sim$100 objects. In Section 5, we discuss these results and use them to provide broad constraints on the architectures of planetary systems forming in Ophiuchus, which we compare to current statistics of exoplanets. Finally, in Section 6, we summarize our main conclusions.

\begin{table*}[!h]
\caption{{Basic disc properties for ODISEA  Band-8 sample}} 
\label{t:basic_disc_prop}
\makebox[\textwidth]{ 
\resizebox{\textwidth}{!}{%
\begin{tabular}{lcccccccccccccccccc}
\hline
\hline
Field & $x_0$ & $y_0$ & $i$ & PA & $R_{\rm out}$ & $F_{\rm B8}$ & $d$ & Beam PA & Beam maj. & Beam min. & Class & Stage & FOV & $R_{95}$ & $M_{\rm disc,B4}$ & $T_{\rm bol}$ & Frank res. & $R_{68}$ \\
 &  &  & (deg) & (deg) & ($^{\prime\prime}$) & (mJy) & (pc) & (deg) & ($10^{-3}\,^{\prime\prime}$) & ($10^{-3}\,^{\prime\prime}$) &  &  & ($^{\prime\prime}$) & (au) & ($M_\oplus$) & (K) & (au) & (au) \\
(1) & (2) & (3) & (4) & (5) & (6) & (7) & (8) & (9) & (10) & (11) & (12) & (13) & (14) & (15) & (16) & (17) & (18) & (19) \\
\hline
ISO-Oph\_123 & 16:27:17.574 & -24.05.14.300 & $25.57\pm18.21$ & $49.00\pm45.00$ & 0.50 & $28.90\pm1.40$ & 136.58 & -88.47 & 65.34 & 54.55 & II & 0 & 0.50 & $19.74\pm0.17$ & $9.35\pm0.44$ & 1000 & $8.664\pm0.052$ & $13.35\pm0.06$ \\
ISO-Oph\_13 & 16:26:07.032 & -24.27.24.800 & $63.53\pm4.77$ & $165.60\pm5.00$ & 0.50 & $25.10\pm2.50$ & 139.40 & 86.28 & 64.55 & 51.94 & II & 0 & 0.60 & $32.75\pm0.34$ & $6.18\pm0.41$ & 890 & $12.615\pm0.099$ & $22.63\pm0.13$ \\
ISO-Oph\_193$^{e,u}$ & 16:28:12.713 & -24.11.36.340 & $77.09\pm11.92$ & $73.00\pm20.00$ & 0.50 & $11.57\pm0.97$ & 151.74 & -86.49 & 67.42 & 55.10 & II & 0 & 0.50 & $18.52\pm0.24$ & $3.88\pm0.23$ & 1300 & $35.859\pm0.465$ & $12.87\pm0.09$ \\
ISO-Oph\_208 & 16:31:59.319 & -24.54.41.140 & $46.72\pm3.73$ & $90.70\pm5.40$ & 0.50 & $57.10\pm2.60$ & 152.16 & 77.19 & 67.74 & 54.03 & II & 2 & 1.00 & $55.57\pm5.38$ & $18.97\pm0.73$ & 1500 & $8.546\pm0.041$ & $32.67\pm2.23$ \\
ODISEA\_C4\_001$^{e,u}$ & 16:21:31.918 & -23.01.40.890 & $81.92\pm6.08$ & $167.10\pm3.80$ & 0.50 & $12.80\pm1.50$ & 137.00 & 79.52 & 64.90 & 52.59 & II & 0 & 1.00 & $38.55\pm0.43$ & $3.14\pm0.19$ & 1400 & $36.088\pm0.675$ & $27.02\pm0.17$ \\
ODISEA\_C4\_007 & 16:22:24.951 & -23.29.55.550 & $38.54\pm7.27$ & $9.70\pm14.10$ & 0.50 & $40.60\pm1.40$ & 138.33 & 79.24 & 65.27 & 52.92 & II & 0 & 0.50 & $17.68\pm0.06$ & $12.49\pm0.47$ & 1400 & $7.831\pm0.034$ & $12.12\pm0.03$ \\
ODISEA\_C4\_009$^{e, m}$ & 16:23:05.416 & -23.02.57.610 & $83.84\pm0.88$ & $73.58\pm0.54$ & 0.50 & $17.70\pm1.10$ & 139.40 & 79.95 & 63.23 & 51.63 & I & 0 & 0.60 & $32.65\pm0.74$ & $4.41\pm0.37$ & 650 & $18.535\pm0.204$ & $22.11\pm0.26$ \\
ODISEA\_C4\_016A$^{b,n}$ & 16:25:02.115 & -24.59.32.952 & $18.15\pm26.40$ & $134.00\pm85.00$ & 0.50 & $35.10\pm3.60$ & 142.04 & 83.70 & 65.17 & 52.02 & II & 5 & 0.50 & $23.56\pm1.21$ & $5.58\pm0.20$ & 1200 & NaN & $16.40\pm0.24$ \\
ODISEA\_C4\_016B$^{b}$ & 16:25:02.008 & -24.59.33.180 & $42.57\pm8.35$ & $130.00\pm14.00$ & 0.25 & $23.70\pm1.80$ & 142.04 & 83.44 & 66.13 & 52.55 & II & 5 & 0.50 & $15.73\pm0.09$ & 5.06 & 1200 & $5.949\pm0.037$ & $11.88\pm0.03$ \\
ODISEA\_C4\_017$^{u}$ & 16:25:06.905 & -23.50.50.970 & $73.72\pm4.31$ & $29.00\pm4.90$ & 0.50 & $14.40\pm1.10$ & 144.45 & 86.66 & 64.26 & 51.85 & II & 0 & 0.50 & $23.68\pm0.32$ & $3.75\pm0.21$ & 1700 & $34.532\pm0.465$ & $16.30\pm0.12$ \\
\hline
\end{tabular}
}
}
\vspace{0.25cm} \\
\noindent 
\tablefoot{ The full table is available in electronic form. {(1) is the name of the disc as used in the ODISEA survey. The superscripts \textit{e}, \textit{u}, {\textit{m}}, \textit{b,} and \textit{n} denote discs that are edge-on ($i=80\deg\pm5\deg$), remain not well resolved moderately resolved (see Section \ref{sec:caveates_resFRANK}), binaries, and where the Frank resolution couldn't be obtained, respectively.} (2), (3), (4), (5), (6), (8),(9), (10), (11) are the RA, DEC, inclination, PA, hyperparameter R$_{out}$, distance of the source, PA of the beam, beam's major axis and beam's minor axis used to produce the \texttt{Frank} radial profiles and models.
 (7) is the Band 8 flux extracted from \cite{2025Dasgupta} and so are (4) and (5), except for ODISEA\_C4\_50B, ODISEA\_C4\_94A, ODISEA\_C4\_94B, ODISEA\_C4\_125B, ODISEA\_C4\_22AB, ODISEA\_C4\_26, for which these parameters were calculated using the \texttt{imfit} function in \texttt{CASA}. (8) and (12) are the distances to the source, and the classes of the discs are from \cite{2019Williams}. (13) is the evolutionary stage of the disc around the targets. (14) is the field of view of the data, models, and residuals shown in Figs. \ref{fig:0+I}-\ref{fig:5+II}). (15) is the disc radius that encloses 95\% of the total integrated intensity. (16) is the disc dust mass as measured in Band 4. (17) is the disc bolometric temperature from {\cite{2009Evans}}. (18) and (19) are the resolutions achieved by \texttt{Frank} models for each disc and the $R_{68}$, the disc radius that encloses 68\% of the total integrated intensity. NaN indicates that the corresponding numerical value is not available.}
\end{table*}

\section{Sample selection, observations, and data reduction} 
\subsection{Sample}
\label{sec:2.1}
\begin{figure}
    \centering
    \includegraphics[width=0.98\linewidth]{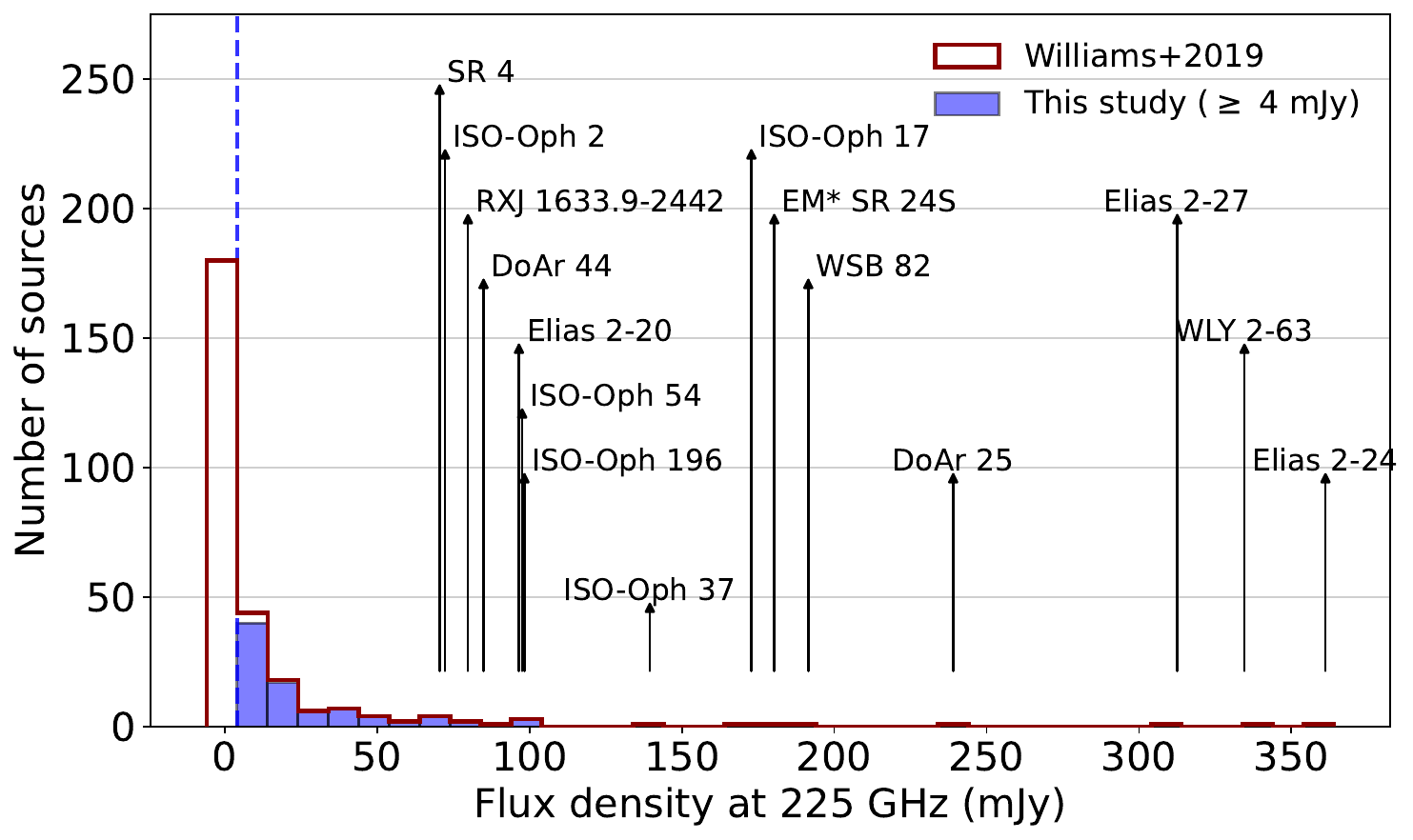}
    \caption{The red bordered histogram depicts the full ODISEA sample from \cite{2019Williams,2019Cieza}. The blue histogram shows the sample for this study, and the arrows with disc names represent the sample presented in \cite{2021Cieza}. The dashed line is the {lower} limit of 4 mJy at 225 GHz for this study.}
    \label{fig:sample}
\end{figure}

 {The original ODISEA survey targeted 289 \emph{Spitzer}-selected Ophiuchus sources; after Gaia vetting and removal of likely background or misclassified objects, the final ODISEA parent sample contains 265 sources, corresponding to 279 discs when multiple systems are counted separately \citep{2019Williams}.} To investigate the incidence of substructures across a broader range of disc masses among discs that are detectable and resolvable with current ALMA capabilities, we performed a Band~8 (410~GHz / 0.7~mm) continuum survey of the $\sim$100 flux-limited ODISEA discs during ALMA Cycles~8 and~9 (see Table 1). This sample includes all objects with Band~6 fluxes $>$4~mJy, spanning nearly two orders of magnitude in flux (up to $\sim$400~mJy). The remaining ODISEA targets consist of $\sim$100 discs with Band~6 fluxes between 0.4 and 4~mJy and a further $\sim$80 sources that were not detected in Band~6, with 3$\sigma$ upper limits of $\sim$0.4~mJy. These populations lie below the sensitivity threshold required for high-fidelity imaging of substructures at the angular resolution achieved in this survey. These statistics illustrate that discs previously targeted by high-resolution observations by \cite{2021Cieza} represent the tail end of the disc population that can be seen in the histogram with black hatches in Fig.~\ref{fig:sample}. We also see in the same Fig.~\ref{fig:sample}, presented in the blue histogram, that the present survey significantly expands the explored parameter space toward lower, yet still detectable, discs. {Therefore, the sample presented here is roughly one-third of the full ODISEA parent population, or $\sim38\%$ of the Gaia-vetted source sample.}

Within our sample, there are two subsets: 45 discs with brightness higher than 20 mJy at 225 GHz/1.3 mm {were} observed at 20 au resolution, and 55 discs with brightness between 4 and 20 mJy were observed at {7 au (0.05'') }resolution. 
This allows for a more meaningful comparison between the incidence of substructures across disc fluxes in Band 8 {\citep[see also][]{2023ASPC..534..423B}}. 
Thus, the present Band~8 sample represents roughly one third of the original ODISEA catalogue, or $\sim38\%$ of the Gaia-vetted ODISEA parent sample of 265 sources \citep{2019Williams}.

\subsection{Observation}
In ALMA Cycle 8, the 45 bright discs were observed with the array configuration C43-4, resulting in a nominal resolution of  0.15'' (PID: 2021.1.00378.S; PI: Lucas Cieza). The total observation time for 45 targets was just 2.1 hrs, highlighting the efficiency of Band-8 observations.  
 These observations were taken on two days, August 4 and 11, 2022, in two execution blocks (EBs) with the same targets. The precipitable water vapour (PWV) was less than 0.6 mm, and the maximum and minimum baselines were 15 m and 1301 m, respectively, for both the EBs. Quasars J1924-2914 and J1427-4206 were used as flux and bandpass calibrators for the two observations, respectively, and J1700-2610 was used as a phase calibrator for both observations in Cycle 8. 

In ALMA Cycle 9, the 55 faint discs were observed with the array configuration C43-7, corresponding to a nominal resolution of  0.05'' (PID: 2022.1.00480.S; PI: Lucas Cieza). The total observing time for these 55 targets was 3.58 hrs. 
In this case, different groups of targets were observed over two nights, May 23 and 29, 2023, in five different EBs. The PWV was $\sim$0.8 mm for both EBs, and the minimum and maximum baselines were 27 m, 3637 m, and 78 m, 3637 m on the two observing nights, respectively. Quasar J1427-4206 is used as a bandpass and flux calibrator for all the observations.  Quasars J1553-2422, J1634-2058, and J1700-2610 were used as phase calibrators on the two nights of observations. All observations used the same correlator set-up in which four spectral windows were centered at 398.0 GHz, 400.0 GHz, 410.0 GHz, and 412.0 GHz, with a bandwidth of 1875.0 MHz.

\subsection{Data reduction}
We concatenated the two Cycle~8 observing nights for the 39 single-disc targets. The six binary systems were re-observed at higher resolution in Cycle~9, and only those data are used. For project 2021.1.00378.S, data were calibrated and imaged using the Common Astronomy Software Applications package (\texttt{CASA}; \cite{2022CASA}) and Pipeline versions 6.2.1.7 and 2021.2.0.128, respectively. For project 2022.1.00480.S, data were calibrated and imaged using \texttt{CASA} and Pipeline versions 6.4.1.12 and 2022.2.0.68, respectively. Data reduction is carried out using a standard multiscale \texttt{CLEAN} with Briggs weighting (robust = 0.5). For centrally peaked single discs, we obtain the phase center, position angle, and major and minor axes using the \texttt{imfit} task, and generate elliptical masks by doubling the measured major and minor axes during the \texttt{tclean} process. For discs with central cavities or binaries, phase centers are set manually during masking, based on values measured interactively in the \texttt{CASA} Viewer after a shallow \texttt{tclean}.  
We also do a phase-only self-calibration for 101 discs, including the binaries.{ The binaries are indicated in Table 1.}  For three faint discs, RA162637.14b, RA162338.52, and RA162413.46, the \texttt{imfit} function and the self-calibration process did not work, which is why we exclude these discs from this study. 
Overall, our final sample includes 101 discs, for which we obtained \texttt{Frank} profiles. 

\section{Data analysis}

\subsection{Visibility plane analysis with \texttt{Frank}}
To identify the substructures, we extract a radial brightness profile directly from the observed visibilities using the \texttt{Frank} code  \citep[hereafter \texttt{Frank}]{2020Jennings}, assuming the discs have axisymmetric structures. {\texttt{Frank} reconstructs the azimuthally averaged radial brightness profile of an axisymmetric disc by fitting the interferometric visibilities directly in the Fourier domain, rather than extracting the profile from a \texttt{tclean} image. The method uses a non-parametric Gaussian-process framework with a discrete Hankel transform to recover sub-beam radial structure while avoiding the beam-convolution limitations of image-plane profiles.} {We perform this analysis on spectrally unaveraged data, as spectral averaging can smooth out or suppress subtle substructures and also may produce wiggles and false substructure signals (see the example in Section A.1)}. For each disc in our sample, we first fix the geometry to the inclination and PA as given in \cite{2025Dasgupta}. This choice is made since it is computationally expensive to run \texttt{Frank} to fit each disc's geometry in our sample. We also fix the hyperparameters N=300, $\alpha$ =1.3, wsmooth= 1e-1, p0 =1e-35, asin\_h=0.02 and bin widths=[2e2,2e3]. {The adopted values of $\alpha$=1.3 and wsmooth = 0.1 suppress noise-dominated features and reduce overfitting (e.g., \cite{2025A&A...695A.147V}). An example of overfitting is presented in Appendix A for the source ODISEA\_C4\_143 (Fig.~\ref{fig:overfitting}).} We adopted a single set of standard hyperparameters for all discs rather than optimizing them individually. {A systematic exploration of hyperparameter space and statistical optimization (e.g., via $\chi^2$ minimization) combined with visual inspection for each of the 101 discs' spectrally unaveraged visibilities is beyond the scope of this work.} Such an analysis is better suited to dedicated studies of individual discs or small subsamples. 
 $R_{out}$ sets the angular distance out to which the fitting is performed, and should be set to contain all disc emission conservatively. For the majority of the discs (66\%), $R_{out}$ =$0.5''$. However, we varied this number after visual inspections for very small binaries such as ODISEA\_C4\_94A \& B and massive discs such as ODISEA\_C4\_41, ODISEA\_C4\_143. The exact values are detailed in Table 1.  
Finally, for the convolution of the radial profiles into models, we use the output of the \texttt{imfit} task to get the beam details, such as the beam's PA, major, and minor axes.

Since \texttt{Frank} extracts the radial profiles assuming an axisymmetric structure, asymmetries leave signatures in the visibility data of the discs \footnote{A systematic analysis of asymmetries in the sample is left for future work.}. 
We also used a noiseless model due to computational limitations.
For the binaries in our sample, we attempt to obtain a radial profile with \texttt{Frank} for each companion. When the separation is large, one disc has little effect on the other (e.g., \cite{2024A&A...682A..55M}).{ We concatenate the R$_{max}$ values of the individual components to minimize contamination from the companion during visibility modeling. We acknowledge that the ideal approach is to subtract the visibility contribution of one component while modeling the other; however, such a visibility-domain decomposition is beyond the scope of this work.} 
{The error bars in the radial profiles include both the $1\sigma$ statistical uncertainty on the reconstructed brightness profile and the 15\% ALMA Band 8 absolute flux calibration uncertainty, added in quadrature. The statistical uncertainty is obtained by converting the log-space variance from the fit into linear brightness uncertainty as
\[
\sigma_I = \sqrt{\left(e^{\mathrm{variance}} - 1\right)e^{2\log(I)}} ,
\]
where $I$ is the fitted surface brightness profile at each radius.}

\subsection{Definition of inflection points, gaps, rings, and cavities}

The description of the evolutionary stages proposed by \cite{2021Cieza} and modeled by \cite{2025ApJ...984L..57O} was highly qualitative. 
With this large sample of ODISEA, we now develop a more systematic classification scheme based on cavities,  inflection points, and local maxima and minima in the \texttt{Frank} brightness profiles. We adopt the following analytical definitions.

Gaps are identified as local minima in the profiles, while rings are defined as the corresponding local maxima, similar to \cite{2021Cieza} and \cite{2020ApJ...891...48H}. The gap depth is quantified as the ratio of the intensities at the adjacent local minimum and maximum, and the gap width is measured at the mean intensity between them. Inflection points are identified as curvature sign changes in the radial intensity profile 
($d^2I/dr^2 = 0$) that coincide with a local extremum in the radial intensity gradient. 
{In practice, these features appear as shoulders or changes in slope in the radial profile, rather than as a well-defined local minimum--maximum pair. }
Curvature changes occurring between adjacent gap-ring pairs are excluded, as they result from the smooth monotonic transition between a gap minimum and the corresponding ring maximum and do not constitute independent disc substructures. Cavities are defined when the peak of the profile is offset from the disc center by at least 10\% (when the peak emission is not at the center of the disc). For discs with cavities, we additionally measure the radii where the intensity reaches 10\%, 50\%, and 90\% of the maximum, as done in \cite{2021Cieza}. 
{To focus on the most robust features, we only search for substructures over the radial range where the normalized intensity profile remains above an empirically chosen threshold. We adopt a threshold of 0.05 for single discs and 0.1 for binaries. These limits are selected after inspecting the full set of reconstructed profiles: below these levels, the profiles are dominated by low-level outer emission, noise, or reconstruction artifacts rather than clearly identifiable disc structure. The higher threshold for binaries is adopted because the visibility reconstruction is more susceptible to artifacts when emission from multiple components is present in the same field. We further restrict the analysis to 95\% of the maximum radial distance ($R_{95}$) and visually inspect all profiles, retaining only features that are consistently identifiable and not associated with obvious reconstruction artifacts.}
To minimize spurious detections of substructures, we visually inspect all profiles and only retain features considered to be robust. 

{{Uncertainties on the feature locations and radial distances were estimated through Monte Carlo perturbations of the normalized radial brightness profiles using the statistical uncertainty ($\sigma_I$). Gap-ring width uncertainties were obtained from the scatter in the midpoint radii between adjacent extrema. Gap-depth uncertainties were propagated from the profile uncertainties at the gap minimum and the reference adjacent ring. For cavities, the uncertainties on $R_{10}$, $R_{50}$, and $R_{90}$ were taken as half of the 16th--84th percentile range of the Monte Carlo distribution.}}

\section{Results}

\subsection{Identification of inflection points, gaps, rings, and cavities. }

From the analysis of the 101 \texttt{Frank} profiles, we have identified a total of 25 inflection points in 20 of our discs. There are also 17 discs with gap-ring pairs.  In these discs, we find 26 gap-ring pairs. Only 5 discs have multiple gap-ring pairs. Finally, we identified {15 discs with cavities.} These features are not mutually exclusive, as several targets show
different combinations of substructures. {All the feature details are provided in Table \ref{t:substructures}. The overall incidence of substructures in our Band~8 sample is broadly consistent with recent visibility-based surveys. In particular, \citet{2025ApJ...989....9V} detected dust substructures in 15 out of 30 AGE-PRO discs, including five systems with large inner dust cavities, and found that Ophiuchus Class~I discs show substructures in $\sim80\%$ of the resolved sources.}

\begin{table*}[htbp]
\caption{Properties of disc Substructures}
\label{t:substructures}
\centering
\resizebox{\textwidth}{!}{%
\begin{tabular}{llccccccc}
\hline
\hline
Target & Feature & Type & $R_{0}$ & Width & Depth & $R_{10}$ & $R_{50}$ & $R_{90}$ \\
       &         &      & (au)    & (au)  &       & (au)     & (au)     & (au)     \\
(1)    & (2)     & (3)  & (4)     & (5)   & (6)   & (7)      & (8)      & (9)      \\
\hline
ISO-Oph\_208 & D-22 & Gap & $22.45\pm0.51$ & $16.95\pm1.97$ & $0.07\pm0.03$ &  &  &  \\
ISO-Oph\_208 & B-33 & Ring & $32.57\pm0.49$ & $12.88\pm1.75$ &  &  &  &  \\
ODISEA\_C4\_016A$^c$ & B-11 & Ring & $10.80\pm0.60$ & $12.29\pm1.48$ &  & $0.18\pm0.00$ & $2.34\pm1.47$ & $7.94\pm1.39$ \\
ODISEA\_C4\_016B$^c$ & B-8 & Ring & $7.76\pm0.57$ & $8.82\pm1.37$ &  & $0.09\pm0.00$ & $0.09\pm0.09$ & $5.00\pm1.29$ \\
ODISEA\_C4\_043 & D-11 & Gap & $10.52\pm0.09$ & $10.31\pm1.42$ & $0.80\pm0.14$ &  &  &  \\
ODISEA\_C4\_043$^c$ & B-21 & Ring & $20.71\pm0.53$ & $9.76\pm1.18$ &  &  &  &  \\
ODISEA\_C4\_045$^c$ & B-21 & Ring & $20.76\pm1.81$ &  &  & $9.22\pm0.20$ & $12.78\pm0.55$ & $17.35\pm1.46$ \\
ODISEA\_C4\_065$^c$ & B-7 & Ring & $6.66\pm1.16$ &  &  & $0.18\pm0.00$ & $0.18\pm0.30$ & $4.57\pm1.05$ \\
ODISEA\_C4\_081$^c$ & B-13 & Ring & $12.69\pm1.27$ &  &  & $2.67\pm0.52$ & $6.22\pm0.47$ & $9.93\pm1.19$ \\
ODISEA\_C4\_104 & I-20 & Inflection & $20.67\pm0.16$ &  &  &  &  &  \\
\hline
\end{tabular}
}
\vspace{0.15cm}
\tablefoot{
The full table is available in electronic form
(1) Target name. The superscript \textit{c} denotes the cavity feature.
(2) Substructure label: gaps are denoted by ``D'', rings by ``B'', and inflection points by ``I'', followed by the feature location in au.
(3) Substructure type.
(4) Radial location of the feature.
(5) Width of the gap or ring feature.
(6) Gap depth, defined as $I_{\rm min}/I_{\rm max}$ for adjacent gap--ring pairs.
(7)--(9) Cavity radii $R_{10}$, $R_{50}$, and $R_{90}$, respectively. Blank entries denote quantities that are not relevant for the corresponding substructure.}

\end{table*}

\begin{figure*}[!h]
    \centering
    \includegraphics[trim={0.5cm 0cm 3cm 0cm},clip,width=0.9\textwidth, angle=270,angle=90]{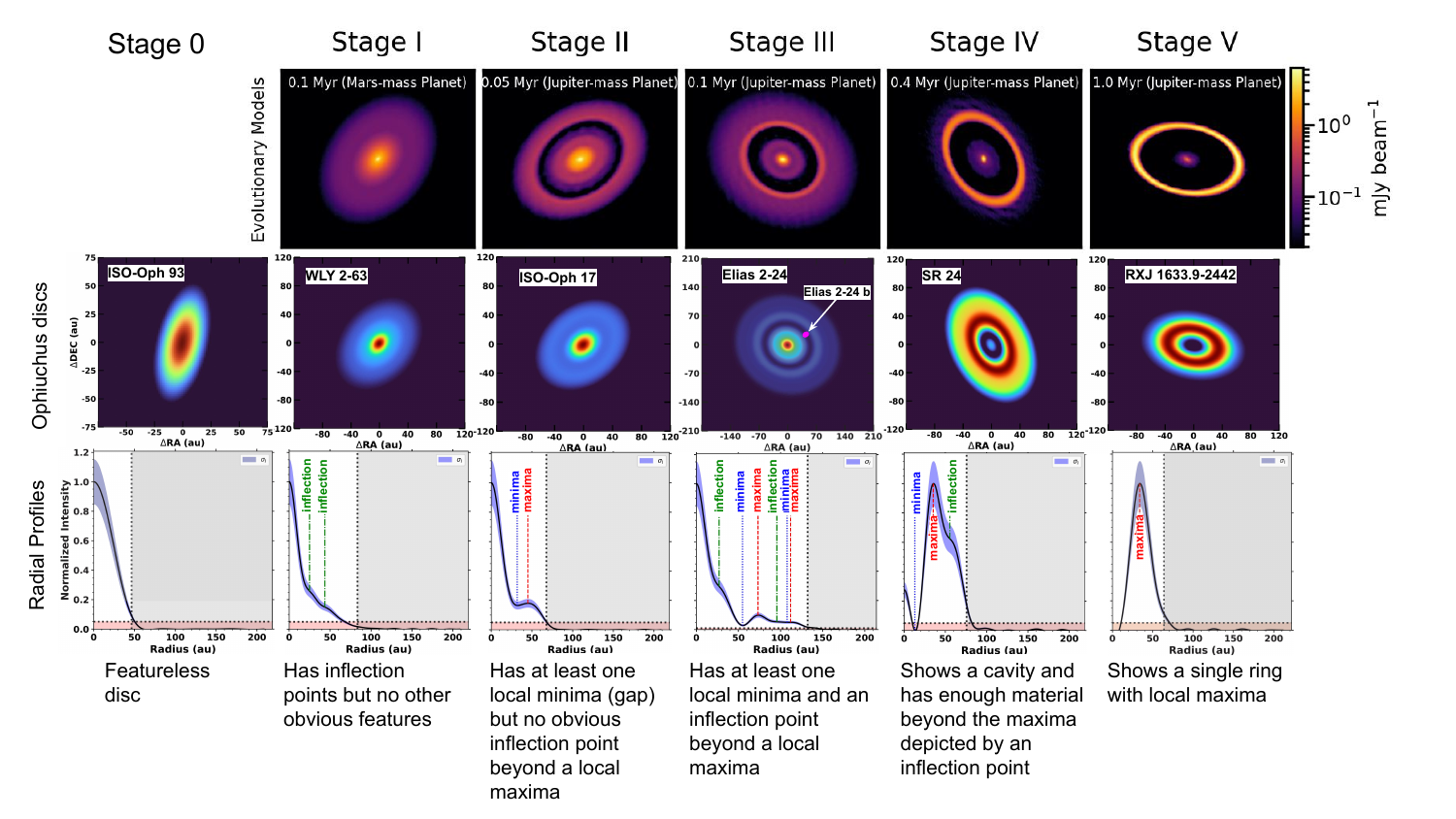}
    \caption{The top panel shows the progression of the disc evolutionary models into the different disc stages under the influence of a 1 M$_{Jup}$ planet \citep{2025ApJ...984L..57O}. The middle and bottom panels are the models and radial profiles of six evolutionary stages in which we have classified the $\sim$100 objects in our sample. We follow the Stage I to Stage V nomenclature introduced by \cite{2021Cieza}, but have now included a Stage 0 to describe discs looking truly featureless, lacking even inflection points.  
    The exact definitions we adopt, based on inflection points and local minima and maxima, are indicated. {The magenta oval in the Stage III panel of Elias 2-24b marks the location of the confirmed planet-mass object detected by \citet{2021AJ....161..146J} and Bernadi et al. (in press).}}
    \label{fig:stages}
\end{figure*}

\subsection{Classification into substructure stages}

In our sample, we have discs classified by their SED classes from \cite{2019Williams} into two categories, {young sources} (Class I and Flat Spectra, denoted as \textit{F}) and Class II objects. 
We further separate the discs into evolutionary stages following \cite{2021Cieza, 2025ApJ...984L..57O}, ranging from Stage I to Stage V, and introduce an additional Stage 0 to represent discs without detectable substructures (Fig.\ref{fig:stages}). 

{We clarify that the SED classes and the physically motivated structural stages are not equivalent classifications. The SED class is used here as a traditional, population-level proxy for the broader evolutionary state of the young stellar object, while the stages describe the morphology of the millimeter dust disc and are intended to trace the development of substructures associated with dust evolution and planet formation. These two evolutionary axes are expected to be broadly correlated statistically, since Class I/flat-spectrum sources are generally younger than Class II sources, but they need not agree for individual objects. In particular, inclination, extinction, and source confusion can affect the infrared SED, while the detectability of millimeter substructures depends on disc size, optical depth, signal-to-noise ratio (S/N), and angular resolution \citep[e.g.][]{2025ApJ...989....2R}. Moreover, our Band 8 data are continuum-only and therefore cannot independently identify envelope material or quantify cloud contamination through molecular-line diagnostics. We therefore use the SED classes only as a complementary population-level reference, while our primary disc-evolution interpretation is based on the structural stages.}

As mentioned above, these proposed stages were mostly qualitative in \cite{2021Cieza} and were meant to link the substructures seen in the brightest 15 discs in Ophiuchus to different stages of giant planet formation in the core accretion paradigm.   
We now implement them by adopting precise definitions based on the presence of cavities,  inflection points, local maxima, and minima, but keeping their original intent.

{The structural stages are physically motivated by the planet–disc interaction models of \cite{2025ApJ...984L..57O}, shown in the top row of Fig. 2, which reproduce a sequence of continuum morphologies as dust evolves in the presence of embedded planets. In this framework, smooth or weakly structured discs correspond to the earliest stages of detectable substructure development, while progressively deeper gaps, dust accumulation at pressure maxima, cavities, and narrow rings arise as the planet–disc system evolves. We therefore use the term “stage” to describe a proposed evolutionary ordering of disc morphologies.} 

In Stage I, the first hints of substructures appear in the form of inflection points, suggesting that a planetary core has already formed. 
The modeling by \cite{2025ApJ...984L..57O} suggests that a planetary core with the mass of Mars is enough to produce a detectable inflection point in the radial profile.  
Stage II is characterized by the presence of a gap (local minimum), although the disc immediately beyond the gap remains smooth, lacking an additional inflection point beyond the local maximum. 
In this Stage, the planetary cores are believed to become massive enough to carve detectable gaps.  
Stage III is characterized by an inflection point beyond the ring, indicating that significant dust accumulation has already occurred at the ring\footnote{The differences between Stages II and III are subtle and usually require very high-resolution observations to be identified.}.
This stage may correspond to the period soon after a giant planet has accreted its envelope, producing a significant pressure bump that acts as a dust trap at the edge of the gap. 
Based on the modeling by \cite{2025ApJ...984L..57O}, the bright ring at the edge of the gap seen in Elias 2-24, the prototypical object for this stage, can be explained by 0.1 Myr of dust accumulation after the formation of a 1 M$_{Jup}$ planet, but we note that the exact mass of the planet strongly depends on the modeling parameters, such as the viscosity of the disc. 
{This scenario is consistent with the direct detection of Elias 2-24 b in the near-IR \citep[][Bernadi et al. in press]{2021AJ....161..146J}}.
In Stage IV, we see the development of a cavity, while the outer disc still contains substantial material beyond an inflection point. 
In the proposed scenario, the development of the mm cavity is a natural consequence of the dust filtration effect of the giant planet, which only allows micron-sized particles to cross the gap.  
By Stage V, dust in the outer disc has drifted significantly, accumulating all mm dust in a single ring.
The above sequence describes the evolution of disc structures driven by the formation of a single giant planet. 
{The fact that it works very well to classify our entire sample (with only four discs as exceptions, e.g., objects showing multiple gaps) is noteworthy and has implications for the architectures of planetary systems expected in Ophiuchus (see Sections \ref{sec:stageII},\ref{sec:stageIII} and \ref{sec:stageIV})}.     

\subsubsection{Stage 0}
\label{sec:stage0}
The newly introduced Stage 0 is  defined by its smooth \texttt{Frank} 
profiles. In most cases,  the objects also look featureless in the corresponding  \texttt{tclean} images. A few exceptions, however, are observed in sources such as ODISEA\_C4\_107 (Fig~\ref{fig:0+I_app}), ODISEA\_C4\_144 (Fig~\ref{fig:0+I_app}), ODISEA\_C4\_071 (Fig~\ref{fig:0+I_app}),  ODISEA\_C4\_060 (Fig~\ref{fig:0+II_app}), ISO-Oph\_123 (Fig~\ref{fig:0+II_app}), and ISO-Oph\_13 (Fig~\ref{fig:0+II_app}). These discs show faint {axisymmetrical} substructures in the image plane that are likely poorly resolved due to limited angular resolution {and asymmetries}, which {prevent} \texttt{Frank} from reproducing these signatures.

A notable case is the very close binary ODISEA\_C4\_94 (Fig.~\ref{fig:0+I_app}), where the separation between the primary ODISEA\_C4\_94A and the secondary ODISEA\_C4\_94B is comparable to our resolution limit ($\sim$7 au, 0.05''). \texttt{Frank} successfully produced a radial profile for both sources by limiting the $R_{out}$ to 0.07'' and 0.05'', respectively.

\texttt{Frank} fitting did not encounter difficulties with other binaries, even in cases with very faint companions (<9 mJy at 410 GHz), such as ODISEA\_C4\_134B, ODISEA\_C4\_53B, ODISEA\_C4\_125B, ODISEA\_C4\_050B, and ODISEA\_C4\_106B 
(Fig.~\ref{fig:0+I_app},\ref{fig:0+II_app}).

\noindent
\begin{minipage}{.49\textwidth}
	 \centering
	 	 \hrulesep
	 	 \includegraphics[width=1\linewidth]{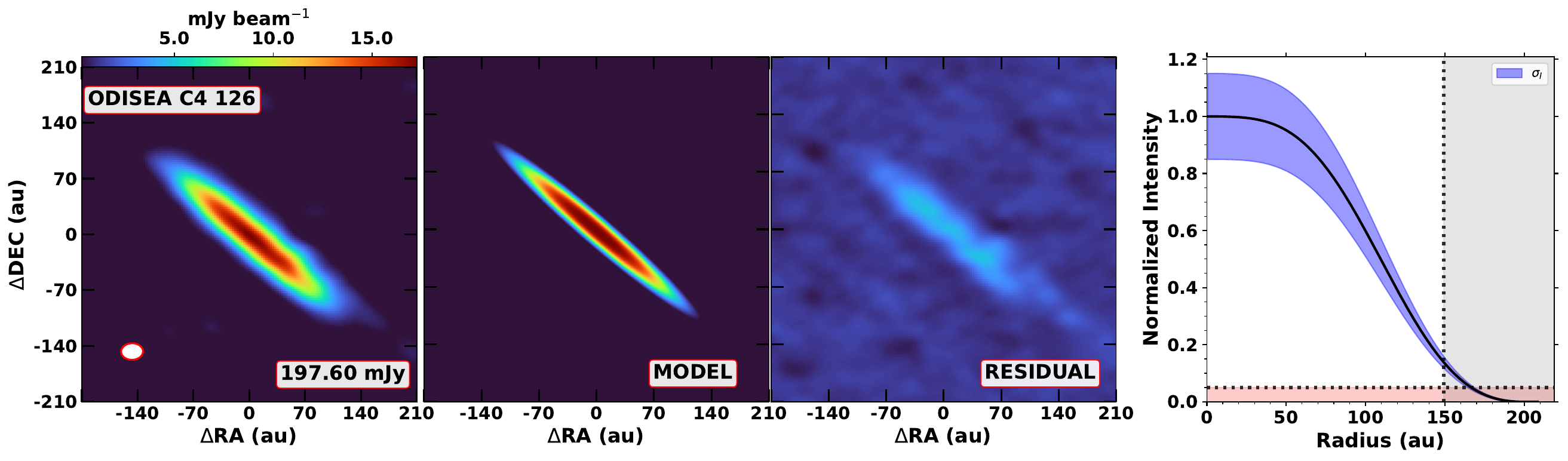}
\end{minipage}%
\vrulesep
\noindent
\begin{minipage}{.49\textwidth}
	 \centering
	 	 \hrulesep
	 	 \includegraphics[width=1\linewidth]{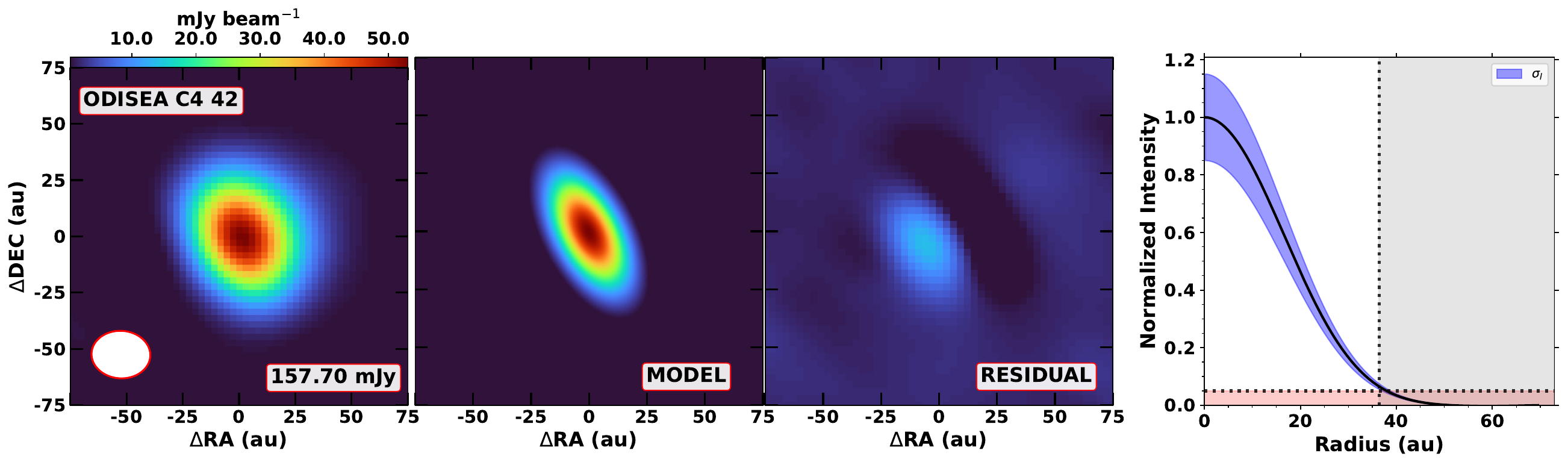}
\end{minipage}%
\vrulesep

\captionof{figure}{
Stage 0 disks in embedded sources (Class I/F). 
{Horizontally, the first images in each panel are those created by \texttt{tclean}; the second, third, and fourth images are the models, residuals, and 1D radial profiles created by \texttt{Frank}. The colorbar indicates the flux scale for both \texttt{tclean} image and the residual.} Only the two brightest sources in this category are shown. The rest are presented in the Appendix.}
\label{fig:0+I}
\vspace{0.8cm}%
\noindent
\begin{minipage}{.49\textwidth}
	 \centering
	 	 \hrulesep
	 	 \includegraphics[width=1\linewidth]{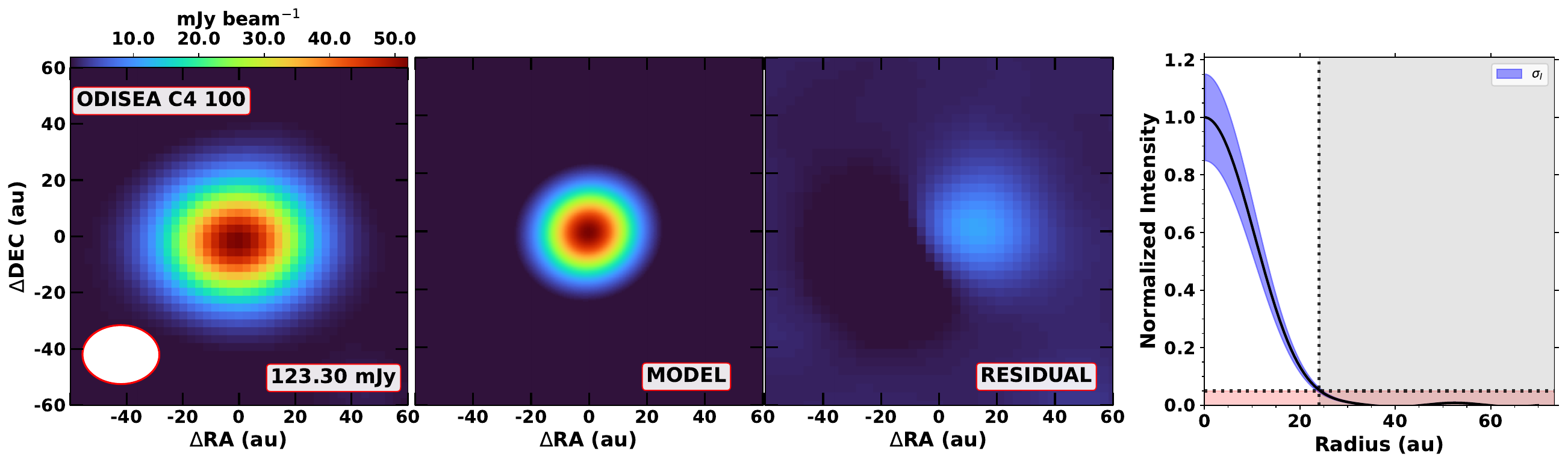}
\end{minipage}%
\vrulesep
\noindent
\begin{minipage}{.49\textwidth}
	 \centering
	 	 \hrulesep
	 	 \includegraphics[width=1\linewidth]{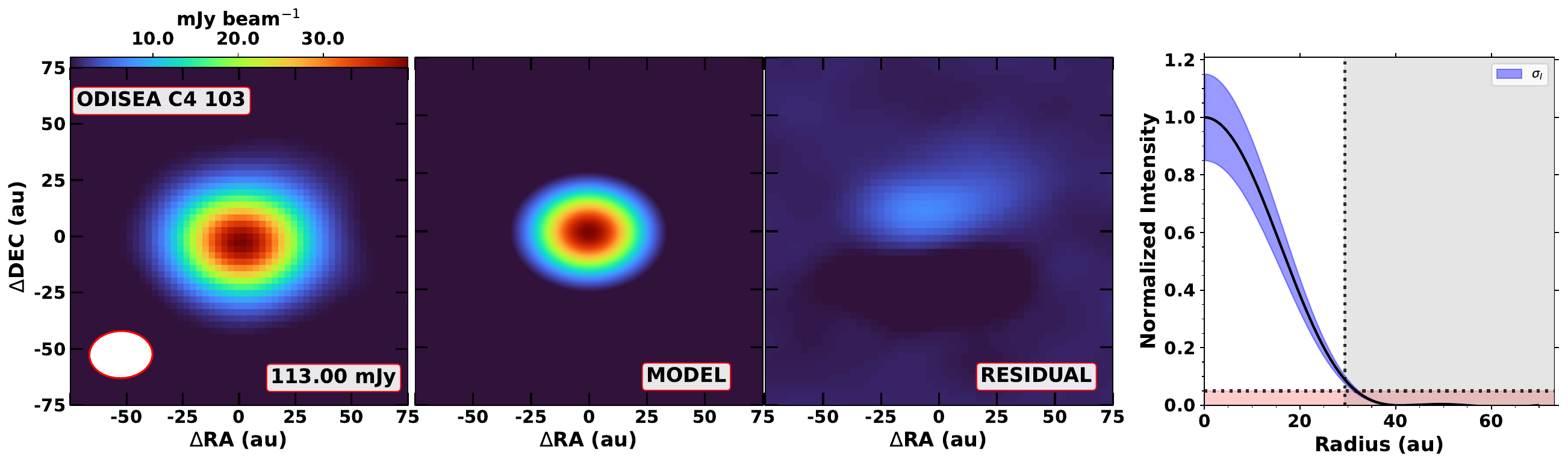}
\end{minipage}%
\vrulesep

\captionof{figure}{
Stage 0 disks in  Class II sources. {Horizontally, the first images in each panel are those created by \texttt{tclean}; the second, third, and fourth images are the models, residuals, and 1D radial profiles created by \texttt{Frank}. The colorbar indicates the flux scale for both \texttt{tclean} image and the residual.} Only the two brightest sources in this category are shown. The rest are presented in the Appendix.}
\label{fig:0+II}
\vspace{0.8cm}%
\subsubsection{Stage I}
\label{sec:stageI}
For most Stage I discs, no structures are visually apparent in the image plane alone. However, fitting the visibilities with \texttt{Frank} reveals clear inflection points in both the radial profiles and the corresponding models (Fig~\ref{fig:1+I},~\ref{fig:1+I_app},\ref{fig:1+II}). The only exception is ODISEA\_C4\_3, which shows an asymmetric extended structure in the image plane; this feature is reflected in the radial profile as an inflection point at 95 au.    

\noindent
\begin{minipage}{.49\textwidth}
	 \centering
	 	 \hrulesep
	 	 \includegraphics[width=1\linewidth]{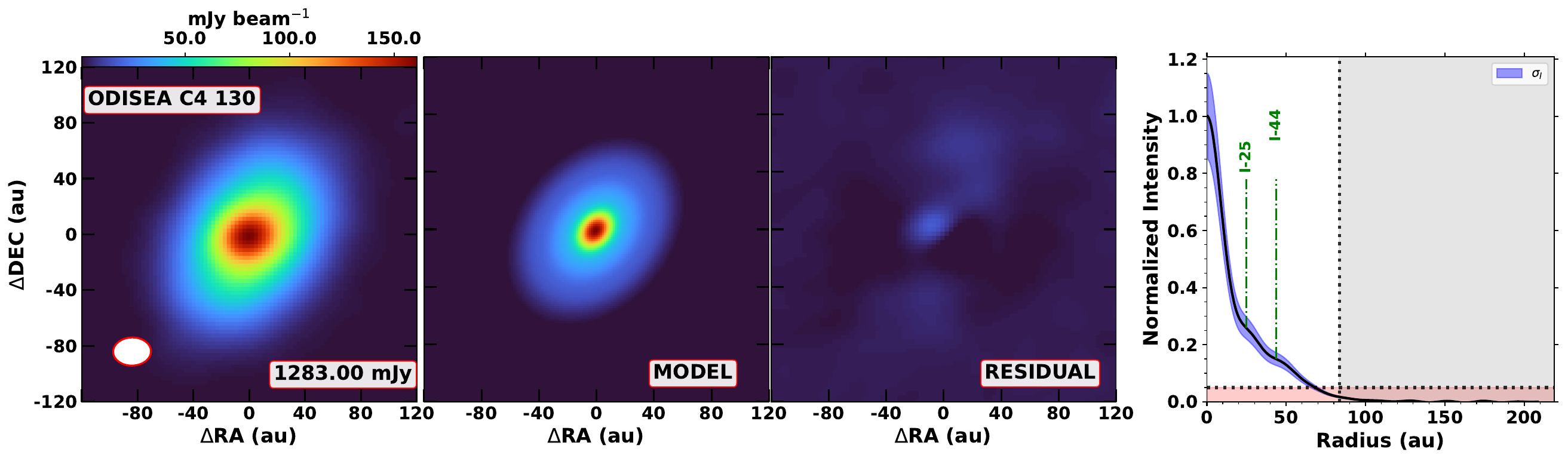}
\end{minipage}%
\vrulesep
\noindent
\begin{minipage}{.49\textwidth}
	 \centering
	 	 \hrulesep
	 	 \includegraphics[width=1\linewidth]{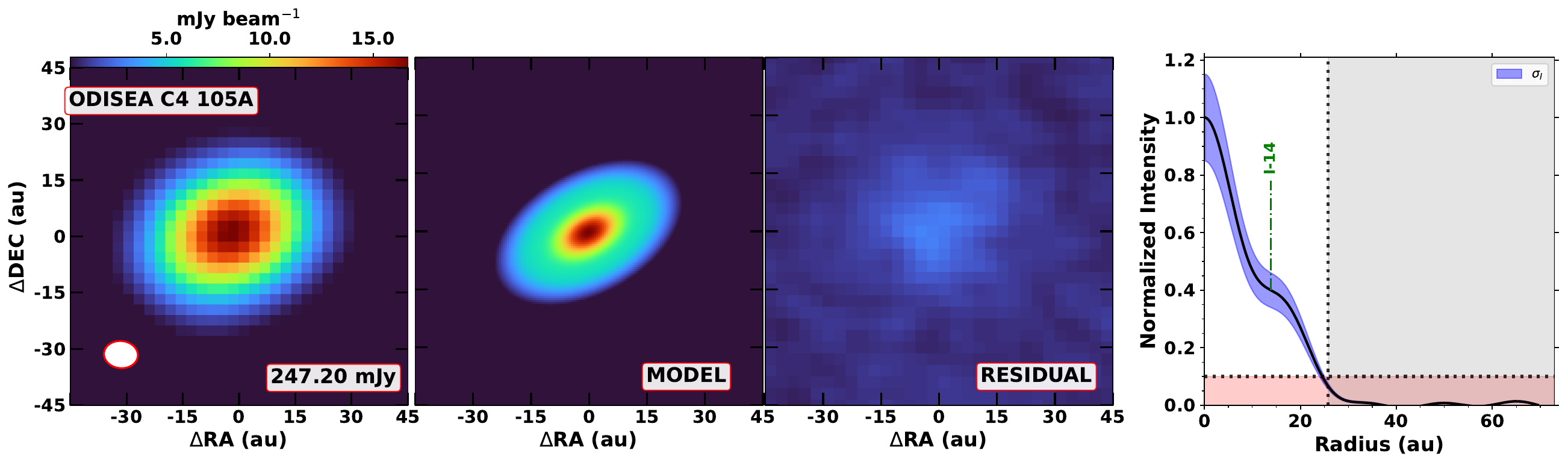}
\end{minipage}%
\vrulesep

\captionof{figure}{Stage I disks in embedded sources (Class I/F). {Horizontally, the first images in each panel are those created by \texttt{tclean}; the second, third, and fourth images are the models, residuals, and 1D radial profiles created by \texttt{Frank}. The colorbar indicates the flux scale for both \texttt{tclean} image and the residual.}}
\label{fig:1+I}
\vspace{0.8cm}
\noindent
\begin{minipage}{.49\textwidth}
	 \centering
	 	 \hrulesep
	 	 \includegraphics[width=1\linewidth]{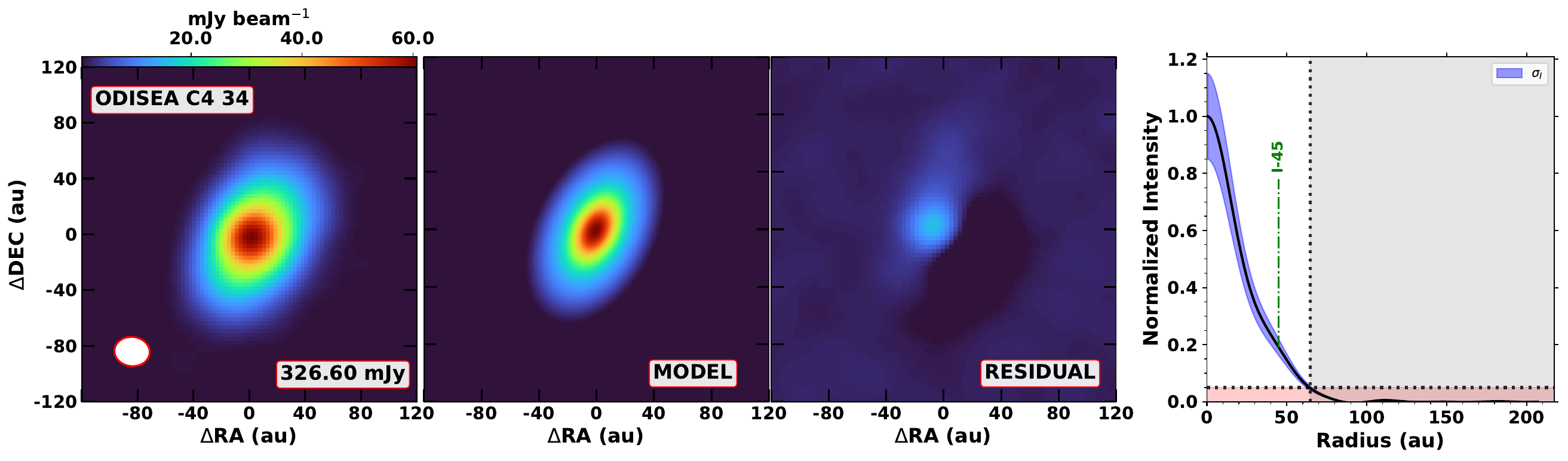}
\end{minipage}%
\vrulesep
\noindent
\begin{minipage}{.49\textwidth}
	 \centering
	 	 \hrulesep
	 	 \includegraphics[width=1\linewidth]{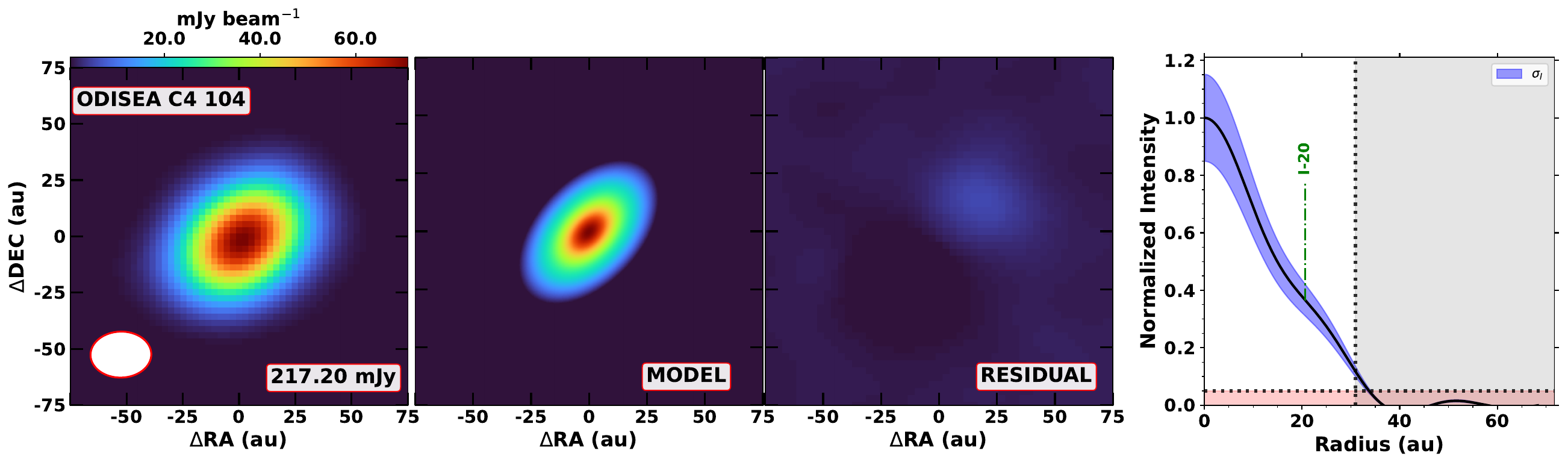}
\end{minipage}%
\vrulesep
\noindent
\begin{minipage}{.49\textwidth}
	 \centering
	 	 \hrulesep
	 	 \includegraphics[width=1\linewidth]{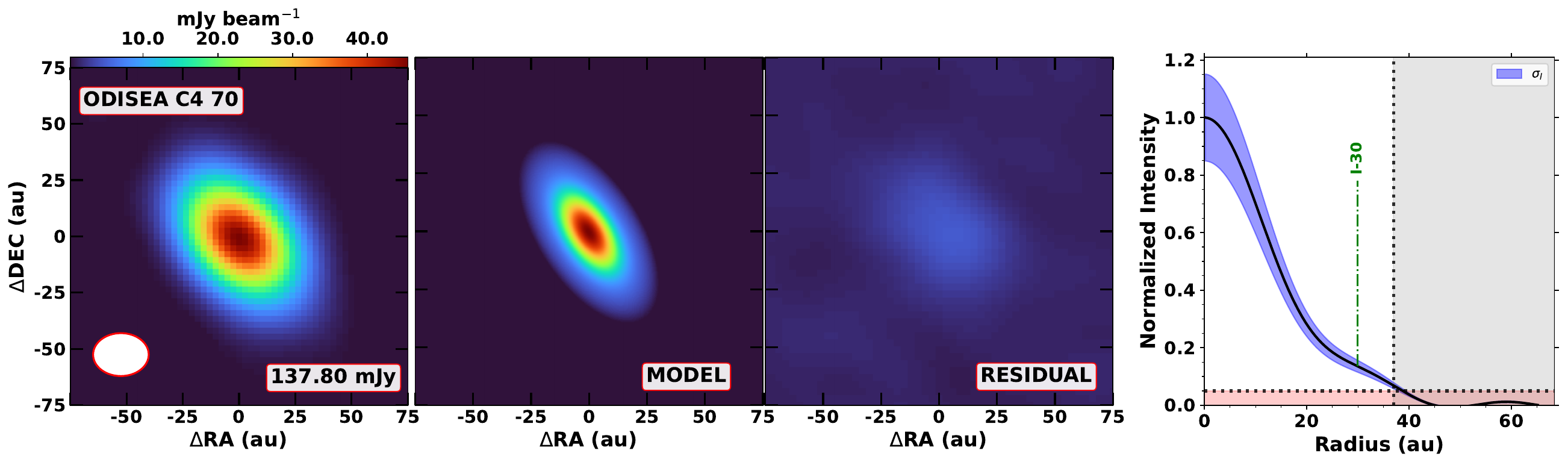}
\end{minipage}%
\vrulesep
\captionof{figure}{Stage I disks in  Class II sources.
{Horizontally, the first images in each panel are those created by \texttt{tclean}; the second, third, and fourth images are the models, residuals, and 1D radial profiles created by \texttt{Frank}. The colorbar indicates the flux scale for both \texttt{tclean} image and the residual.}}
\label{fig:1+II}
\vspace{0.8cm}

\subsubsection{Stage II}
\label{sec:stageII}

In Stage II, the radial profiles and \texttt{Frank} models clearly reveal gap-ring pairs. These structures are also evident in most of the \texttt{tclean} images, except for ODISEA\_C4\_83 (Fig~\ref{fig:2+I}), and ODISEA\_C4\_27 (Fig~\ref{fig:2+II_app}). 

\noindent
\begin{minipage}{.49\textwidth}
	 \centering
	 	 \hrulesep
	 	 \includegraphics[width=1\linewidth]{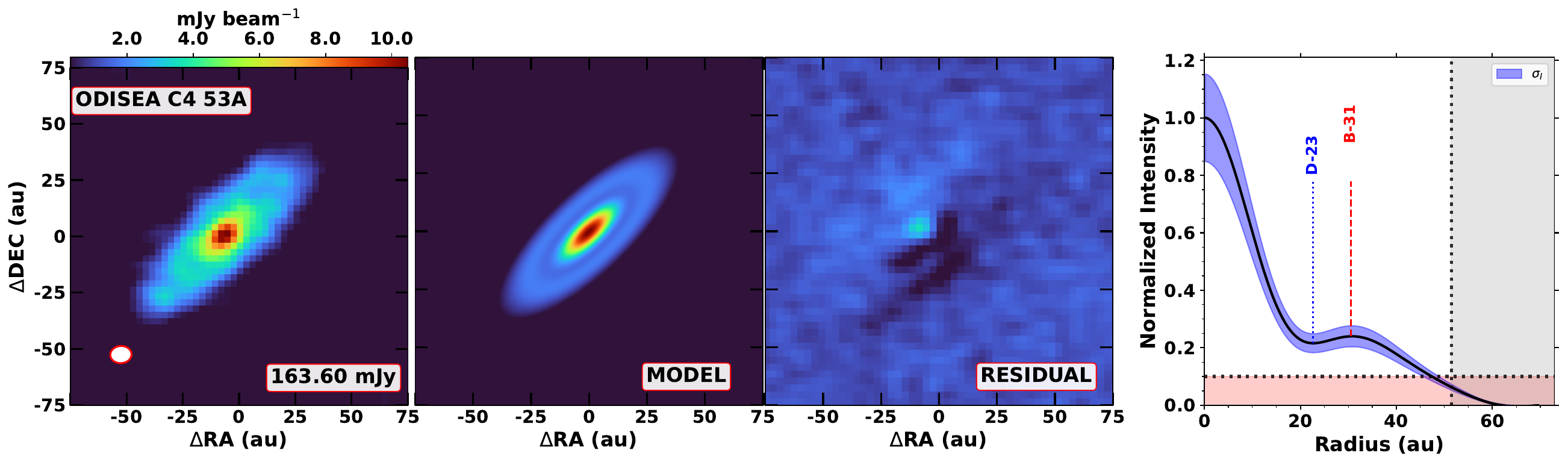}
\end{minipage}%
\vrulesep
\noindent
\begin{minipage}{.49\textwidth}
	 \centering
	 	 \hrulesep
	 	 \includegraphics[width=1\linewidth]{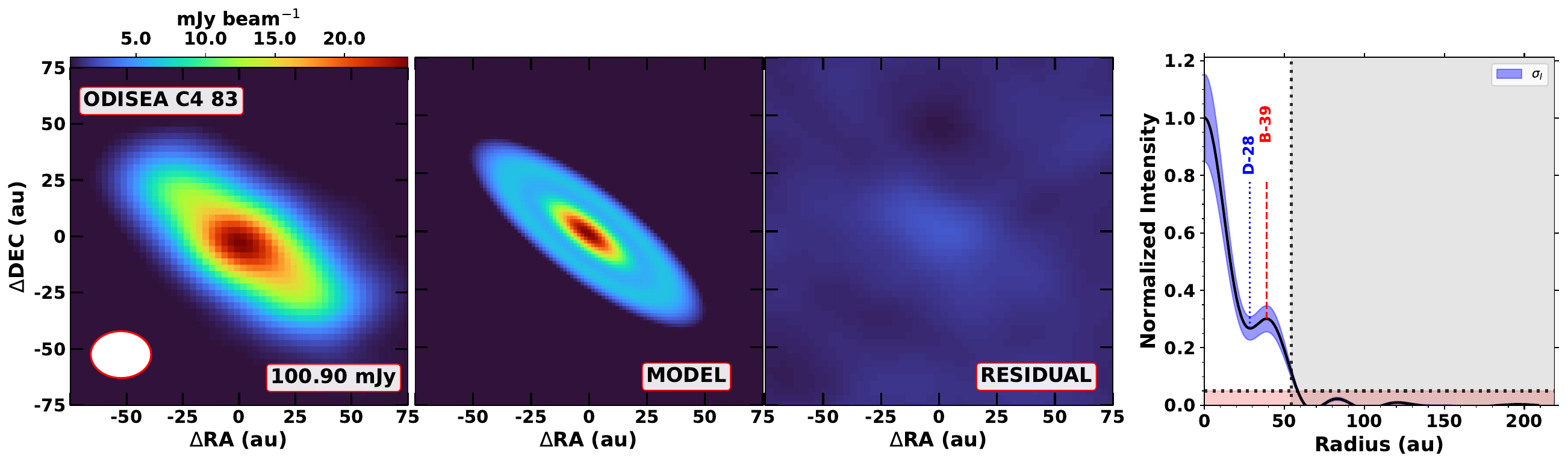}
\end{minipage}%
\vrulesep
\noindent
\begin{minipage}{.49\textwidth}
	 \centering
	 	 \hrulesep
	 	 \includegraphics[width=1\linewidth]{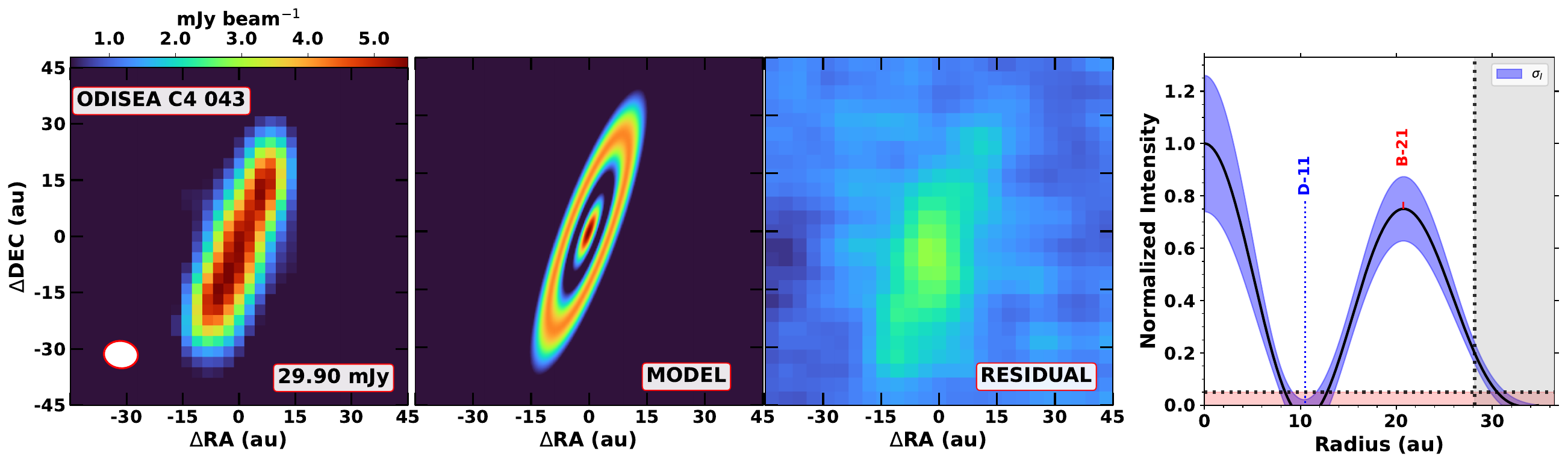}
\end{minipage}%
\vrulesep
\captionof{figure}{Stage II disks in embedded sources (Class I/F). {Horizontally, the first images in each panel are those created by \texttt{tclean}; the second, third, and fourth images are the models, residuals, and 1D radial profiles created by \texttt{Frank}. The colorbar indicates the flux scale for both \texttt{tclean} image and the residual.}}
\label{fig:2+I}
\vspace{0.8cm}
\noindent
\begin{minipage}{.49\textwidth}
	 \centering
	 	 \hrulesep
	 	 \includegraphics[width=1\linewidth]{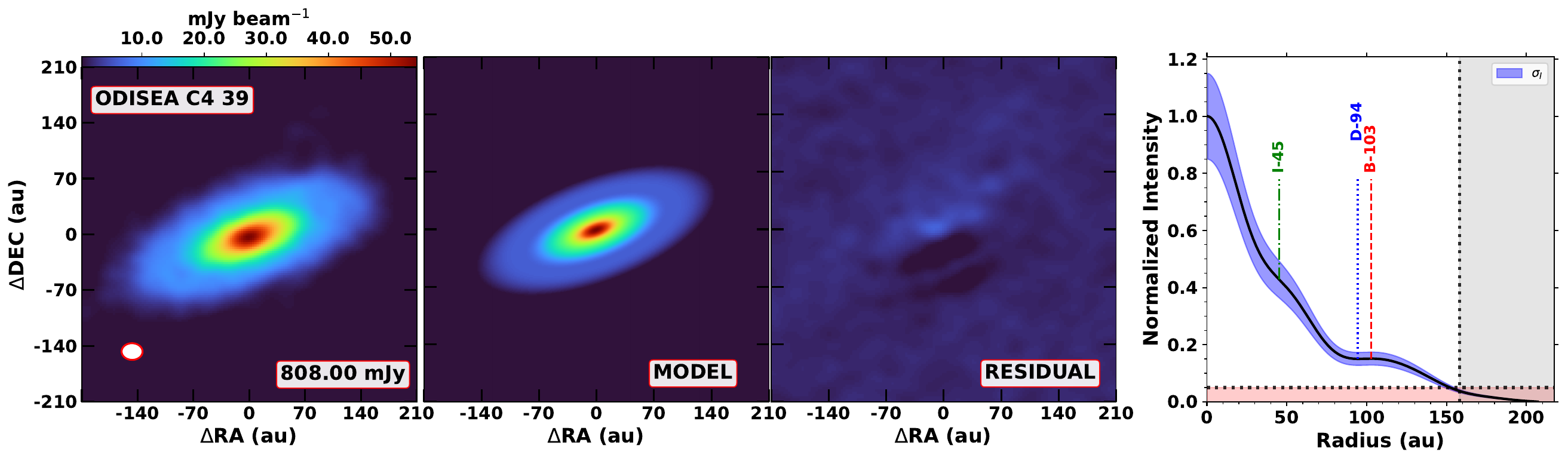}
\end{minipage}%
\vrulesep
\noindent
\begin{minipage}{.49\textwidth}
	 \centering
	 	 \hrulesep
	 	 \includegraphics[width=1\linewidth]{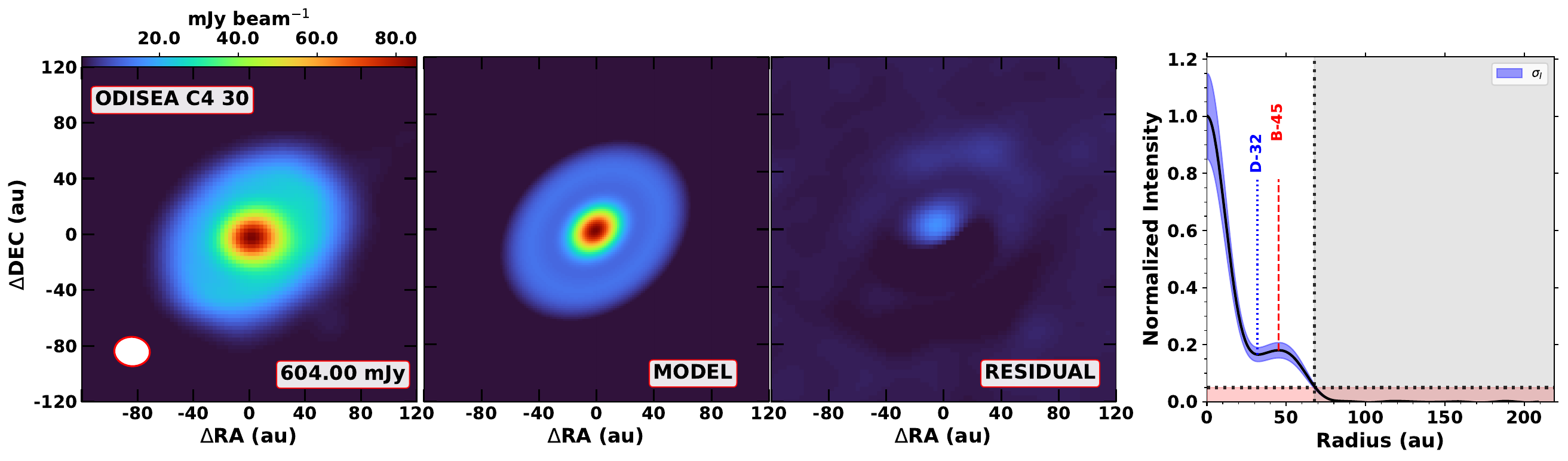}
\end{minipage}%
\vrulesep
\captionof{figure}{Stage II disks in Class II sources.
{Horizontally, the first images in each panel are those created by \texttt{tclean}; the second, third, and fourth images are the models, residuals, and 1D radial profiles created by \texttt{Frank}. The colorbar indicates the flux scale for both \texttt{tclean} image and the residual. Only the two brightest sources in this category are shown. The rest are presented in the Appendix.}}
\label{fig:2+II}
\vspace{0.8cm}

\subsubsection{Stage III}
\label{sec:stageIII}
Stage III has only five objects, where we see an inflection point after a gap-ring pair. The gap-ring pair is not visible in the \texttt{tclean} image in ODISEA\_C4\_38 (Fig~\ref{fig:3+I}). Compared to other discs in this stage (Fig~\ref{fig:3+I},~\ref{fig:3+II}), the radial profile of ODISEA\_C4\_38 shows little material beyond the gap–ring pair; however, the hint of an inflection point leads us to classify this target as Stage III.
We note that Stage III is very subtle and that its unambiguous identification requires very
 high {S/N} is the case for Elias 2-24, which is the prototype.   
\noindent
\begin{minipage}{.49\textwidth}
	 \centering
	 	 \hrulesep
	 	 \includegraphics[width=1\linewidth]{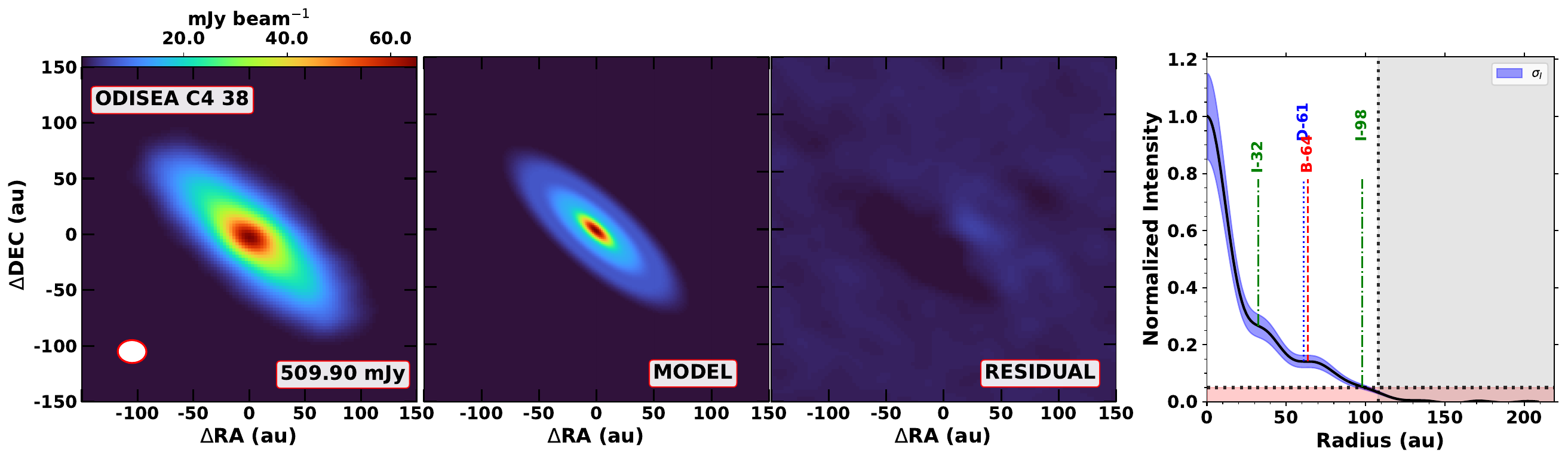}
\end{minipage}%
\vrulesep
\noindent
\begin{minipage}{.49\textwidth}
	 \centering
	 	 \hrulesep
	 	 \includegraphics[width=1\linewidth]{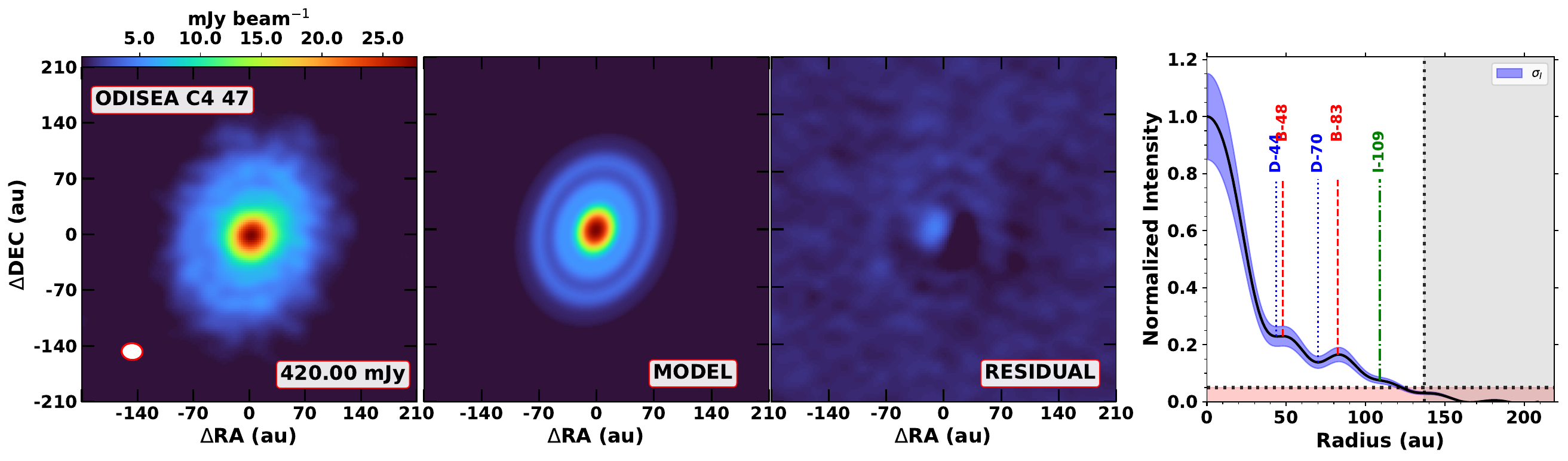}
\end{minipage}%
\vrulesep
\captionof{figure}{Stage III disks in Class I/F sources.
{Horizontally, the first images in each panel are those created by \texttt{tclean}; the second, third, and fourth images are the models, residuals, and 1D radial profiles created by \texttt{Frank}. The colorbar indicates the flux scale for both \texttt{tclean} image and the residual.}}
\label{fig:3+I}
\vspace{0.8cm}
\noindent
\begin{minipage}{.49\textwidth}
	 \centering
	 	 \hrulesep
	 	 \includegraphics[width=1\linewidth]{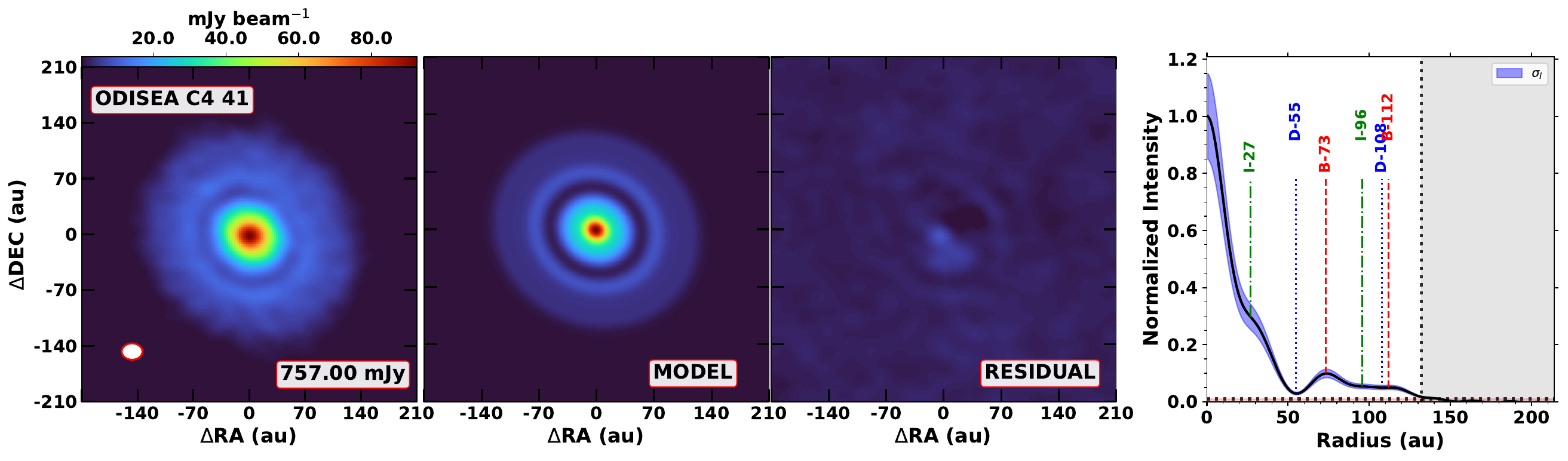}
\end{minipage}%
\vrulesep
\noindent
\begin{minipage}{.49\textwidth}
	 \centering
	 	 \hrulesep
	 	 \includegraphics[width=1\linewidth]{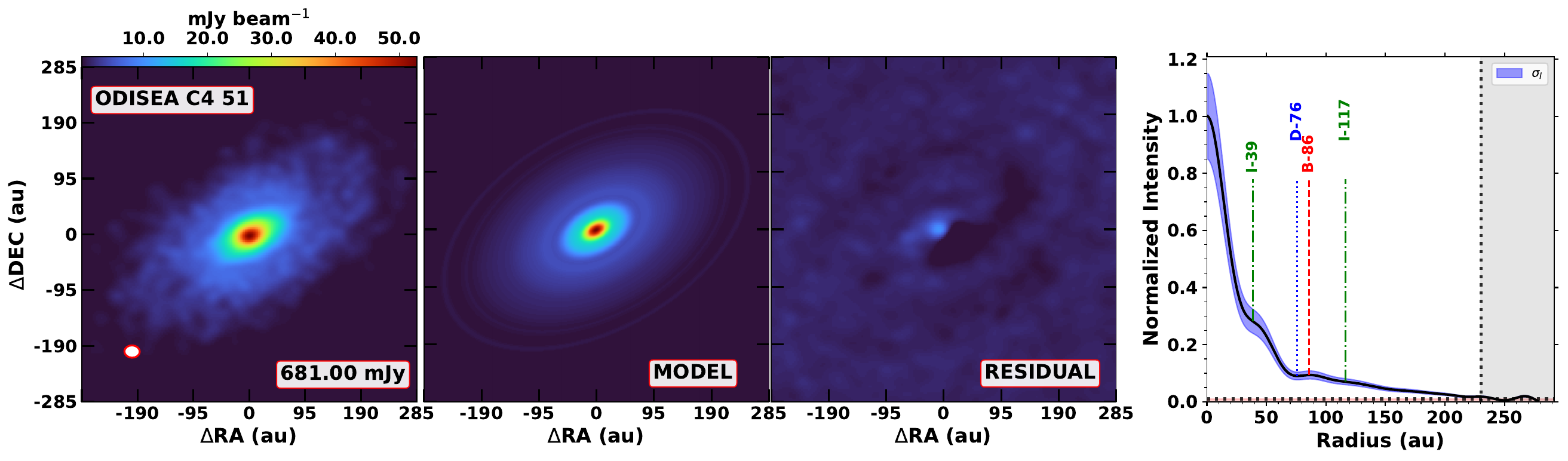}
\end{minipage}%
\vrulesep
\noindent
\begin{minipage}{.49\textwidth}
	 \centering
	 	 \hrulesep
	 	 \includegraphics[width=1\linewidth]{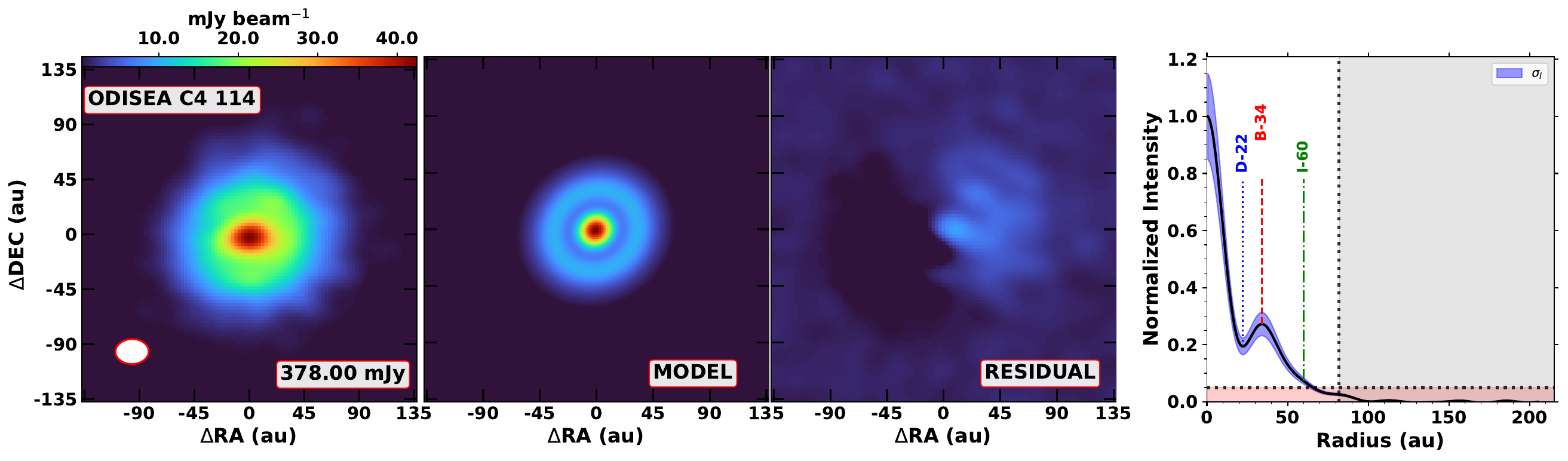}
\end{minipage}%
\vrulesep
\captionof{figure}{Stage III disks in Class II sources.
{Horizontally, the first images in each panel are those created by \texttt{tclean}; the second, third, and fourth images are the models, residuals, and 1D radial profiles created by \texttt{Frank}. The colorbar indicates the flux scale for both \texttt{tclean} image and the residual.}}
\label{fig:3+II}
\vspace{0.8cm}

\subsubsection{Stage IV}
\label{sec:stageIV}
This stage includes only three Class II objects (Fig.~\ref{fig:4+II}). The discs show a central cavity accompanied by inflection points at larger radii, implying the presence of significant material at large distances and leaving room for the formation of additional planets before the disc evolves into a single-ring morphology.

\noindent
\begin{minipage}{.49\textwidth}
	 \centering
	 	 \hrulesep
	 	 \includegraphics[width=1\linewidth]{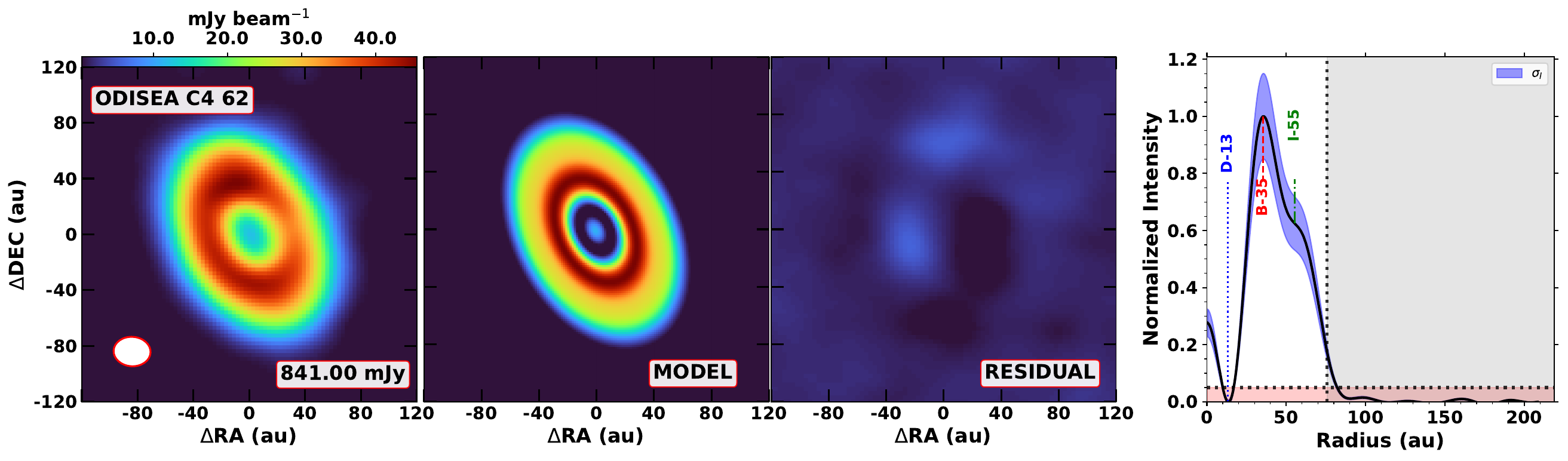}
\end{minipage}%
\vrulesep
\noindent
\begin{minipage}{.49\textwidth}
	 \centering
	 	 \hrulesep
	 	 \includegraphics[width=1\linewidth]{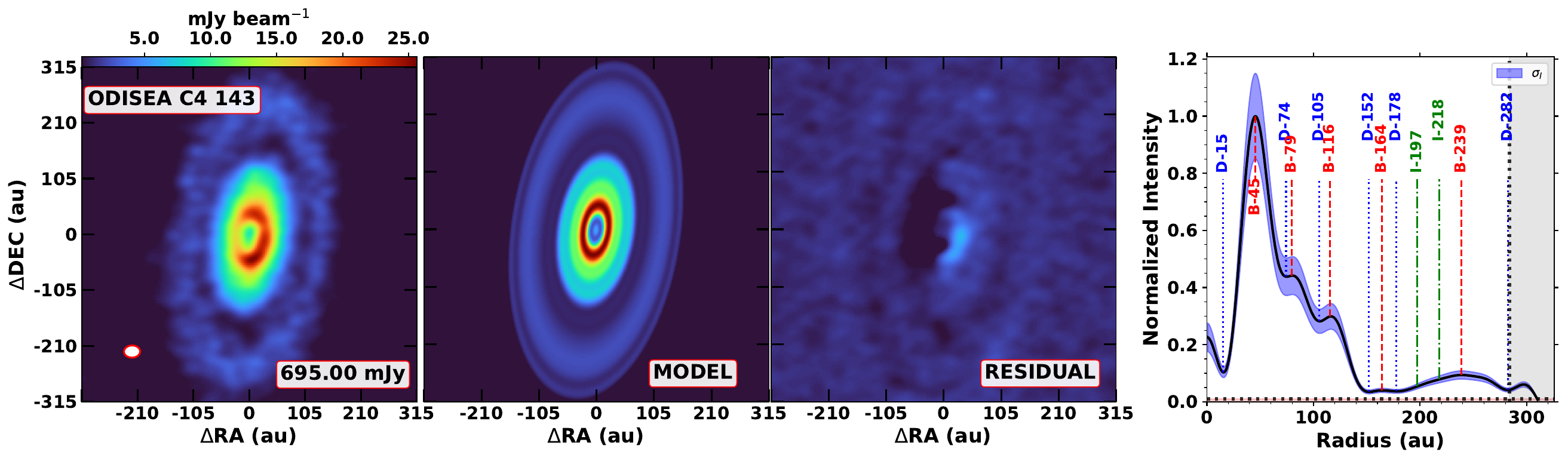}
\end{minipage}%
\vrulesep
\noindent
\begin{minipage}{.49\textwidth}
	 \centering
	 	 \hrulesep
	 	 \includegraphics[width=1\linewidth]{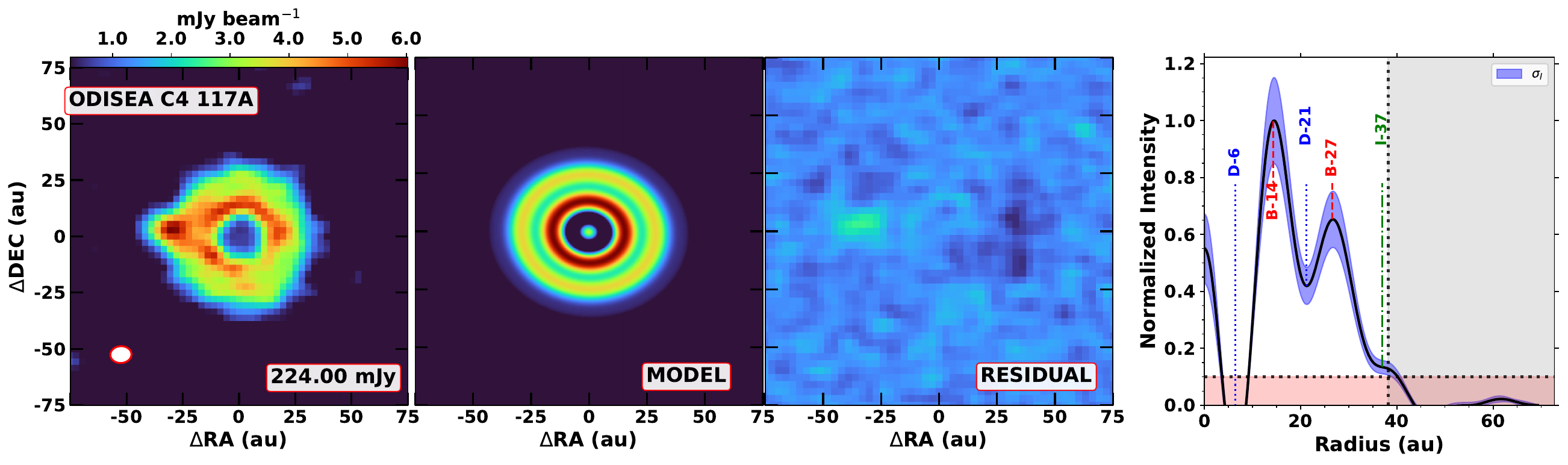}
\end{minipage}%
\vrulesep
\captionof{figure}{Stage IV disks in Class II sources.
{Horizontally, the first images in each panel are those created by \texttt{tclean}; the second, third, and fourth images are the models, residuals, and 1D radial profiles created by \texttt{Frank}. The colorbar indicates the flux scale for both \texttt{tclean} image and the residual.}}
\label{fig:4+II}
\vspace{0.8cm}

\subsubsection{Stage V}
\label{sec:stageV}
In the final stage, the majority of discs display a clear cavity followed by a single ring, which is evident in the image plane, the \texttt{Frank} models, and the radial profiles (Fig.~\ref{fig:5+I},~\ref{fig:5+I_app},~\ref{fig:5+II},~\ref{fig:5+II_app}). The exceptions are ODISEA\_C4\_68 (Fig.~\ref{fig:5+I}) and ODISEA\_C4\_75 (Fig.~\ref{fig:5+II_app}), where the cavities are hinted but not fully resolved in the image plane; however, in the visibility plane, the \texttt{Frank} models clearly reveal their presence.  

\noindent
\begin{minipage}{.49\textwidth}
	 \centering
	 	 \hrulesep
	 	 \includegraphics[width=1\linewidth]{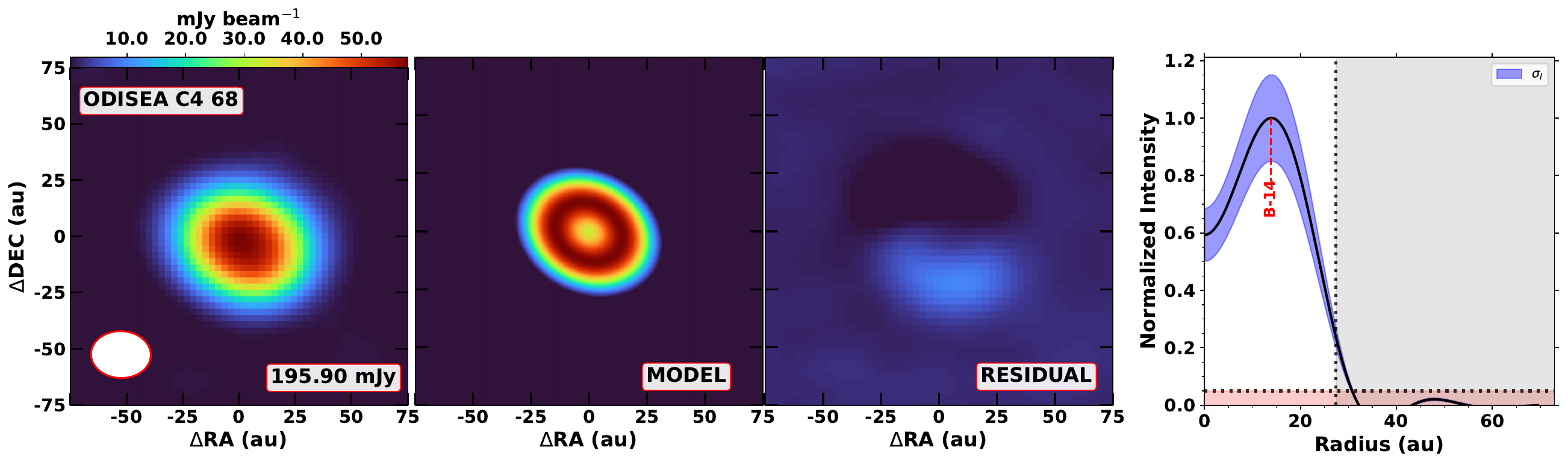}
\end{minipage}%
\vrulesep
\noindent
\begin{minipage}{.49\textwidth}
	 \centering
	 	 \hrulesep
	 	 \includegraphics[width=1\linewidth]{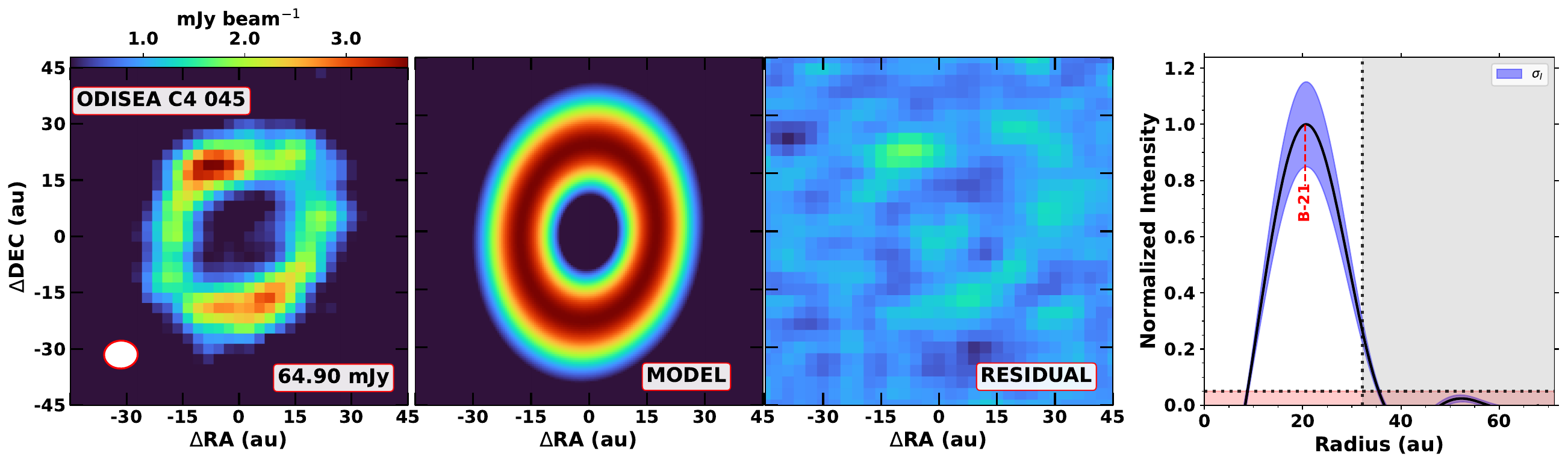}
\end{minipage}%
\vrulesep

\captionof{figure}{Stage V disks in embedded sources (Class I/F). {Horizontally, the first images in each panel are those created by \texttt{tclean}; the second, third, and fourth images are the models, residuals, and 1D radial profiles created by \texttt{Frank}. The colorbar indicates the flux scale for both \texttt{tclean} image and the residual. Only the two brightest sources in this category are shown. The rest are presented in the Appendix.}}
\label{fig:5+I}
\vspace{0.8cm}
\noindent
\begin{minipage}{.49\textwidth}
	 \centering
	 	 \hrulesep
	 	 \includegraphics[width=1\linewidth]{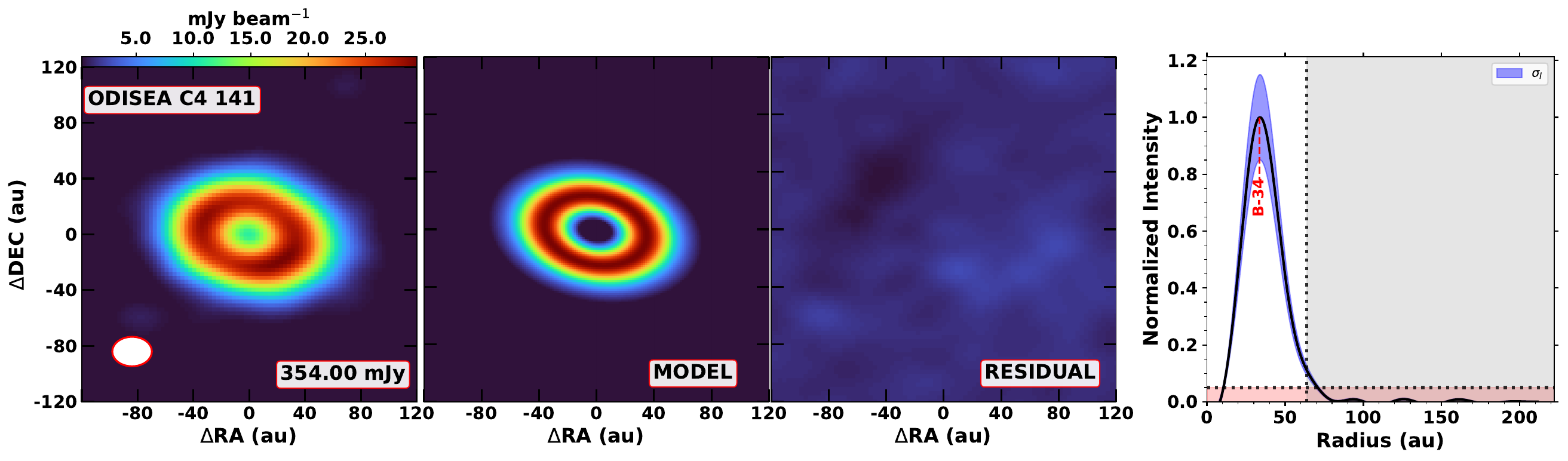}
\end{minipage}%
\vrulesep
\noindent
\begin{minipage}{.49\textwidth}
	 \centering
	 	 \hrulesep
	 	 \includegraphics[width=1\linewidth]{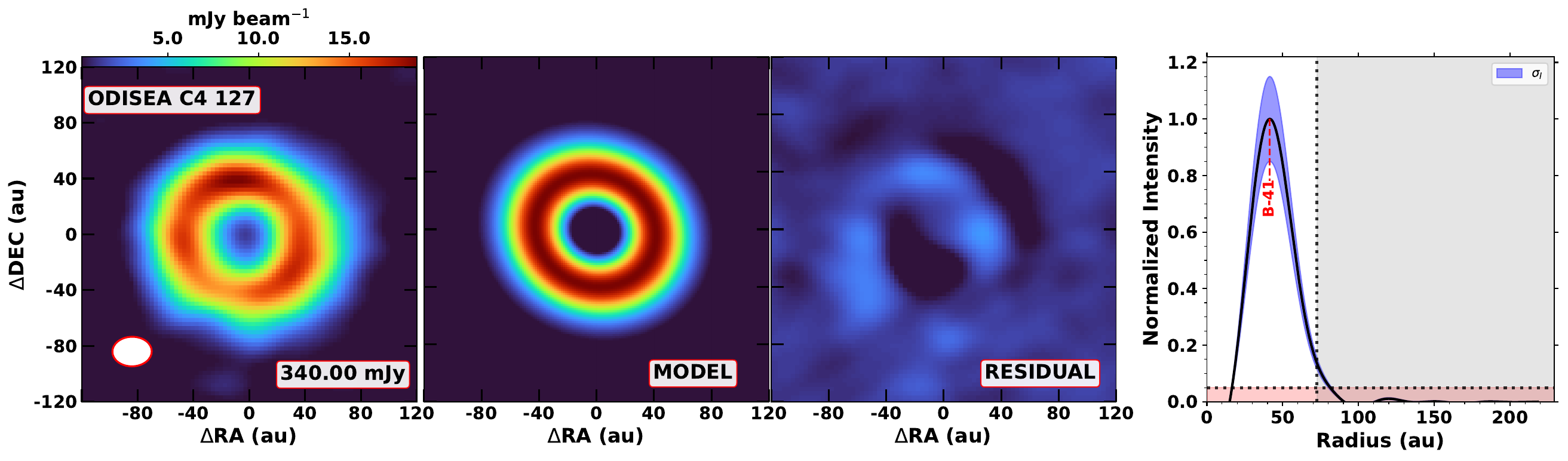}
\end{minipage}%
\vrulesep

\captionof{figure}{Stage V disks in Class II sources.
{Horizontally, the first images in each panel are those created by \texttt{tclean}; the second, third, and fourth images are the models, residuals, and 1D radial profiles created by \texttt{Frank}. The colorbar indicates the flux scale for both \texttt{tclean} image and the residual. Only the two brightest sources in this category are shown. The rest are presented in the Appendix.}}
\label{fig:5+II}
\vspace{0.8cm}

 \subsection{Caveats of resolution achieved by \texttt{Frank} models.}
\label{sec:caveates_resFRANK}

\begin{figure*}
    \centering
    \includegraphics[width=0.49\linewidth]{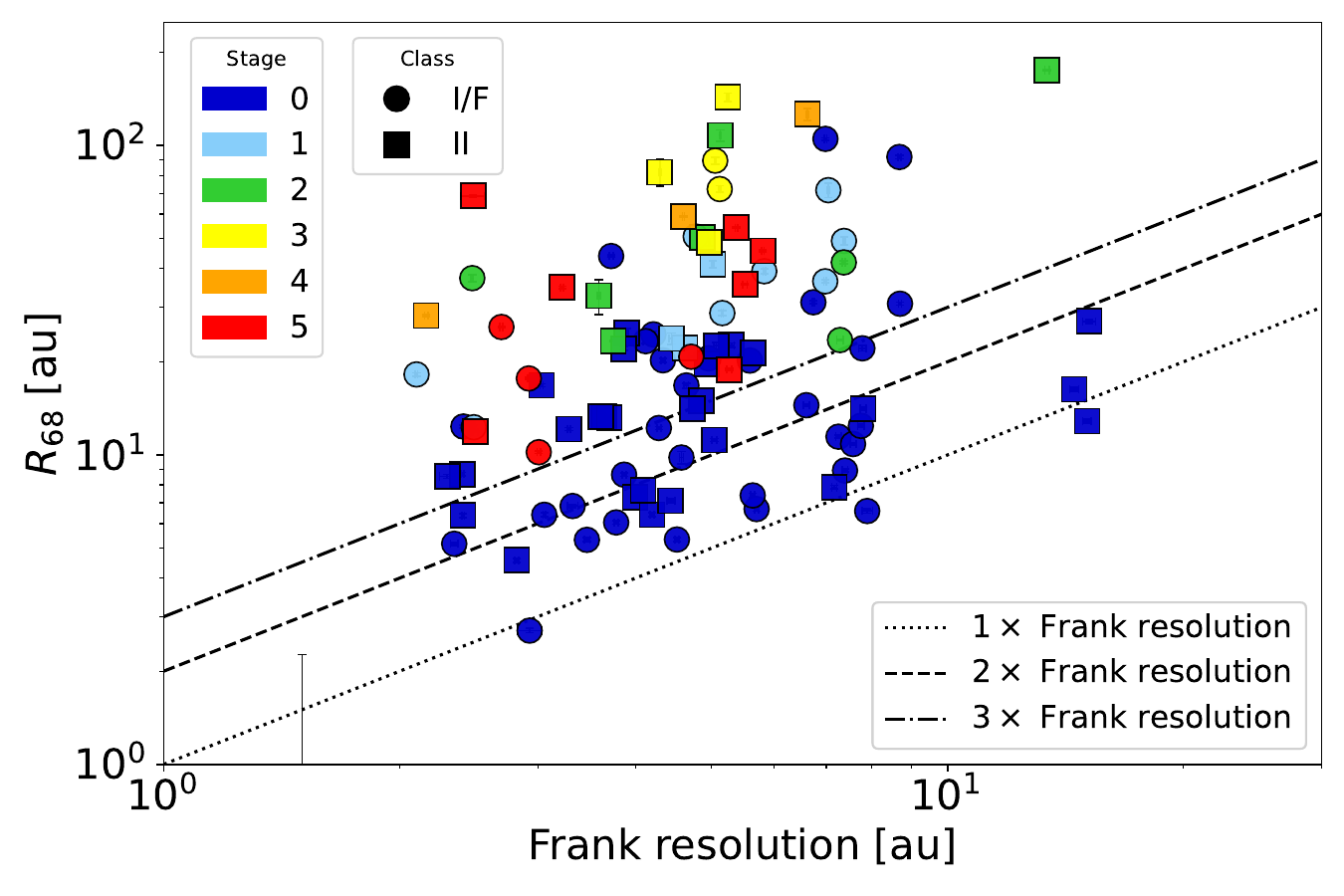}
    \includegraphics[width=0.49\linewidth]{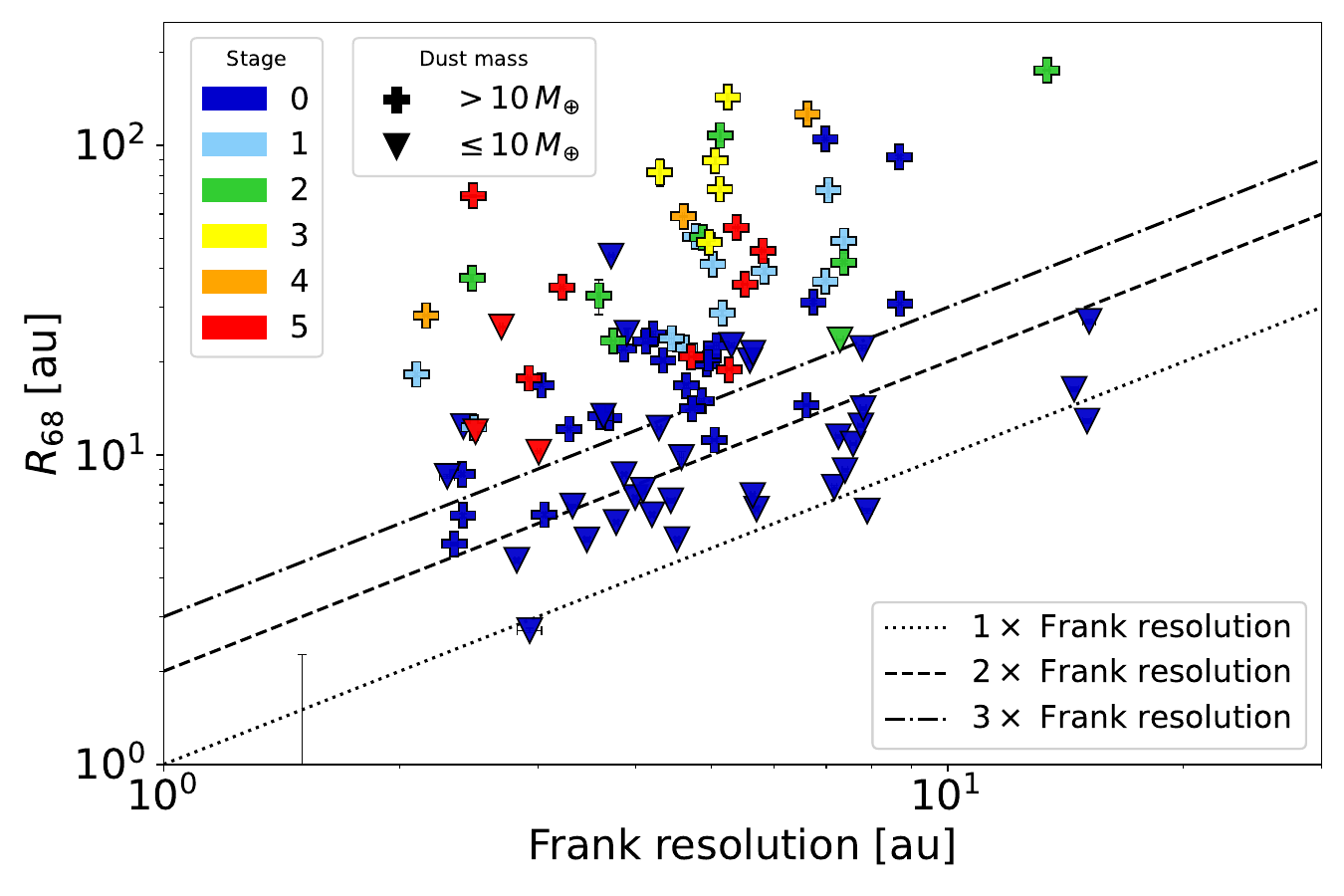}
    \caption{ $R_{68}$ as a function of the {\texttt{Frank}} resolution, with diagonal dashed lines indicating the 1$\times$, 2$\times$, and 3$\times$ resolution limits. Color denotes the morphological stage. The dot-dashed line is 3 $\times$ the \texttt{Frank} resolution, the dashed line is 2 $\times$ the \texttt{Frank} resolution, and the dotted line is 3 $\times$ the \texttt{Frank} resolution.
Left: Symbols differentiate the two SED classes (Class~I/F and Class~II). Right:Symbols differentiate the two mass regime ($M_{\mathrm{d}} \leq 10~M_{\oplus}$ and $M_{\mathrm{d}} > 10~M_{\oplus}$).
}
    \label{fig:frank_res}
\end{figure*}

In order to investigate the effect of spatial resolution on our results, we estimate the effective spatial resolution of each using synthetic visibility experiments. For each disc, we generate visibilities of a narrow delta ring sampled with the same uv-coverage as the observations and scaled to match the total flux of the disc. These synthetic visibilities are then fitted with \texttt{Frank} using the same hyperparameters as the real data.

The smallest radial scale that can be reliably recovered in this procedure defines the effective \texttt{Frank} resolution. Because \texttt{Frank} discards low-S/N long baselines depending on the dataset and hyperparameters, this resolution depends on both the uv-coverage and the S/N of each disc. We emphasize that this is not the synthesized beam resolution, but rather the smallest radial scale recoverable by \texttt{Frank} given the data quality. The resulting angular resolution is converted to physical units using the target distance. \footnote{{For the discs \ODISEA{16A} and \ODISEA{125B} we do not report the resolutions because the reconstructed radial profiles contained non-finite values after normalization.}}

{To assess how well the discs are resolved, we compared the \texttt{Frank} resolution to the 68\% flux radius ($R_{68}$), derived from the \texttt{Frank} radial brightness profile as the radius enclosing 68\% of the integrated continuum flux. The \texttt{Frank} resolution is estimated by fitting a Gaussian function to the radial point-spread function of the \texttt{Frank} reconstruction, with its full width at half maximum (FWHM) adopted as the characteristic resolution scale. The details of extracting the FWHM from the  \texttt{Frank} reconstruction are detailed in \cite{2024ApJ...974..306S}.  Assuming a Gaussian profile,
\begin{equation}
I(r)=I_0\exp\left(-\frac{r^2}{2\sigma^2}\right),
\end{equation}
where I(r) is the radial intensity profile and $\sigma$ is the Gaussian width, the radius enclosing 68\% of the total flux is equivalent to
\begin{equation}
R_{68}\iff0.42\times\mathrm{FWHM},
\end{equation}
with
\begin{equation}
\mathrm{FWHM}=2\sqrt{2\ln2},\sigma.
\end{equation}
We therefore convert the \texttt{Frank} FWHM to the equivalent $R_{68}$ by applying the factor of 0.42, allowing a direct comparison between the measured $R_{68}$ and the effective spatial resolution of the reconstruction.}

We find that nearly 93\% of the discs lie above the 1$\times$-resolution line, indicating that the majority of the sample is formally resolved by \texttt{Frank}. Only three discs fall below this threshold, implying marginal resolution, while three additional sources appear as outliers that could not be fitted due to incomplete or low S/N visibilities. 
{Discs located above the $3\times$-resolution line are the best resolved, while those between the $2\times$ and $3\times$-resolution lines are moderately resolved and flagged with the superscript $m$ in Table~1, and those below the $2\times$-resolution line should be interpreted with caution and are flagged with the superscript $u$. Discs above the $2\times$-resolution line are sufficiently resolved for our statistical analysis and exhibit well-constrained substructures such as gaps, rings, and inflection points.}
Overall, this comparison demonstrates that \texttt{Frank} reconstructs reliable radial brightness profiles for approximately 93\% of the sample. The majority of discs are resolved at least twice the model resolution, providing robust constraints on disc morphology and confirming the validity of the inferred substructures across different evolutionary stages. 

When we divide the sample by dust mass (Fig.~\ref{fig:frank_res}, \textit{Right panel}), we find that the majority of less-resolved discs correspond to the low-mass population ($M_{\mathrm{d}} \le 10~M_{\oplus}$). These sources typically have smaller emitting areas and lower S/N ratios, making it more difficult to identify faint or narrow substructures. In contrast, high-mass discs ($M_{\mathrm{d}} > 10~M_{\oplus}$) are generally well resolved and display multiple features across several resolution elements. This suggests that the apparent absence of substructures in many low-mass discs may be driven by limited spatial resolution and sensitivity, not necessarily intrinsic smoothness.

To quantify the role of angular resolution, we performed controlled degradation tests on a subset of well-resolved discs by progressively removing the longest baselines from the calibrated visibilities and re-fitting the profiles with \texttt{Frank} (Fig.~\ref{fig:frank_degraded}). For ODISEA\_C4\_114 and ODISEA\_C4\_081, the original high-resolution profiles reveal a clear gap, ring, and inflection point in the radial intensity distribution (Fig.~\ref{fig:frank_degraded_overplot}). When the baselines are truncated to emulate coarser angular resolutions, these substructures are gradually washed out: narrow gaps become shallower, broad rings merge into a single smooth component, and in the most degraded case, the profiles would be classified as featureless according to our criteria, even though the underlying high-resolution data demonstrate the presence of substructures. Importantly, continued degradation drives the discs into a regime where $R_{68} \le 2\times$ the Frank resolution, comparable to many low-mass systems in our sample, highlighting how limited angular resolution alone can bias morphological classification.

These tests therefore demonstrate that discs located below the 2–3$\times$ \texttt{Frank}-resolution thresholds may still host detectable substructures if observed at sufficiently high angular resolution. We emphasize that this experiment isolates the impact of spatial resolution, as the total flux and noise properties of the data are otherwise preserved. In reality, low-mass discs are also fainter and thus observed at lower S/N, which likely further suppresses the detectability of weak or narrow features. The boundary between “structured” and “smooth” discs in our low-mass disc sample ($M_{\mathrm{d}} \le 10~M_{\oplus}$) is therefore shaped by a combination of angular resolution and S/N, rather than purely by intrinsic disc physics. Such discs are compact and faint, and thus most susceptible to beam smearing and noise. Consequently, while we robustly characterize the incidence and morphology of substructures in the more massive, well-resolved discs, our conclusions on the structural evolution of the lowest-mass discs should be regarded as conservative and likely incomplete. Higher-resolution, higher-sensitivity ALMA observations will be required to uncover potential fine-scale structures in these compact, low-mass systems and to fully map the structural evolution across the entire dust-mass distribution.

{A potential limitation concerns highly inclined ($i = 80^\circ \pm 5^\circ$) discs. Among these, \ODISEA{82}, \ODISEA{131}, \ODISEA{125}, \ODISEA{126}, \ODISEA{26}, \ODISEA{144}, \ODISEA{001}, and \ODISEA{009} with no substructures are classified in Stage~0, while \ODISEA{3} and \ODISEA{64} with inflection points are in Stage~I. Their apparent lack of evolved substructures like gap-ring pairs may result from the combined effects of high inclination and limited angular resolution. This interpretation is supported by \citet{2022ApJ...930...11V}, who resolved gap--ring pairs in \ODISEA{126} using $\sim$3 au ($0.02''$) resolution observations and \cite{2026A&A...708A.143B} who suggest the presence of inner cavity in \ODISEA{144}. Furthermore, \cite{2026A&A...708A.143B} show that the high inclination of \ODISEA{26} makes accurate disc modeling challenging, preventing a conclusive identification of substructures.
 Conversely, we identify gap--ring pairs in the edge-on discs \ODISEA{043} and RA162813.74. However, their residuals exhibit asymmetric emission that cannot be reproduced by an axisymmetric model, while the corresponding \texttt{tclean} images appear patchy. Therefore, the presence of these substructures should be confirmed with higher-resolution observations. }

{ In terms of the residuals seen in Fig. \ref{fig:0+I}-\ref{fig:5+II}, these can be further improved when geometrical parameters, hyperparameters, and centering offsets are allowed to vary. The off-centering visible in the current residual images could be reduced by exploring a grid of $\delta$RA and $\delta$DEC values, computing the residuals for each combination, and selecting the model that minimizes them. Implementing this procedure is beyond the scope of the present work due to the high computational cost. Nevertheless, we strongly encourage such refinements, particularly in small-sample analyses or detailed studies of individual discs.}

\section{Discussion}

\subsection{Stage of the structures}  

\subsubsection{Structures as a function of SED class}

{{Here on all our analysis is done for discs that are resolved; i.e we restrict the sample to discs located above the dashed line in Fig.~\ref{fig:frank_res}, i.e., $R_{68} \ge 2\times$ the \texttt{Frank} resolution.}

Considering the sizes of the samples and to improve the statistics, we divide our targets into four evolutionary groups: Stage~0, Stage~I, Stage~II/III, and Stage~IV/V, corresponding respectively to 1) featureless discs, 2) discs with inflection points,  3) discs with gaps, and 4) discs with cavities.}

{We emphasize that the evolutionary sequence presented here is defined solely by disc morphology and is independent of SED classification. The Ophiuchus molecular cloud has a significant age spread, containing both embedded Class I/F sources, whose population ages are typically \(\lesssim 1\) Myr in nearby star-forming regions \citep{2009Evans}, and more evolved Class II systems with estimated ages of a few Myr, including the Stage V archetype RXJ1633.9–2442 \citep{2012ApJ...752...75C, 2025ApJ...984L..57O}. This age spread allows disc morphologies at different evolutionary states to be compared within the same star-forming region. In addition, the onset of substructure formation is expected to vary from source to source depending on the initial disc conditions and on the location and formation timescale of the planet responsible for the feature. Therefore, a larger fraction of evolved morphologies (Stages II/III–IV/V) may be expected among Class II sources because they have had more time for planets to grow and carve gaps and cavities.}

{The onset of structure formation will vary from source to source depending on both the initial conditions of the disc and the location of the planet responsible for the feature. However,  a higher fraction of evolved structures (Stages II/III–IV/V) among Class II sources could be expected given the longer timescales available for the formation of the planets carving the gaps and cavities.
}

As shown in Fig.~\ref{fig:pie_chart_resolved} (left column), the Class~I/F population broadly follows this expectation, with most discs remaining featureless, consistent with their young ages and limited time for substructure growth. At the same time, the presence of structured discs already within the Class~I/F population suggests that planet–disc interactions can begin early, before envelope dissipation, as proposed by \citet{2021Cieza} and modeled by \citet{2025ApJ...984L..57O}.

On the other hand, a substantial fraction of the Class~II population (46\%) still exhibits featureless morphologies. This diversity may reflect intrinsic physical differences in the timescales of dust evolution and planet formation.

Although these smooth discs are resolved as per Fig.\ref{fig:frank_res}, part of this smooth appearance may still arise from observational limitations: several discs are very compact (see Section \ref{sec:5.1.2}), possibly having substructures that are unresolved at the current spatial resolution.

This motivates a complementary analysis in which we consider the disc population in separate mass regimes, allowing potential substructures in the lowest-mass and lowest-S/N discs to be examined independently (Section \ref{sec:5.1.2}). 

\subsubsection{Structures as a function of SED class and disc mass}
\label{sec:5.1.2}

\begin{figure*}
    \centering
    \includegraphics[ clip,width=0.9\linewidth]{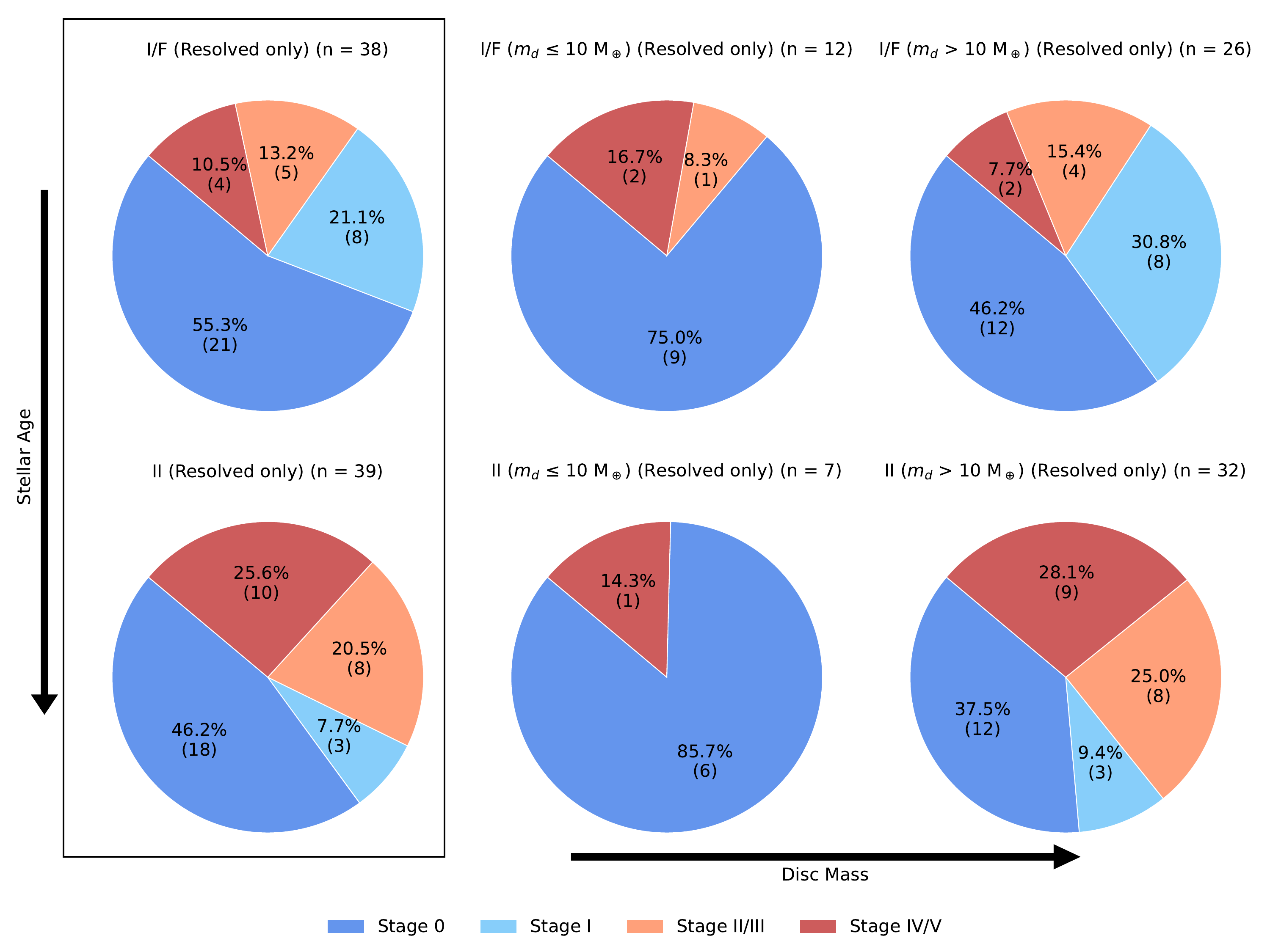}
    \caption{Top and bottom rows show pie charts of the sample divided by SED class: Class~I/F (top) and Class~II (bottom) for the sample of targets that are above two times the Frank resolution.
The first column presents the resolved sample, the second column includes discs with $M_{\mathrm{d}} \le 10~M_{\oplus}$, and the third column includes discs with $M_{\mathrm{d}} > 10~M_{\oplus}$.
Each pie chart indicates the fraction of discs in different morphological stages, Stage~0 (featureless), Stage~I (inflection), Stage~II/III (gaps), and Stage~IV/V (cavities), represented in different colors.}
    \label{fig:pie_chart_resolved}
\end{figure*}
 
 \begin{figure}
    \centering
    \includegraphics[width=0.49\textwidth]{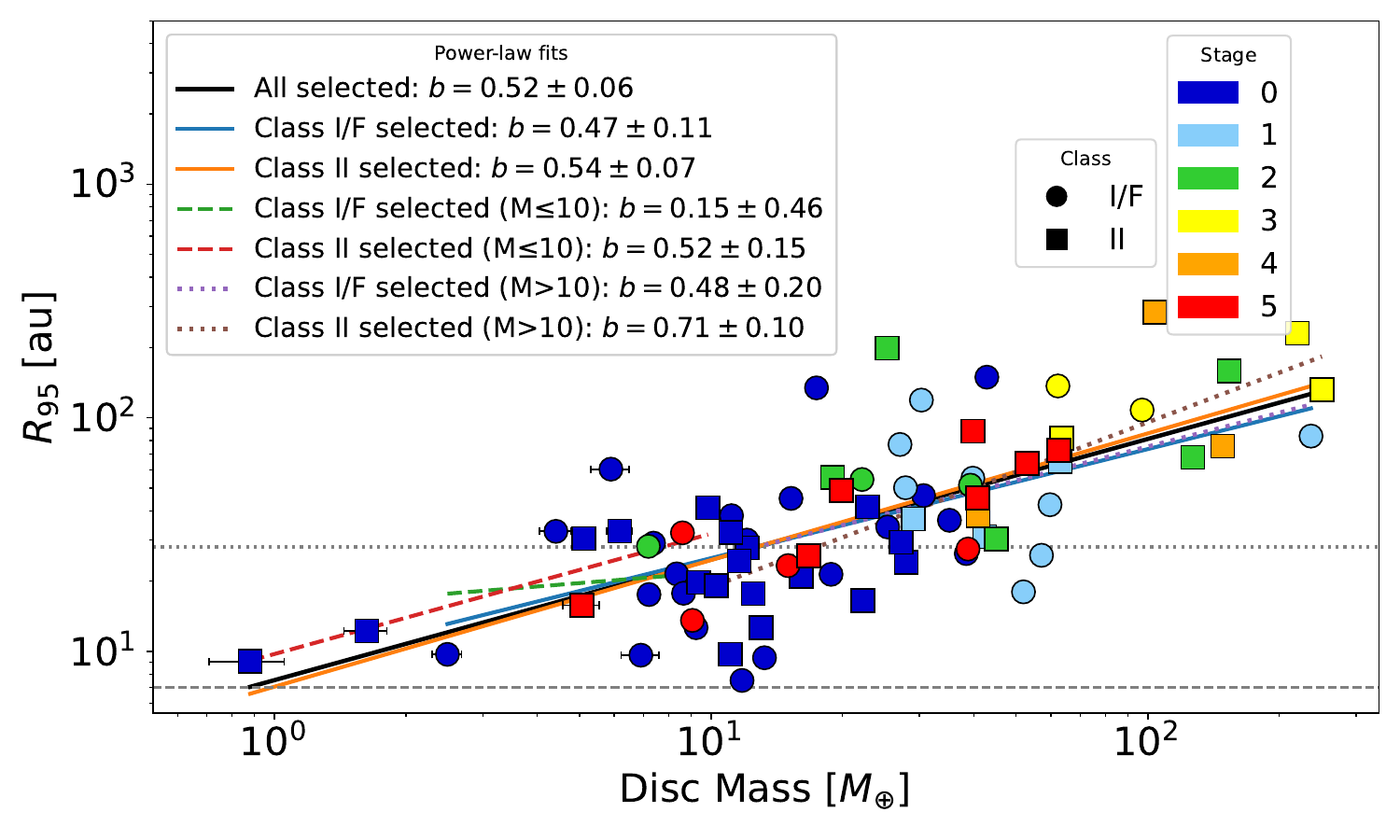}
    \caption{Log–log plot of the $R_{95}$ versus disc mass ($M_{\mathrm{d}}$) in units of $M_{\oplus}$.
The five stages of disc evolution are shown in different colors, while the two SED classes, Class~I/F and Class~II, are in different plotting symbols.
Best-fit power-law relations for each subset are overplotted, indicating the dependence of disc size on mass and evolutionary stage. The dotted and dashed horizontal lines mark
the 22 au (0.2'') and 7 au (0.05'') beam resolution limits of our sample.   }
    \label{fig:fl_radii}
\end{figure}

\begin{figure}
    \centering
    \includegraphics[width=0.49\textwidth]{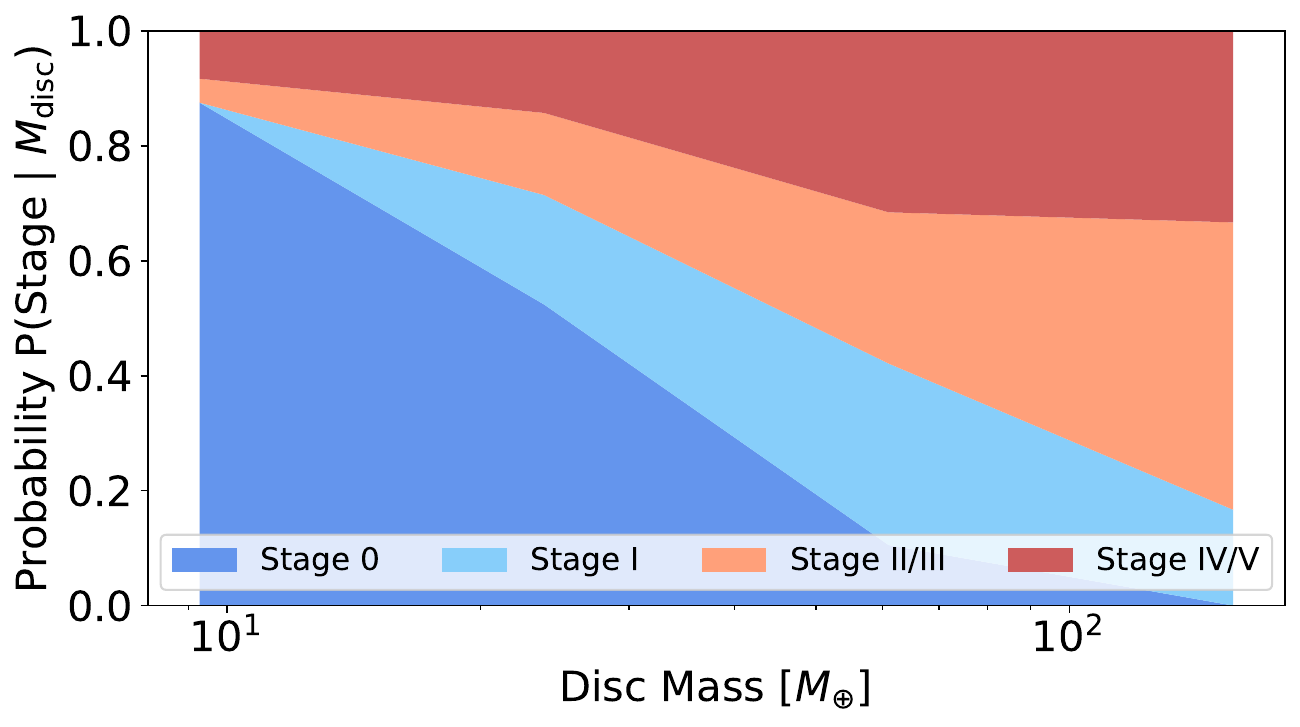}
    \caption{Conditional probability of grouped disc evolutionary stages as a function of disc mass. The mass range is divided into six logarithmically spaced bins (nbins = 6), requiring a minimum of five discs per bin (minN = 5).}
    \label{fig:prob_dist}
\end{figure}
For the first time, we have high-resolution observations capable of resolving a large sample of low-mass discs in Ophiuchus, down to radii of about $\sim7\mathrm{au}$, allowing a statistical analysis of a more complete
and representative population, including discs that may only be massive enough to form terrestrial-planet systems. We use ODISEA Band~4 data rather than Band~6 observations to estimate disc masses, as Band~4–based masses are more consistent with those derived from multi-frequency analyses (see Chavan et al., in prep., for details).
To provide a reference single-frequency estimate of the dust mass, we assume optically thin continuum emission and an isothermal characteristic dust temperature $T_{\rm dust}=20$~K, and compute
\begin{equation}
M_{\rm dust} = \frac{F_{\nu}\, d^{2}}{\kappa_{\nu}\, B_{\nu}(T_{\rm dust})},
\end{equation}
where $F_{\nu}$ is the integrated flux density at frequency $\nu$, $d$ is the source distance, $\kappa_{\nu}$ is the dust mass absorption opacity, and $B_{\nu}$ is the Planck function. We adopt a power-law dust opacity prescription
\begin{equation}
\kappa_{\nu} = 2.3\left(\frac{\nu}{230~{\rm GHz}}\right)^{\beta}~{\rm cm^{2}\,g^{-1}},
\end{equation}
with $\beta=1$, corresponding to $\kappa_{100~{\rm GHz}} = 1.0~{\rm cm^{2}\,g^{-1}}$, following \cite{2021ApJS..257...14S}.

Using the disc masses so derived, we divided the sample into two groups: discs with dust masses smaller and larger than 10 M$_{\oplus}$. This boundary results in two subsamples {with 20 and 57 discs}, respectively, and is physically meaningful as 10 M$_{\oplus}$ is the canonical mass for the core of a Jupiter-mass planet in the core-accretion model.     

Dividing the sample at a dust-mass threshold of $10 M_{\oplus}$ (Fig.~\ref{fig:pie_chart_resolved}, second and third columns) reveals that among low-mass discs ($M_{\mathrm{disc}} \le 10 M_{\oplus}$), both Class I/F and Class II populations are featureless: {over 80\% of the discs belong to Stage 0}. This suggests that such compact, low-mass discs rarely develop detectable substructures. However, at a fixed angular resolution, the small emitting areas of low-mass discs make them more susceptible to beam smearing, limiting the detectability of substructures. As we show through our resolution-degradation tests (Section \ref{sec:caveates_resFRANK}), even intrinsically structured discs can appear featureless when the angular resolution is progressively degraded. This interpretation is consistent with recent high-resolution ALMA observations in Lupus by \cite{2025A&A...696A.232G}, who find that substructures are preferentially detected in larger and more massive discs, while compact, low-mass systems remain largely smooth even at high resolution. Therefore, the smooth appearance of many low-mass discs likely reflects a combination of intrinsic compactness and observational limitations.

In contrast, high-mass discs ($M_{\mathrm{disc}} > 10 M_{\oplus}$) exhibit a different behavior: the fraction of evolved morphologies (Stages II/III and IV/V) rises to nearly {65\% in Class II and 58\%} in Class I/F systems. This emphasizes that disc mass is a key driver of substructure formation, consistent with models in which more massive discs are more prone to developing pressure maxima and planet-carved gaps \citep{2018Zhang,2012Pinilla,2007Brauer}. 

However, a larger disc mass also significantly enhances the detectability of substructures even when discs are observed at modest resolution. 
This is due to the steep correlation between disc mass and radius shown in Fig.~\ref{fig:fl_radii}.  
A power-law relation of the form $R_{95} \propto M_{\mathrm{d}}^{b}$ yields a slope $b = 0.56\pm0.05$ for the full sample. This slope is consistent with the value $b\sim0.5$ found by \cite{2017Tripathi} for a subsample of $\sim$50 protoplanetary discs mostly in the Taurus and Ophiuchus regions. When separated by SED class, the slopes remain consistent within the uncertainties for both {Class I/F ($b = 0.47\pm0.11$) and Class II ($b = 0.54\pm0.07$). Low-mass discs show: Class I/F systems $M_{\mathrm{d}} \le 10,M_{\oplus}$ have a slope ($b = 0.15\pm0.46$), and Class II discs of similar mass ($b = 0.52\pm0.15$). For massive discs ($M_{\mathrm{d}} > 10,M_{\oplus}$), both classes converge to $b = 0.48\pm0.20$–$0.67\pm0.1$}. Within uncertainties, all subsamples are consistent with $b\sim0.5$–0.6, indicating that disc size scales approximately as $R_{95}\propto M_{\mathrm{d}}^{1/2}$ across evolutionary stages and mass regimes.

The probability distribution reveals a clear and systematic mass-dependent transition in the observed disc morphology (Fig.\ref{fig:prob_dist}). At low disc masses ($<$10 M$_\oplus$), the population is overwhelmingly dominated by Stage 0 systems (80\%), indicating that smooth radial profiles are strongly favored in low-mass discs. As disc mass increases into the intermediate regime ($\sim$15–40 M$_\oplus$), the fraction of Stage 0 discs declines steadily, while Stage I and Stage II/III systems become increasingly common, marking the progressive emergence of substructures. Above $\sim$60–70 M$_\oplus$, Stage 0 discs significantly decrease (<10\%), and the population becomes dominated by Stage II/III and Stage IV/V systems, indicating that prominent gaps and cavities are preferentially associated with higher-mass discs.

Together, these results demonstrate that while substructures become more common as discs evolve, disc mass sets a prominent condition for their development and detectability. Because more massive discs are systematically larger, potential substructures occur at wider and better-resolved radii, increasing the likelihood that they are both formed and observationally detected. In contrast, low-mass discs are intrinsically compact and therefore more susceptible to beam smearing, which may contribute to their predominantly smooth appearance. However, as shown in Section \ref{sec:caveates_resFRANK}, limited spatial resolution can conceal faint or narrow structures in this population. High-resolution, high-S/N ALMA observations will therefore be crucial for fully assessing substructure demographics in the lowest-mass regime.

\subsubsection{Radial distribution of structures}

\begin{figure}
    \centering
    \includegraphics[width=0.49\textwidth]{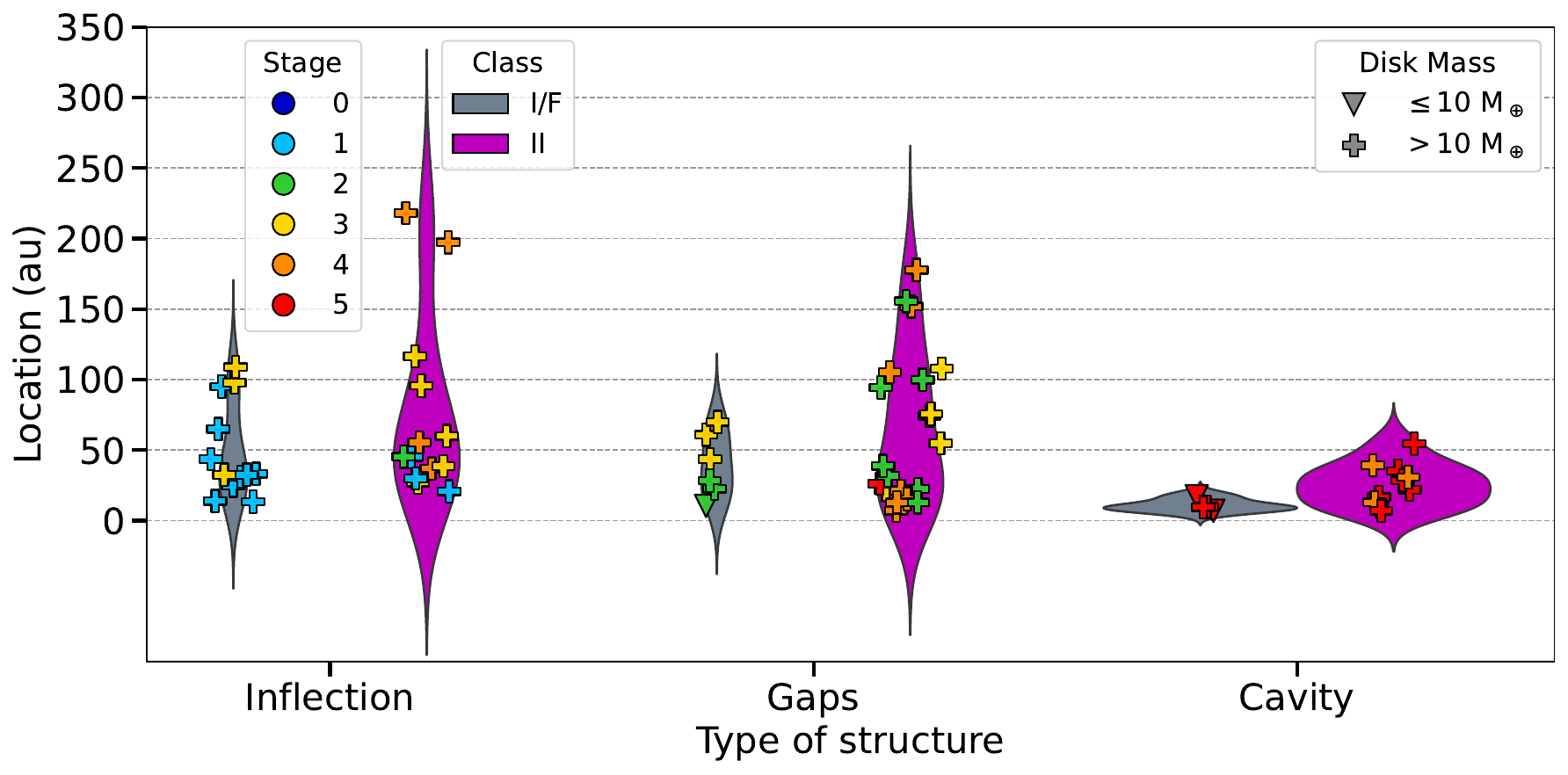}
    \caption{Violin plot of the radial location of the identified features: inflection points, gaps, and cavities, as a function of SED class and disc mass. Each point is color-coded by evolutionary stage, with symbol size representing disc mass.}
    \label{fig:violin_combined}
\end{figure}

Figure~\ref{fig:violin_combined} reveals that inflection points are typically located within $\lesssim50~\mathrm{au}$, often marking the first detectable departure from a smooth radial profile. Gaps are generally found at larger radii, between $\sim30$ and $200~\mathrm{au}$, consistent with the presence of partial dust depletion or ring–gap pairs produced by pressure perturbations.

For cavities, the characteristic radius shown here corresponds to $R_{Cav,90}$, the radial distance where the normalized intensity recovers to 90\% of the peak value just outside the cleared region. Physically, this represents the inner edge of the outer disc.
$R_{Cav,90}$ can thus be interpreted as the approximate location of 
a giant planet inducing a strong pressure gradient that halts inward dust drift \citep[e.g.,][]{2018Zhang,2012Pinilla,2025ApJ...984L..57O}. The clustering of these $R_{90}$ values between $\sim20$ and {$50~\mathrm{au}$} indicates that the majority of the observed cavities can be explained by planet–disc interactions operating at tens of astronomical units, consistent with predictions from hydrodynamical simulations \citep{2018Zhang,2012Pinilla}.

Interestingly, we also find cavities in some low-mass discs ($M_{\mathrm{disc}} \le 10~M_{\oplus}$), and these systems exhibit relatively small cavity radii, consistent with the formation of giant planets at smaller radii.   
However, the small number of such systems suggests that most low-mass discs instead evolve toward compact, smooth profiles.

In contrast, high-mass discs ($M_{\mathrm{d}} > 10~M_{\oplus}$) display a wider distribution of cavity radii, with $R_{90}$ extending up to {$\sim50~\mathrm{au}$} and suggesting that massive discs develop Jupiter-mass planets at a wider range of separations.

{Figure~\ref{fig:violin_combined} suggests that, in the resolved sample, substructures in Class I/F sources are preferentially detected at smaller radii than in Class II objects, for inflection points, gaps, and cavities. However, this trend should be interpreted with caution, as the detectable location of substructures is coupled to the overall disc size, dust mass, angular resolution, and sensitivity; larger discs naturally provide a wider radial range over which substructures can be identified. A possible contributing factor is that structures at larger orbital radii evolve on longer dynamical timescales and may therefore require more time to form, deepen, or become readily detectable.}

 \subsection{Comparisons to previous work}
\subsubsection{Substructures as seen in previous studies}

 In Stage~0 (see Figs:\ref{fig:0+I},\ref{fig:0+II}, \ref{fig:0+I_app},\ref{fig:0+II_app}), we resolve the closest binary in our sample, ODISEA\_C4\_94A+B (ISO Oph 171){\footnote{We include the commonly used names of the discs in parentheses.}}, in the image plane, with a separation of $\sim$7~au (0.05'') between the primary and companion.
This binary was also resolved in \cite{2025Shoshi}, although their analysis relied on sparse modeling of the archival Band~6 ODISEA data at 200~mas resolution.
\cite{2025Shoshi} additionally reports tentative substructures in ODISEA\_C4\_102 (ISO Oph 165), ODISEA\_C4\_26 (2MASS J16254662-2423361), ODISEA\_C4\_63 (ISO Oph 93), ODISEA\_C4\_144 (2MASS J16395292-2419314), and ODISEA\_C4\_125A (ISO Oph 105), none of which are confirmed in our analysis. 
{As explained in Section \ref{sec:caveates_resFRANK},  \cite{2026A&A...708A.143B} find \ODISEA{144} to have an inner cavity and \cite{2022ApJ...930...11V} have multiple gap-ring pairs in \ODISEA{126} (Oph 163131), however due to a featureless radial profile in our analysis we categorize these discs in Stage 0.
\cite{2026A&A...708A.143B} also confirms the featureless smooth profile of \ODISEA{056} (SSTc2d J162738.3-235732) and \ODISEA{101} (SSTc2d J162738.3-235732) similar to our analysis.}

In Stage~I (see Fig:\ref{fig:1+I}, \ref{fig:1+I_app}), our detection of inflection points in ODISEA\_C4\_130 (WLY 2-63) is consistent with the range reported by \cite{2021Cieza} using higher resolution Band 6 data.  
Conversely, \cite{2025Shoshi} obtain an edge-on geometry and bumpy brightness profiles of ODISEA\_C4\_67 (ISO Oph 99), ODISEA\_C4\_72 (ISO Oph 112), ODISEA\_C4\_3 (2MASS J16214513-2342316), and ODISEA\_C4\_64 (ISO Oph 94), classifying them as more evolved discs with gap–ring pairs.
{\cite{2026A&A...708A.143B} suggest a gap-ring pair and an inflection point for \ODISEA{3} (SSTc2d J162145.1-234232), but we see only the later inflection point.}

In Stage~II (see Figs:\ref{fig:2+I}, \ref{fig:2+II}, \ref{fig:2+II_app}), we identify a system where the primary, ODISEA\_C4\_53A (WL2), hosts a clear gap–ring pair, also seen by \cite{2025Shoshi}.
Although the \texttt{tclean} image shows structural variations that are not fully resolved, \texttt{Frank} reproduces these features as a well-defined gap–ring pair.
The central region within 20~au appears slightly bulged, similar to ODISEA\_C4\_83 (ISO Oph 127), which \texttt{Frank} interprets as a possible unresolved gap-like feature also suggested by \cite{2025Shoshi}. {\cite{2026A&A...708A.143B} also identifies the same gap-ring pair in \ODISEA{83} (SSTc2d J162718.4-243915).} 
For ODISEA\_C4\_39 (DoAr 25), we detect an inflection point at 45~au, which was not identified by \cite{2018Huang}, although the position of the gap–ring pair agrees with their expected range.
We also confirm the second gap–ring pair reported in ODISEA\_C4\_30 (ISO Oph 17) by \cite{2021Cieza}. \cite{2018Huang} reported a gap–ring pair in ODISEA\_C4\_27 (SR 4) also reproduced by \texttt{Frank} but remains unresolved in our image plane.

In Stage~III (see Figs.~\ref{fig:3+I},\ref{fig:3+II}), ODISEA\_C4\_47 (ISO Oph 54) displays the same three substructures—two gap–ring pairs and an inflection point—originally seen by \cite{2021Cieza}.
We recover all the gap-ring pairs in ODISEA\_C4\_41 (Elias 2 24) and ODISEA\_C4\_51 (Elias 2 27) as seen in \cite{2018Huang}.
ODISEA\_C4\_114 (ISO Oph 196) shows a single gap–ring pair with sufficient material beyond it to classify the disc as Stage~III.
\cite{2021Cieza} resolve this pair at higher angular resolution, within our resolution limit.
We do not detect any central cavity but confirm one inflection point beyond the gap–ring pair, whereas the second weaker one noted by \cite{2021Cieza} falls below our 5\% normalized intensity threshold (Fig.~\ref{fig:3+II}).

In Stage~IV (see Fig.~\ref{fig:4+II}), we detect the same substructures in ODISEA\_C4\_62 (SR 24)  as reported by \cite{2021Cieza}.
While they identified an inflection point at 37~au and noted small variations in the radial profile between 37–50~au, our analysis places the outermost inflection point at 55~au.
In our \texttt{Frank} profile, this appears as a smooth shoulder extending from 52 to 57~au, where the curvature change is most evident. Although we don't resolve any inner disc in the image plane, in the \texttt{Frank} radial profile, we see non-zero emission within the cavity.  
ODISEA\_C4\_143 (WSB 82) shows all previously identified features in \cite{2021Cieza}—one inflection point, two gaps, and a central cavity (Fig.~\ref{fig:4+II}).
This is the largest disc in our sample, with $R_{95} = 283.8$~au.
We locate six gaps, compared to two in \cite{2021Cieza}.
The inflection point at 80~au noted by that study corresponds to a possible gap–ring pair in our \texttt{Frank} model.
Strong emission from the inner disc ($r \le 15~\mathrm{au}$) is also seen in their radial profile but remains unresolved in the image planes.
A detailed study of the inner disc is beyond the scope of this work.
Finally, we clearly resolve the cavity in ODISEA\_C4\_117A (SR 13), as indicated in \cite{2025Shoshi}. SR 13 is a hierarchical triple system in which the central binary (SR 13 Aa–Ab) is orbited by the tertiary component SR 13 B. The millimeter emission primarily traces a circumbinary disc around the central pair, while the tertiary is detected as a localized emission peak to the north-east. In the image plane, we spatially resolve the cavity and detect diffuse emission within it, denoting an inner disc component. This inner emission is reproduced in the \texttt{Frank} model.
The first ring at 8~au is more distinct than the second at 16~au; however, faint traces of the latter remain visible in the image plane. This supports our interpretation of the second ring, even though the \texttt{Frank} model and radial profile are partially affected by contamination from the tertiary component along the line of sight.

In Stage~V (see Fig.~\ref{fig:5+I},~\ref{fig:5+I_app},~\ref{fig:5+II},~\ref{fig:5+II_app}), we detect a clear cavity in ODISEA\_C4\_045 (ISO Oph 51) with a ring at 21~au, consistent with \cite{2025Shoshi}.
We also note mild asymmetries along the semi-minor axis, which could be due to the limited S/N ratio of our data.
{The cavity in \ODISEA{116} (SSTc2d J162823.3-242241) is also resolved in \cite{2026A&A...708A.143B}.}
We do not detect an inner disc in ODISEA\_C4\_127 (DoAr 44), but the first ring after the cavity appears at the same location as that reported by \cite{2021Cieza}. 
{
That is probably because the central beam is dominated by free-free emission, not an inner dust disc, as shown by recently delivered Band 1 (7 mm) data of the system (Bhowmik et al. in prep.).     
}
An extra ring at 21~au is observed in ODISEA\_C4\_22AB (ISO Oph 2) before the main peak ring.
{This secondary feature most likely corresponds to the non-axisymmetric inner ring studied by \cite{2020Gonzalez-Ruilova}}, although we measure a larger separation of $\sim$40~au rather than 20~au between the two rings.
Finally, the cavity in ODISEA\_C4\_12 (IRAS 16201-2410) was originally suggested by \cite{2019Cieza}, who identified two emission blobs symmetrically placed along the disc’s major axis.
This cavity is clearly visible in our image plane, well reproduced by the \texttt{Frank} model, and independently confirmed by \cite{2025Shoshi}.

Overall, we conclude that despite the snapshot nature of our Band 8 observations, we still recover the substructures reported in previous Band~6 studies using the higher-resolution observations from \cite{2021Cieza} or the sparse visibility modeling approach from \cite{2025Shoshi}.  
We also note that differences with previous Band~6 studies may not arise solely from angular resolution or sensitivity effects, as observations at different wavelengths probe different dust grain populations and can therefore produce small variations in the observed morphology. Some discrepancies between studies may thus reflect intrinsic, wavelength-dependent disc structure rather than purely observational limitations.

 \subsubsection{Campos II, Tbol}

While the SED Classes based on the slope of the IR emission are a commonly used proxy for the youth of the circumstellar disc, it is not universal. Another proxy commonly used is the bolometric temperature ($T_{\mathrm{bol}}$), which is especially useful to identify the {earliest protostellar stages} \citep{2025A&A...700A.235H}. {Bolometric temperature ($T_{\mathrm{bol}}$) is the temperature of a blackbody with the same mean frequency as the observed SED. In this work, we adopt the $T_{\mathrm{bol}}$ values compiled by \citet{2011ARA&A..49...67W} for the Ophiuchus sources, which were estimated from their observed multi-wavelength SEDs.}
To connect our morphological classification with the evolutionary sequence proposed in the CAMPOS survey \citep{2025A&A...700A.235H}, we divided our {resolved} sample according to bolometric temperature at a break of $T_{\mathrm{bol}} = 650~\mathrm{K}$, separating {colder sources} ($T_{\mathrm{bol}} < 650~\mathrm{K}$) from {warmer systems} ($T_{\mathrm{bol}} \ge 650~\mathrm{K}$).
This boundary approximately corresponds to the transition between Class I/F and Class II sources as defined in \cite{2009Evans,2011ARA&A..49...67W}.

Our sample contains relatively fewer cold discs than CAMPOS, which includes many low-temperature ($T_{\mathrm{bol}} < 400~\mathrm{K}$) and Class 0 objects spanning a wide range of masses and radii. In contrast, the ODISEA sample is dominated by warmer sources with only a small subset near the $T_{\mathrm{bol}} = 650~\mathrm{K}$ boundary. Consequently, our analysis primarily traces the later stages of disc evolution, roughly corresponding to the end of the Class I/F phase through mature Class II discs rather than the earliest embedded phase sampled by CAMPOS.

In Fig.~\ref{fig:tbol_pie_chart}, for the cold population ($T_{\mathrm{bol}} < 650~\mathrm{K}$), we find that {37\%} of the discs remain featureless (Stage 0), while {63\% show some level of substructure, including inflection points, gaps, or cavities}. 
{For the warm population ($T_{\mathrm{bol}} \ge 650~\mathrm{K}$), 54\% of the discs are featureless, while 46\% are structured. Therefore, although substructures are present in both $T_{\mathrm{bol}}$ regimes, we do not find a statistically significant difference in the structured fraction across the $650~\mathrm{K}$ boundary.}
Here, {structured discs are defined as all sources classified as Stage I, Stage II/III, or Stage IV/V, while Stage 0 discs are treated as featureless}. 
{A Fisher's exact test comparing featureless and structured discs in the two $T_{\mathrm{bol}}$ bins gives $p=0.29$, indicating that the difference between the two populations is not statistically significant.}

When considering disc mass, there are only four discs in the low-mass regime ($M_{\mathrm{d}} \le 10~M_{\oplus}$) for $T_{\mathrm{bol}} < 650~\mathrm{K}$; one disc shows a cavity-like morphology, while the remaining three are featureless. Given this very small number, no robust statistical conclusion can be drawn for the low-mass, low-$T_{\mathrm{bol}}$ population. 
{For the low-mass warmer population ($T_{\mathrm{bol}} \ge 650~\mathrm{K}$ and $M_{\mathrm{d}} \le 10~M_{\oplus}$), the sample contains 15 discs, of which 80\% are featureless and 20\% are structured. Thus, low-mass discs are dominated by Stage~0 morphologies in both $T_{\mathrm{bol}}$ regimes, although the colder bin is too small for a meaningful statistical comparison.}
{In contrast, the high-mass bins show larger structured fractions. In the high-mass cold bin} ($M_{\mathrm{d}} > 10~M_{\oplus}$), {structured discs are more common, with 40\% showing inflection points and 33\% showing gaps or cavities. Similarly, in the high-mass warm bin, 44\% of the discs are featureless, while 56\% are structured, including 12\% with inflection points, 22\% with gaps, and 22\% with cavities. This is qualitatively consistent with the CAMPOS result that detectable substructures tend to appear preferentially in more massive and extended discs, but our sample size remains too limited to establish a strong trend.}

{However, as shown in Section~\ref{sec:caveates_resFRANK}, even within the resolved sample, compact low-mass discs remain more vulnerable to beam smearing and limited S/N. In such cases, the global disc emission may be resolved, while narrow or low-contrast substructures could remain undetected, causing some intrinsically structured discs to appear featureless. }

{We further examined the dependence of disc properties on $T_{\mathrm{bol}}$ in Fig.~\ref{fig:tbol} and Fig.~\ref{fig:tbol_mass}, which present the relationships between bolometric temperature and disc radius $R_{95}$, \texttt{Frank} resolution limits, integrated flux, and disc dust mass. These diagrams show substantial overlap between the colder and warmer populations. The \texttt{Frank} model resolution remains broadly similar across both regimes, suggesting that the comparison is not dominated by a systematic change in angular resolution. Nevertheless, for the lowest-mass and smallest discs, particularly those near or below our adopted resolution threshold, resolution effects remain important when interpreting the absence of detected features.}

{Together, these results are consistent with the broader evolutionary picture outlined by CAMPOS, in which substructures are already present by the late embedded/Class I--flat-spectrum phase. However, within the resolved ODISEA sample, we do not find statistically significant evidence for a change in the structured-disc fraction across $T_{\mathrm{bol}} = 650~\mathrm{K}$. Our results therefore support a cautious interpretation: substructures are found in both colder and warmer discs, but their detectability depends not only on evolutionary stage, but also on disc mass, size, S/N, and spatial resolution.}

 \subsection{Implication for planet formation}
 \label{sec:5.5}

 \begin{figure*}
    \sidecaption
    \centering
    \includegraphics[width=0.70\linewidth]{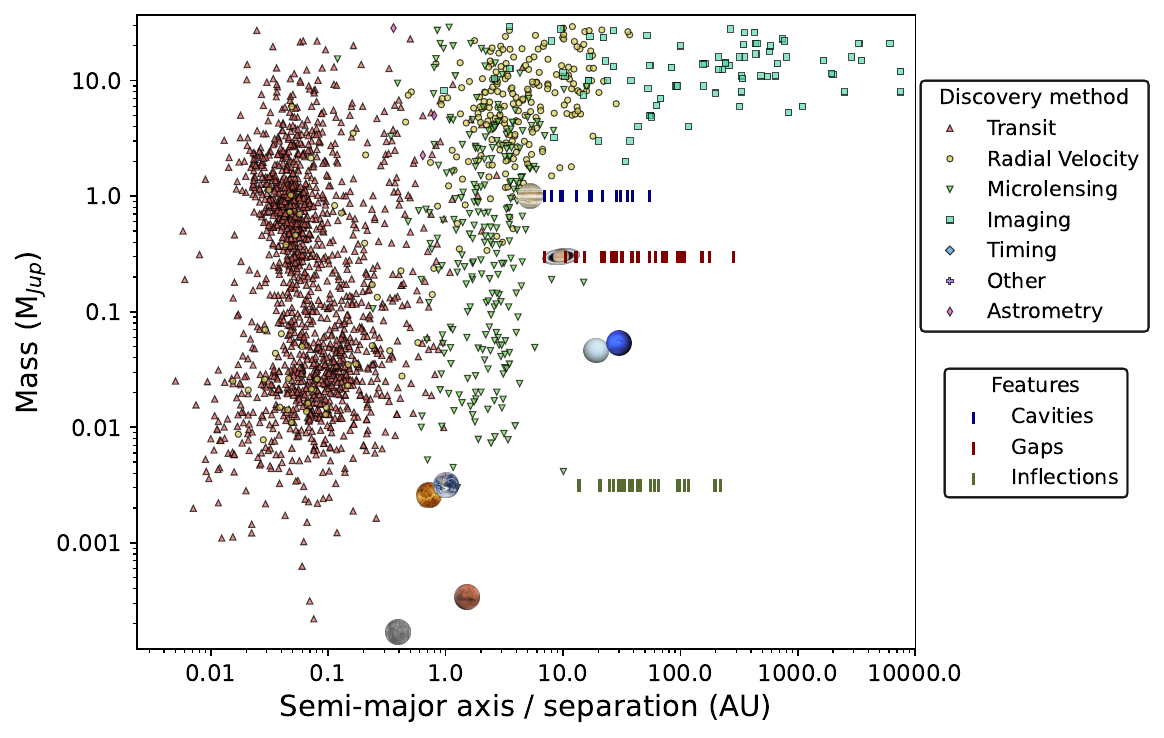}
    \caption{Confirmed exoplanet demographics from the NASA Exoplanet Archive compared with the disc substructures identified in this study. Exoplanets are colored by discovery method, and cavities, gaps, and inflection points are shown as vertical markers at their measured radii along the x-axis (semi-major axis/separation). Gap locations correspond to the location of local minima in the radial profiles, while cavities are the radial distance where the peak intensity reaches 90\%, and inflection positions are as given in Table 2. Each feature is placed on the y-axis {based on very rough estimates of the mass of the planet needed to generate the feature}: 1 Jupiter mass for cavities, 1 Saturn mass for gaps, and 1 Earth mass for inflections. Solar System planets are included to illustrate the corresponding region of parameter space.
    {Even though the mass estimates of the inferred (proto)planets are highly uncertain, the fact that they are  $\lesssim$ 1 M$_{Jup}$ at  $\gtrsim$ 10 au implies that they are, with very few exceptions, beyond current detection limits.} 
    }
    \label{fig:demog}
\end{figure*}

As shown in Fig.~\ref{fig:frank_res}, discs with dust masses $\gtrsim$10~M${\oplus}$ in our sample are typically well resolved, with $R{68} > 3\times$ the \texttt{Frank} resolution. These objects show a high incidence of substructures ($>50\%$ including inflection points) and a clear evolution of structure “stages” when comparing embedded targets and Class~II sources. In particular, the fraction of discs hosting prominent gaps or cavities (Stages~II–V) increases from $\sim$23\% to $\sim$53\% between these two subsamples (see Fig.\ref{fig:pie_chart_resolved}). Since a critical core mass of order 10 M$_{\oplus}$ is often cited as the characteristic mass required to trigger runaway gas accretion in core-accretion models \citep{2025ARA&A..63..217I}, discs with solid reservoirs of this magnitude may contain sufficient material, in principle, to assemble at least one giant planet core, although the efficiency of this process is uncertain. If the observed substructures are primarily planet-induced, then their high incidence in this mass range is consistent with efficient (and possibly common) formation of planetary-mass companions capable of producing detectable pressure perturbations. Moreover, we note that a single planet can, under some disc conditions, generate multiple observable gaps (e.g., through secondary gaps and associated pressure structure), so our “stage” classification should be interpreted as requiring at least one planet rather than uniquely implying a one-planet architecture.

In the full ODISEA sample of $\sim$300 discs, there are only $\sim$60 objects with dust masses $\gtrsim$10 M$_{\oplus}$. 
This suggests that $\sim$20 $\%$  of the discs in Ophiuchus might result in planetary systems with at least one giant planet at separations $>$ 10 au, a parameter space that remains poorly explored by standard planet detection techniques (see Fig.\ref{fig:demog}). 
{However, the total number of members in Ophiuchus (including pre-main-sequence stars lacking discs) could be around $\sim$400-600 based on proper motion studies \citep{2020Esplin}; the overall incidence of these types of planets might be closer to 
$\sim$10-15$\%$. 
}

The other 40 of the 100 objects studied in this paper have dust masses ranging from 2 to 10 M${\oplus}$ and $R{68}$ disc sizes of $\sim$3 to $\sim$10 au. In the absence of a mechanism to halt the radial drift of mm-sized particles, these discs would likely be even more compact. If planet-induced pressure bumps are the dominant process preventing drift, disc size may trace the location of the outermost planet in a system \citep{2025Dasgupta}. Discs in the 2–10 M${\oplus}$ range are unlikely to form gas giants but could still form lower-mass giants similar to Uranus and Neptune, potentially generating substructures. As shown in Fig.~\ref{fig:frank_res}, current observations of discs with dust masses $\lesssim$10 M${\oplus}$ do not yet achieve the same relative resolution as higher-mass counterparts (i.e., $R_{68} > 3\times$ the \texttt{Frank} resolution), limiting our ability to test this scenario. Disc surveys have also not fully exploited ALMA’s highest angular resolution capabilities: the Band~8 data presented here used the C-7 configuration with maximum baselines of 3.6 km, reaching a nominal resolution of 50 mas (7 au at the distance of Ophiuchus), whereas the C-10 configuration provides baselines up to 16 km and roughly a factor of five higher resolution (9 mas, or 1.3 au).
   
According to the modeling results by \cite{2025ApJ...984L..57O}, a Mars-mass planet can generate inflection points in the continuum profile, whereas  Jupiter-mass planets can produce progressively deeper gaps and large cavities. Considering this, we performed a first-order approximation in which we placed the locations of our detected substructures, inflection points, gap minima, and cavity radii onto the confirmed exoplanet mass–separation diagram (Fig.\ref{fig:demog}). For simplicity, and for illustration purposes, here we assume that the location of inflection points would contain a Mars-mass perturber, gap minima would be created by a Saturn-mass planet, and cavity edges (defined at 90\% of the peak intensity) would have a Jupiter-mass planet.
{Within this working interpretation, only five resolved discs show multiple gap-ring pairs, corresponding to $\sim6.5\%$ of the resolved sample. If interpreted as signatures of multiple wide-orbit perturbers, this suggests that such architectures are uncommon in ODISEA; dedicated simulations with multiple embedded planets would be valuable to test this scenario \citep[e.g.,][]{2018Zhang,2025ApJ...984L..57O}}.

This exercise is intended as a qualitative comparison only. A full quantitative derivation of planet masses, such as that carried out by \cite{2018Zhang}, requires detailed modeling. 
Implementing such an analysis is therefore beyond the scope of this work.

Although many discs in our sample exhibit clear substructures, the corresponding feature locations, if interpreted as indicative planet positions, occupy a mass–separation parameter space that remains largely unexplored by current planet-detection techniques. Jupiter- and Saturn-mass planets at tens to hundreds of au lie below or at the edge of the sensitivity limits of existing high-contrast imagers, and accessing this regime will require the capabilities of the next generation of large telescopes. At the lower end, Mars-to-Earth-mass or lower-mass gas planetary cores, which may generate only subtle inflection points, remain far below the detection capabilities of current instruments. Detecting such low-mass, cold cores at large orbital distances will likely require substantial technological advances and may remain out of reach for several decades.
{
We also note that, while the substructures are likely due to the presence of forming planets, these protoplanets are still embedded within a gaseous disc and thus subject to migration.  Therefore, the final configuration of the planetary systems is likely to evolve with time. 
}

\section{Conclusions}
We have carried out a uniform analysis of $\sim$100 protoplanetary discs in Ophiuchus using \texttt{Frank} deprojected radial profiles, providing a consistent characterization of disc sizes, substructures, and effective spatial resolutions across the entire Band 8 continuum sample. This represents the most complete structural demographic study of the region to date, and establishes a homogeneous framework for interpreting disc evolution and the possible signatures of emerging planets.

\begin{enumerate}

    \item We find that Band~8 observations are highly efficient for identifying disc 
    substructures, {despite the higher optical depths expected at shorter 
    wavelengths}. In several cases, Band~8 recovers the same gaps and cavities 
    previously identified at longer wavelengths, while requiring significantly 
    shorter integration times. Across the 101 Frank profiles, we identify 
    25 inflection points in 20 discs, 26 gap--ring pairs in 17 discs, and 15 discs 
    with cavities, including 20 previously unreported substructures.
    
    \item We find that substructures are present across all SED classes. Even among
    the youngest Class~I/F objects, we identify discs with clear gaps and cavities,
    showing that substructure formation can begin early. At the same time, some
    Class~II discs remain smooth. For the better-resolved, higher-mass discs, the
    incidence of evolved substructures, namely gaps and cavities rather than only
    inflection points, increases from the Class~I/F to Class~II population.
    {When bolometric temperature is used as an alternative evolutionary
    tracer, substructures are also found in both colder and warmer discs; however,
    within the resolved ODISEA sample we do not find a statistically significant
    change in the structured-disc fraction across $T_{\rm bol}=650$~K.}
    
    \item When splitting the sample by dust mass, discs above {$10~M_{\oplus}$}
    provide the clearest support for the proposed morphological sequence {which progresses from featureless discs (Stage 0), through weak inflection-point structures (Stage I) and gap–ring systems (Stages II–III), to cavity-bearing discs (Stage IV) and finally single dominant rings (Stage V).}
    {This sequence, originally proposed by \citet{2021Cieza}, is
    physically motivated by the planet--disc interaction models of
    \citet{2025ApJ...984L..57O}, which reproduce a progression from weak
    substructures to gap--ring pairs, cavities, and narrow rings as dust evolves in
    the presence of embedded planets.} The higher-mass discs are generally better
    resolved and show a higher incidence of gaps and cavities, while the
    interpretation of compact, low-mass discs remains limited by spatial resolution
    and sensitivity.
    
    \item {Many feature locations fall within the parameter space expected for
    Saturn--Jupiter mass planets at $\sim$10--100~au, while weaker
    inflection-point-like features may trace lower-mass planetary cores at earlier
    stages. This suggests that Ophiuchus may host a population of embedded planets
    that current direct-imaging surveys are not yet sensitive to.}
    
    \item The structural catalogue presented here provides a benchmark for selecting
    high-priority targets for future high-resolution observations. Because many
    compact, faint discs currently sit at the {1--2$\times$ Frank
    resolution limit}, the sample naturally identifies systems where deeper,
    longer-baseline ALMA observations, especially in Band~8, could reveal previously
    unresolved features.
   
\end{enumerate}

\section*{Data availability}
Tables 1 and 2 are only available in electronic form at the CDS via anonymous ftp to cdsarc.u-strasbg.fr (130.79.128.5) or via \url{http://cdsweb.u-strasbg.fr/cgi-bin/qcat?J/A+A/.}

\begin{acknowledgement}

The authors acknowledge support from ANID -- Millennium Science Initiative Program -- Center Code NCN2024\_001.
T.B. acknowledges financial support from the FONDECYT postdoctorado project number 3230470. 
L.C. acknowledges support from the FONDECYT Regular grant number 1241056.
A.Z. acknowledges support from the FONDECYT Regular grant number 1250249.
S.C. acknowledges support from Agencia Nacional de Investigaci\'on y Desarrollo de Chile (ANID) given by FONDECYT Regular grant 1251456, and ANID project Data Observatory Foundation DO210001. S.P. acknowledges support from FONDECYT 1231663 and ANID FIUF137139-USACH
This paper makes use of the following ALMA data: ADS/JAO.ALMA\#2021.1.00378.S and ADS/JAO.ALMA\#2021.1.00480.S. ALMA is a partnership of ESO (representing its member states), NSF (USA) and NINS (Japan), together with NRC (Canada), NSTC and ASIAA (Taiwan), and KASI (Republic of Korea), in cooperation with the Republic of Chile. The Joint ALMA Observatory is operated by ESO, AUI/NRAO and NAOJ.
\end{acknowledgement}

\bibliography{reference}

\begin{appendix}

\section{Frank fitting spectrally unaveraged visibilities}

{For \ODISEA{143}, we explored a grid of \texttt{frank} hyperparameters ($\alpha = 1.05,\,1.1,\,1.2,\,1.3$ and $w_{\rm smooth} = 10^{-1},\,10^{-2},\,10^{-3},\,10^{-4}$). For each fit, we computed the mean absolute value of the real component of the uv-binned residual visibilities and their standard deviation. The model with the lowest mean absolute value of the residual visibilities was adopted as the best-fit solution, while the standard deviation was used as a diagnostic of the residual scatter.}

\begin{figure*}[!h]
    \includegraphics[width=0.98\linewidth]{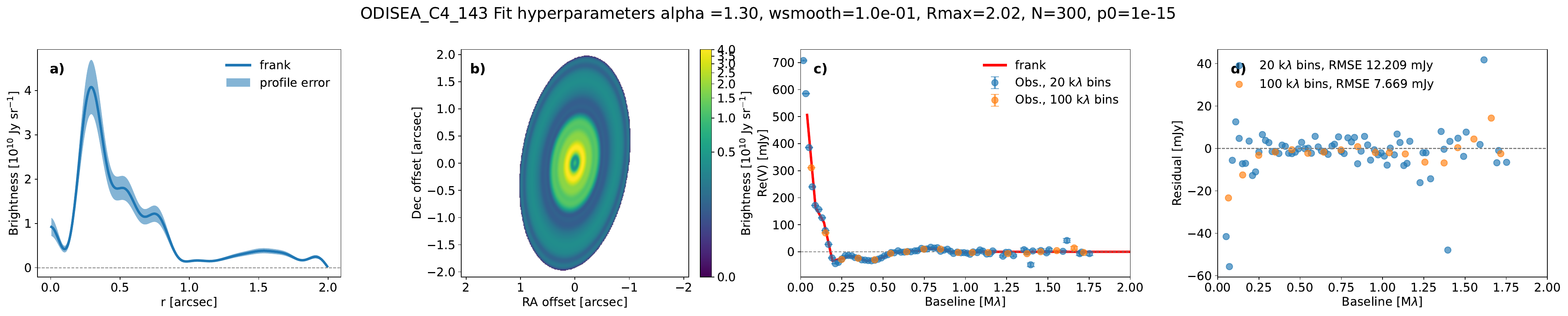}\\
    \includegraphics[width=0.98\linewidth]{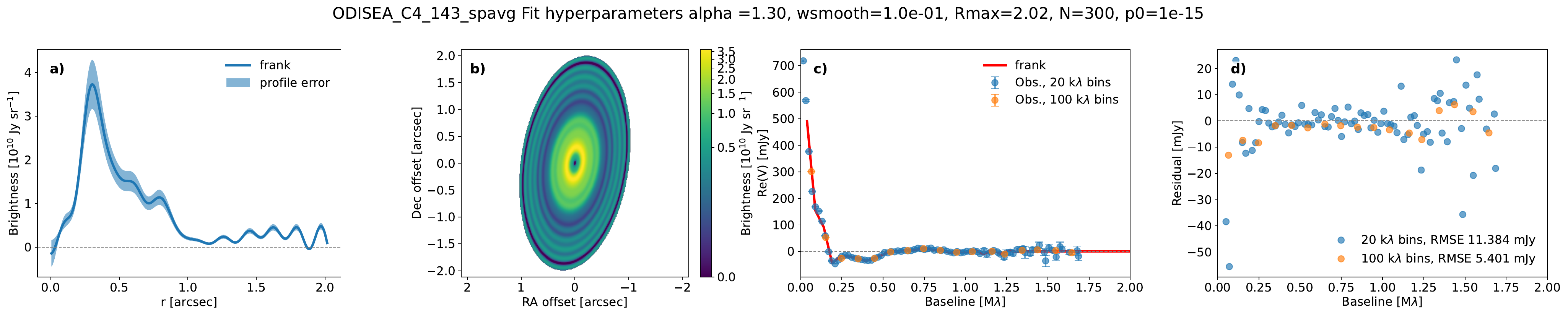}\\
    \includegraphics[width=0.98\linewidth]{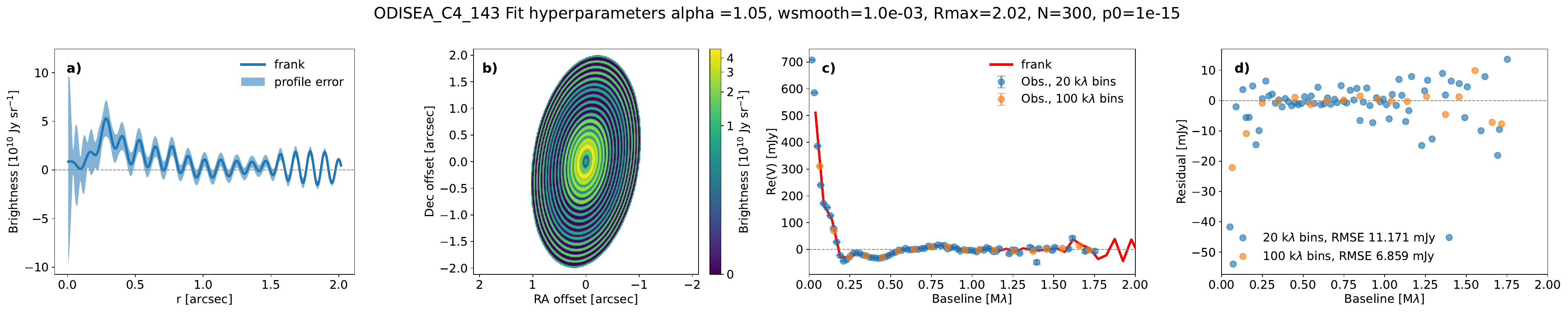}
    \caption{Comparison of the \texttt{frank} fits for \ODISEA{143}. From left to right, the panels show: (a) the radial brightness profile, with the shaded region indicating the uncertainty including the 15\% Band~8 flux-calibration uncertainty added in quadrature; (b) the axisymmetric model image generated by sweeping the radial profile over $2\pi$ and reprojecting it using the adopted inclination and position angle; (c) the fit to the real part of the deprojected visibilities; and (d) the residuals between the binned visibilities and the \texttt{frank} model. The rows correspond to (Top) the spectrally unaveraged visibilities of \ODISEA{143} using $\alpha=1.30$ and $w_{\rm smooth}=1.0\times10^{-1}$, (Middle) the spectrally averaged visibilities with $\alpha=1.30$ and $w_{\rm smooth}=1.0\times10^{-1}$, and (Bottom) the statistically preferred model for the spectrally unaveraged visibilities with $\alpha=1.05$ and $w_{\rm smooth}=1.0\times10^{-3}$ (mean and standard deviation of the residuals: $0.00455\pm0.00738$).}
    \label{fig:overfitting}
\end{figure*}

\section{Remaining discs in each stage}

\subsection{Stage 0}
\noindent
\begin{minipage}{.49\textwidth}
	 \centering
	 	 \hrulesep
	 	 \includegraphics[width=1\linewidth]{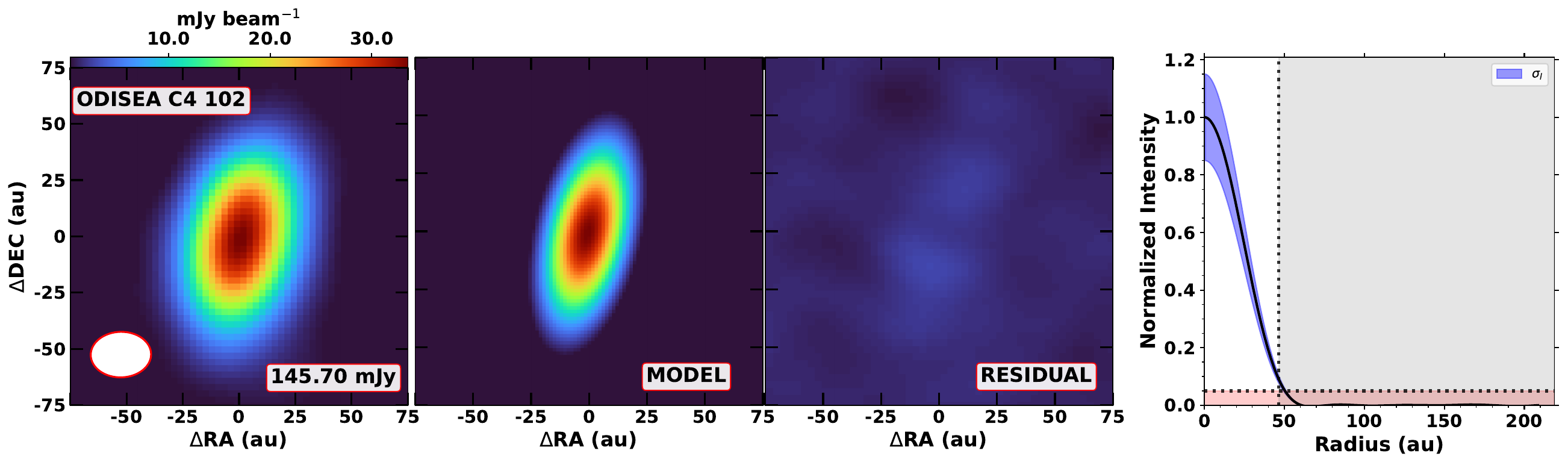}
\end{minipage}%
\vrulesep
\noindent
\begin{minipage}{.49\textwidth}
	 \centering
	 	 \hrulesep
	 	 \includegraphics[width=1\linewidth]{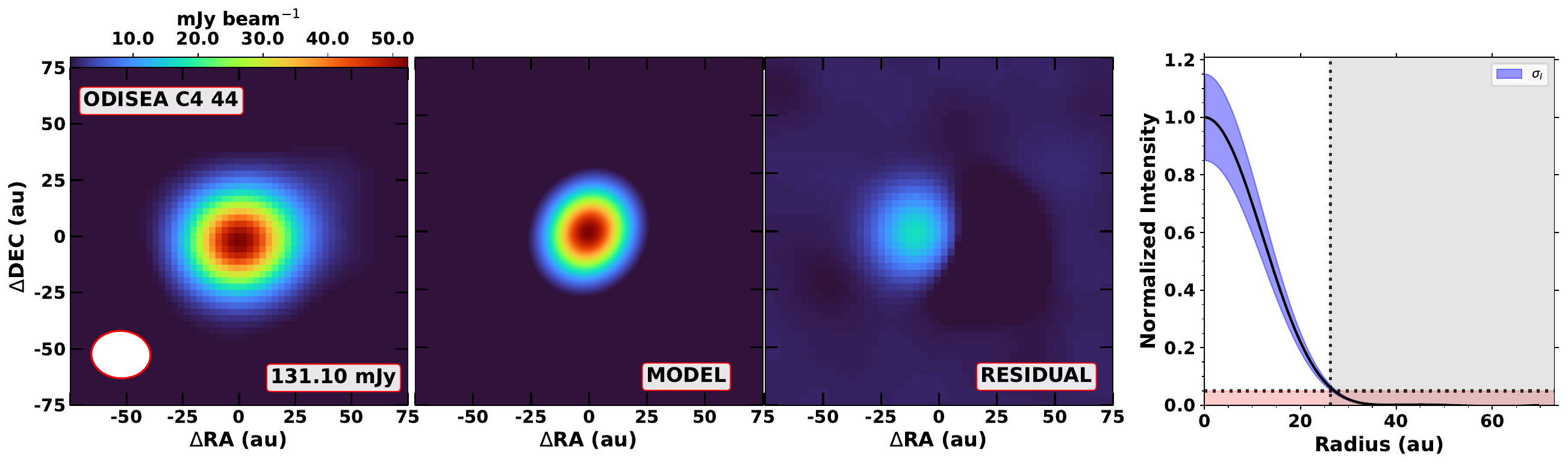}
\end{minipage}%
\vrulesep
\noindent
\begin{minipage}{.49\textwidth}
	 \centering
	 	 \hrulesep
	 	 \includegraphics[width=1\linewidth]{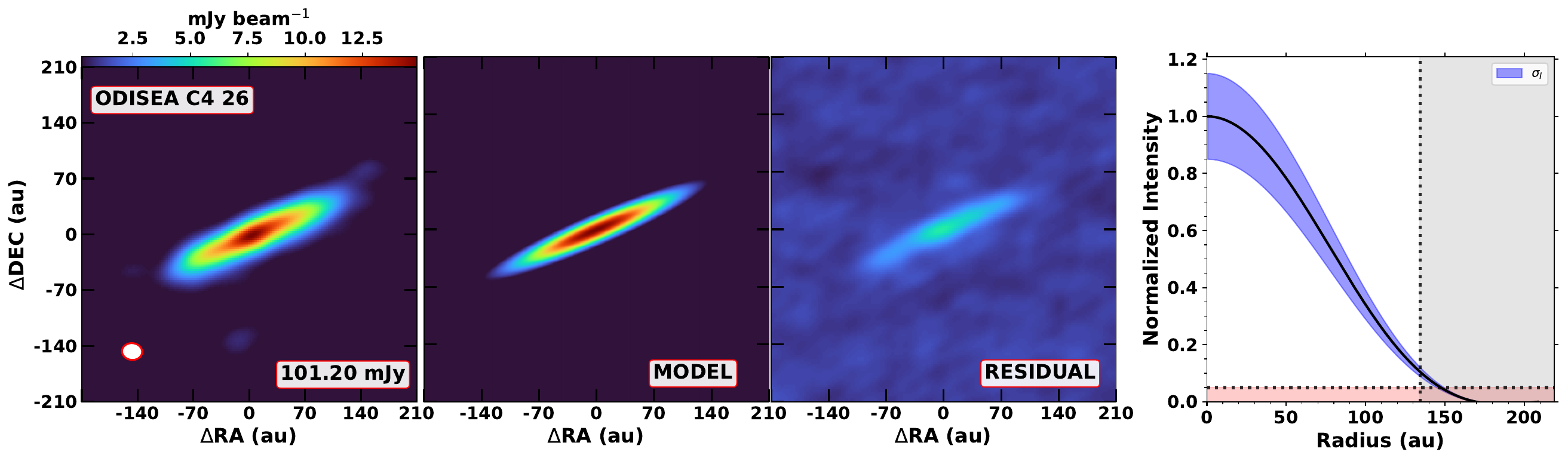}
\end{minipage}%
\vrulesep
\noindent
\begin{minipage}{.49\textwidth}
	 \centering
	 	 \hrulesep
	 	 \includegraphics[width=1\linewidth]{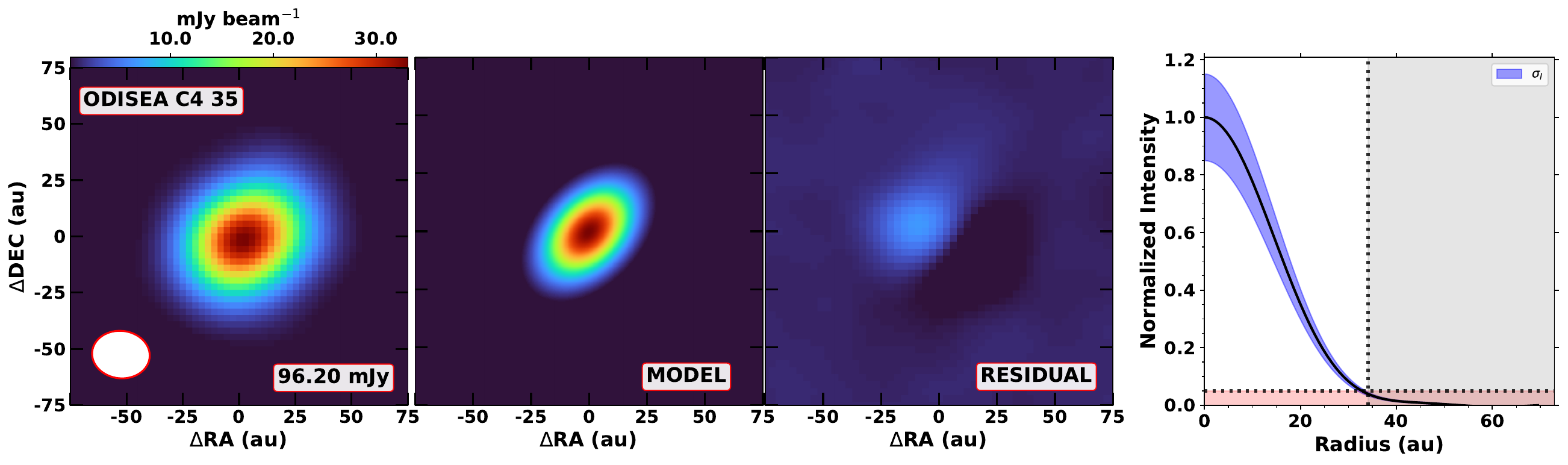}
\end{minipage}%
\vrulesep
\noindent
\begin{minipage}{.49\textwidth}
	 \centering
	 	 \hrulesep
	 	 \includegraphics[width=1\linewidth]{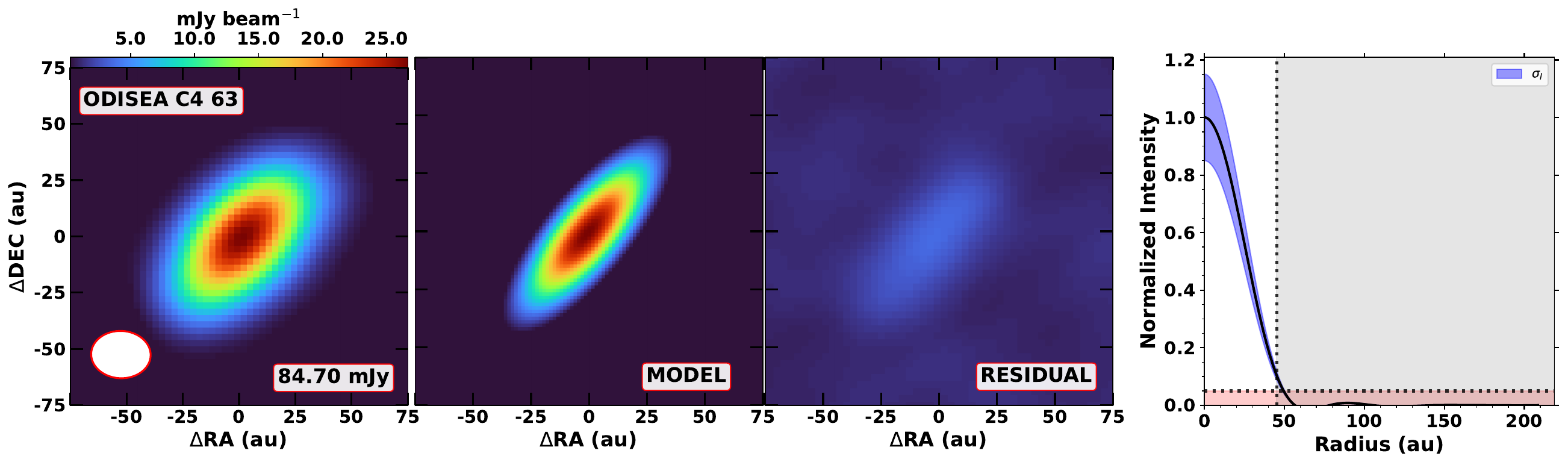}
\end{minipage}%
\vrulesep
\noindent
\begin{minipage}{.49\textwidth}
	 \centering
	 	 \hrulesep
	 	 \includegraphics[width=1\linewidth]{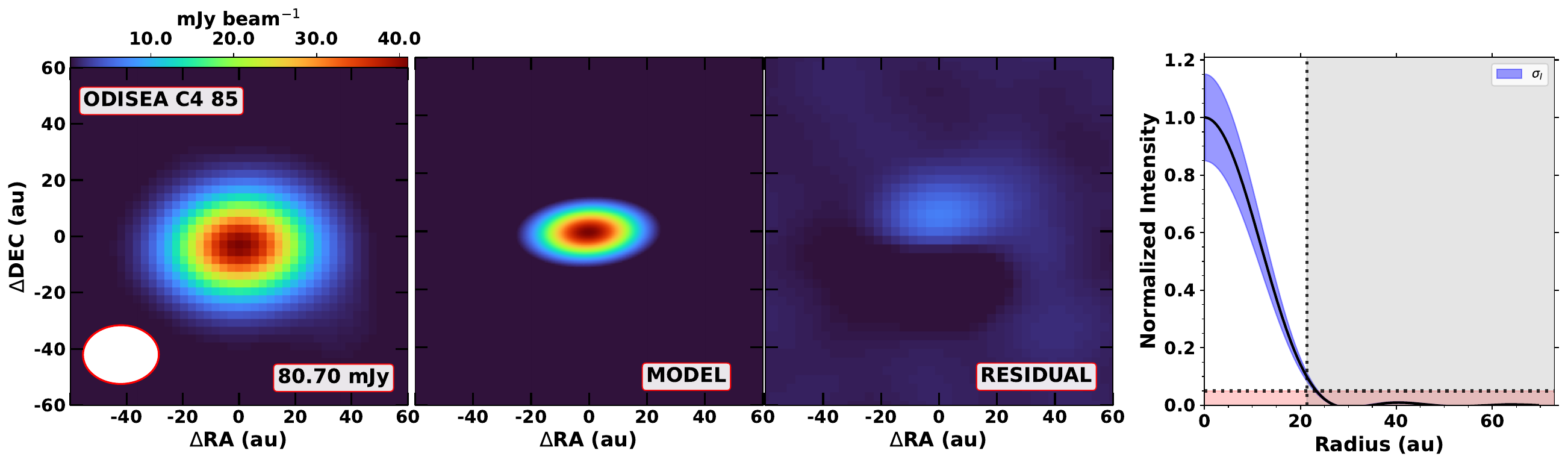}
\end{minipage}%
\vrulesep
\noindent
\begin{minipage}{.49\textwidth}
	 \centering
	 	 \hrulesep
	 	 \includegraphics[width=1\linewidth]{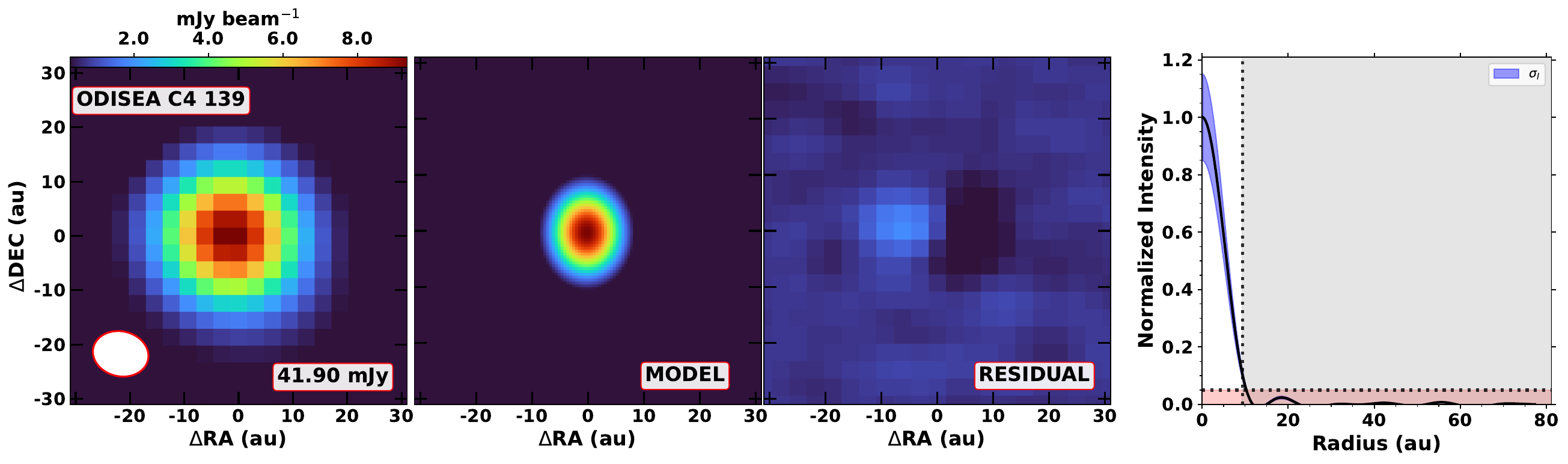}
\end{minipage}%
\vrulesep
\noindent
\begin{minipage}{.49\textwidth}
	 \centering
	 	 \hrulesep
	 	 \includegraphics[width=1\linewidth]{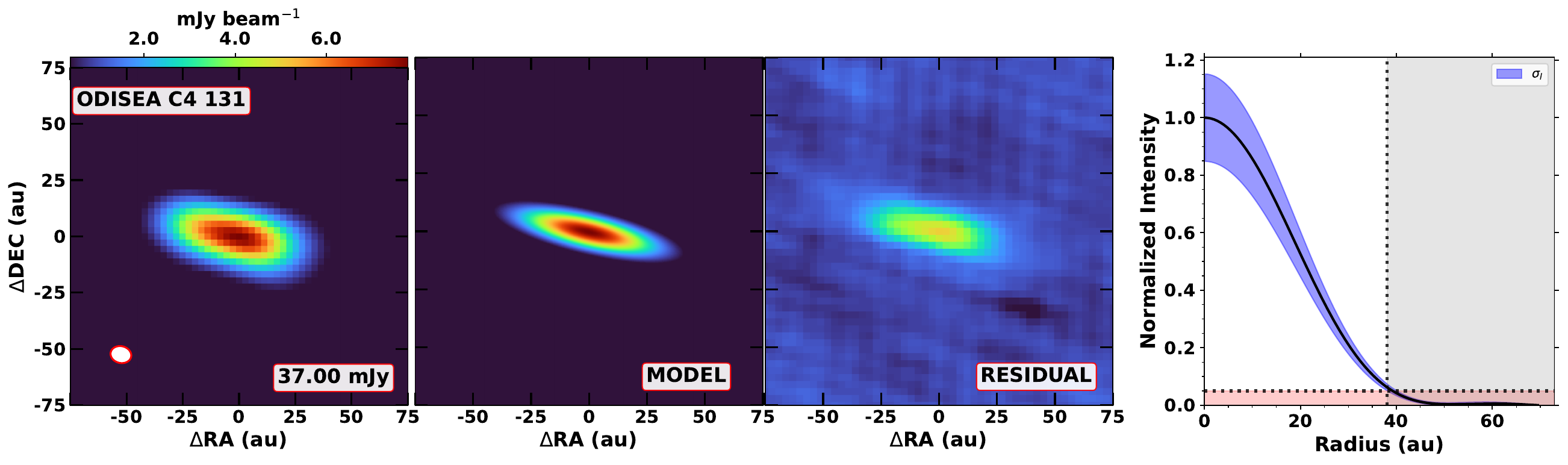}
\end{minipage}%
\vrulesep
\noindent
\begin{minipage}{.49\textwidth}
	 \centering
	 	 \hrulesep
	 	 \includegraphics[width=1\linewidth]{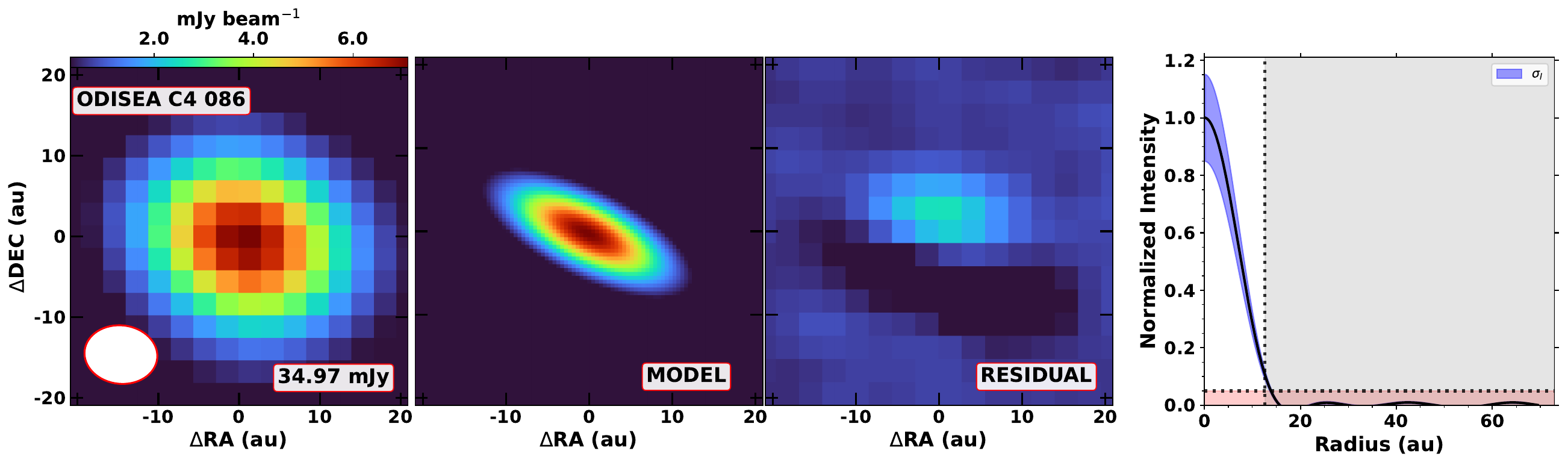}
\end{minipage}%
\vrulesep
\noindent
\begin{minipage}{.49\textwidth}
	 \centering
	 	 \hrulesep
	 	 \includegraphics[width=1\linewidth]{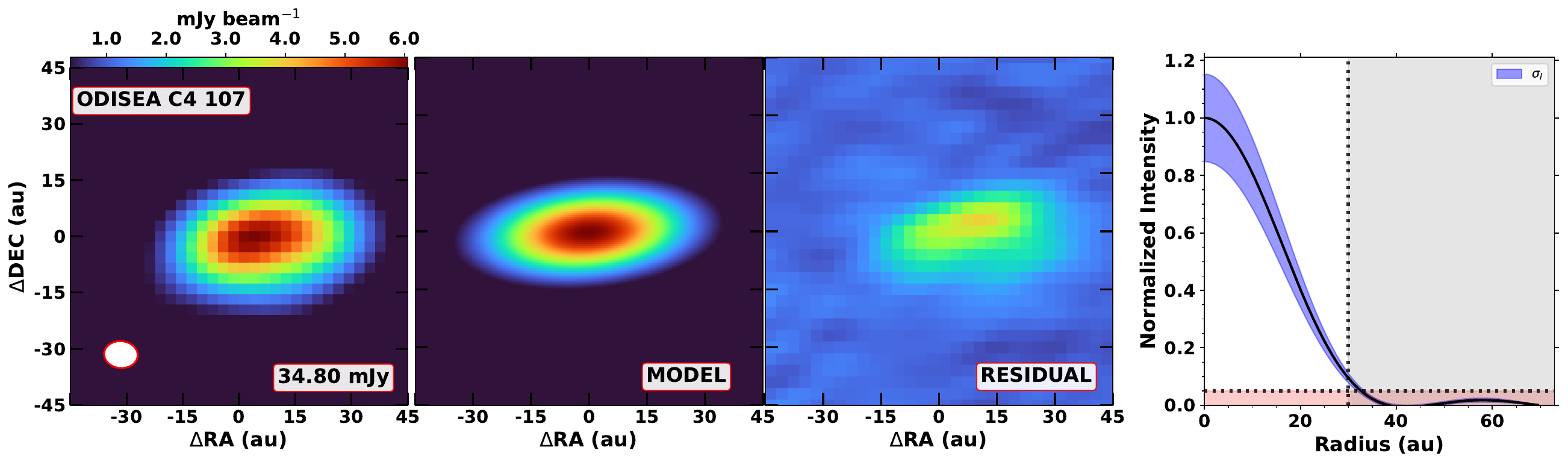}
\end{minipage}%
\vrulesep
\noindent
\begin{minipage}{.49\textwidth}
	 \centering
	 	 \hrulesep
	 	 \includegraphics[width=1\linewidth]{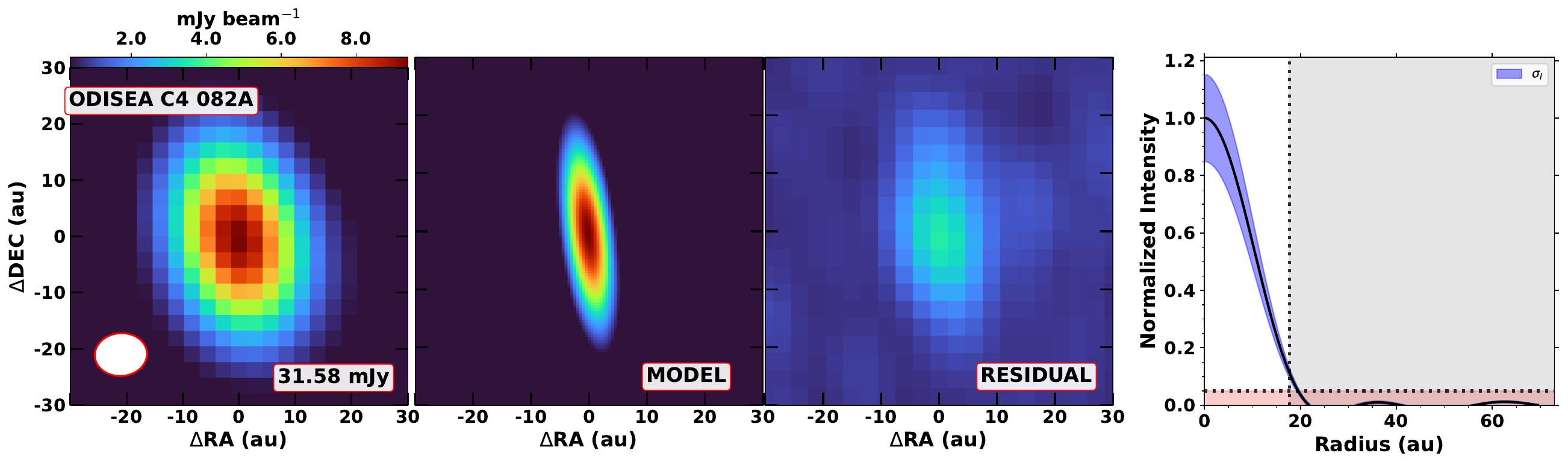}
\end{minipage}%
\vrulesep
\noindent
\begin{minipage}{.49\textwidth}
	 \centering
	 	 \hrulesep
	 	 \includegraphics[width=1\linewidth]{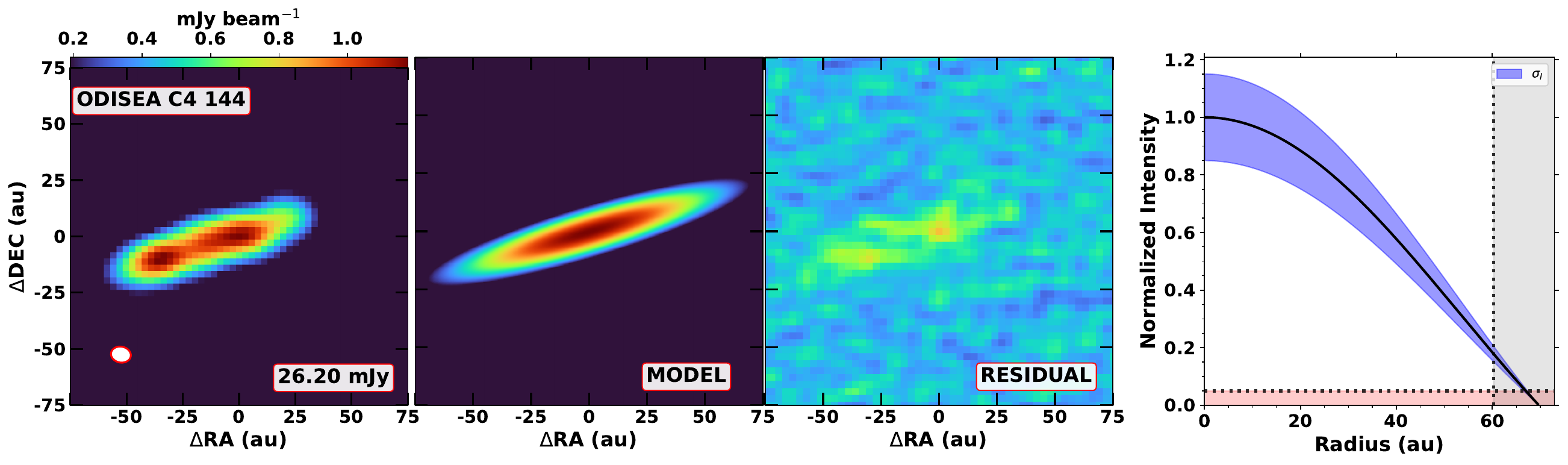}
\end{minipage}%
\vrulesep
\noindent
\begin{minipage}{.49\textwidth}
	 \centering
	 	 \hrulesep
	 	 \includegraphics[width=1\linewidth]{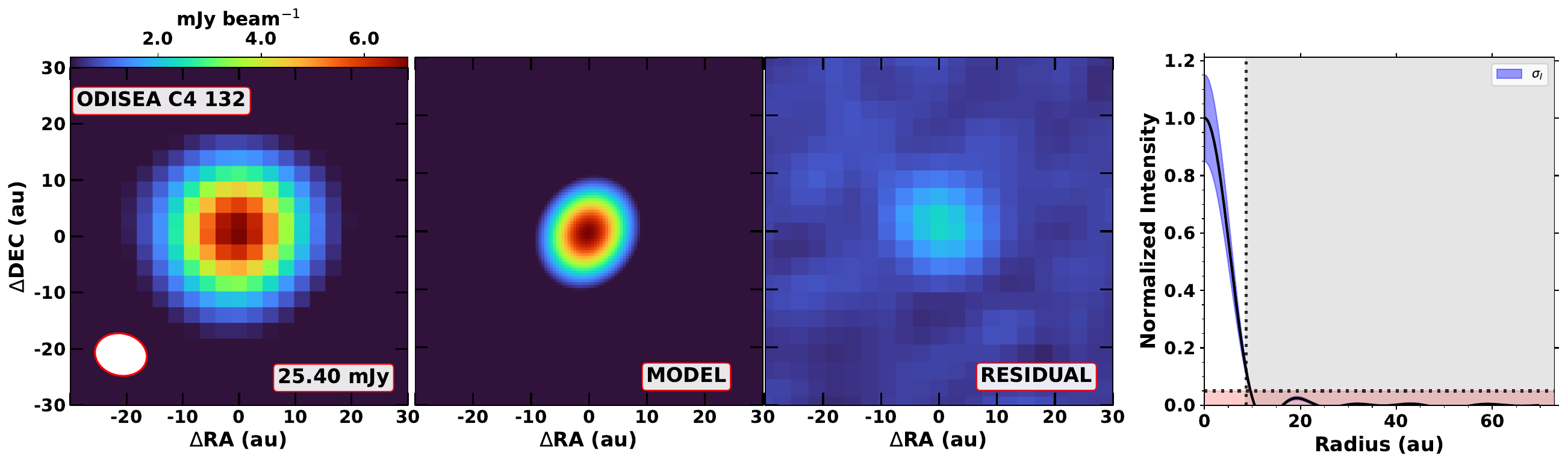}
\end{minipage}%
\vrulesep
\noindent
\begin{minipage}{.49\textwidth}
	 \centering
	 	 \hrulesep
	 	 \includegraphics[width=1\linewidth]{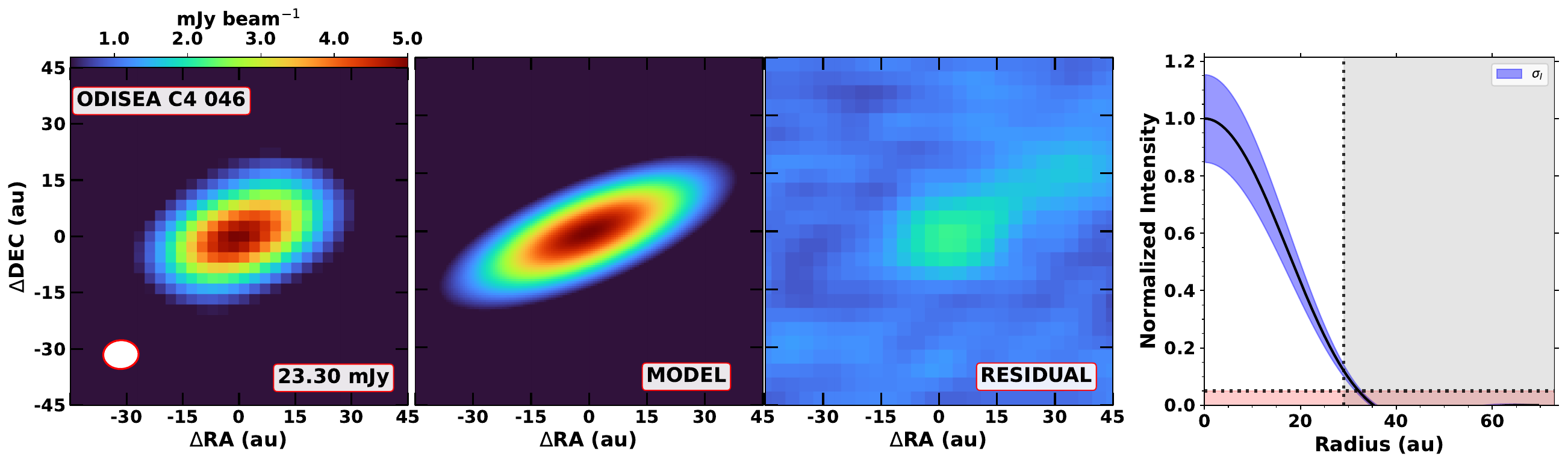}
\end{minipage}%
\vrulesep
\noindent
\begin{minipage}{.49\textwidth}
	 \centering
	 	 \hrulesep
	 	 \includegraphics[width=1\linewidth]{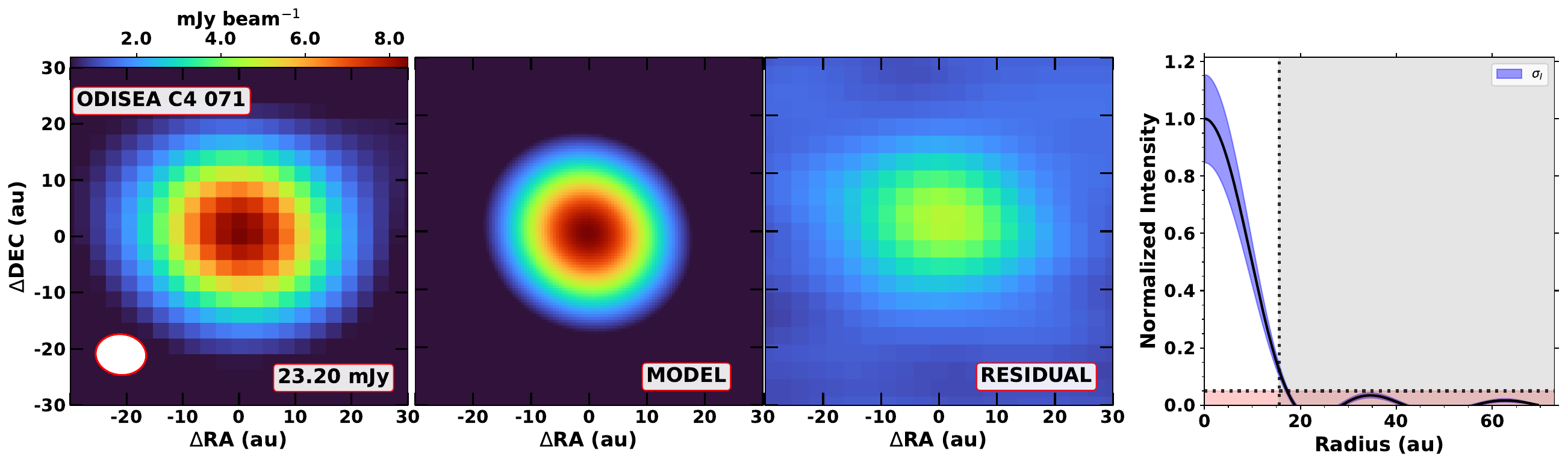}
\end{minipage}%
\vrulesep
\noindent
\begin{minipage}{.49\textwidth}
	 \centering
	 	 \hrulesep
	 	 \includegraphics[width=1\linewidth]{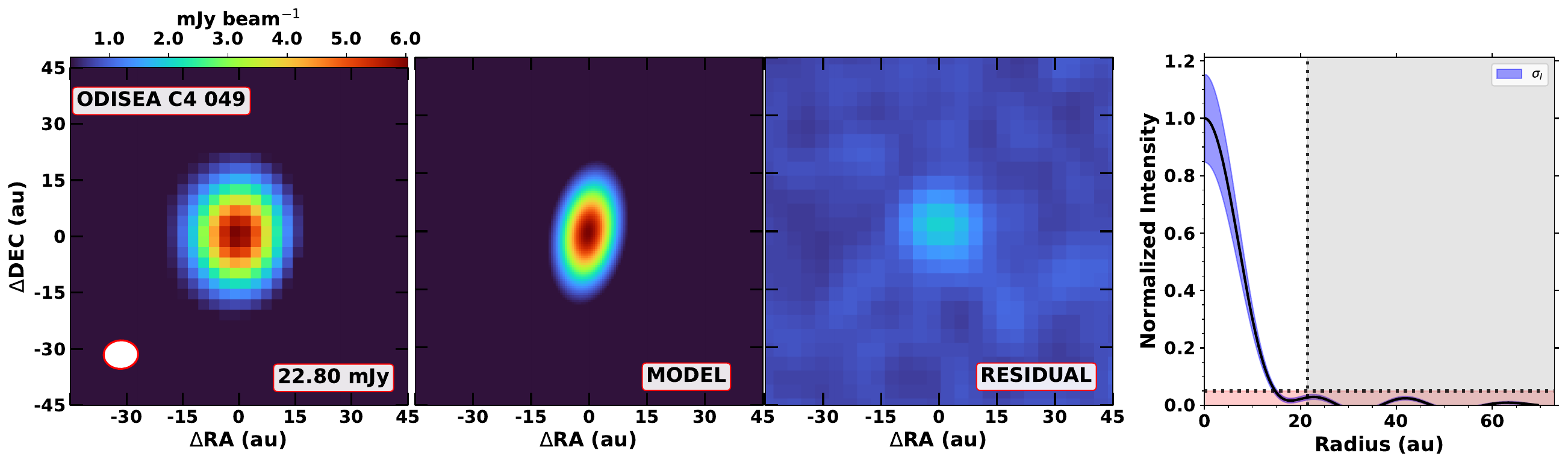}
\end{minipage}%
\vrulesep
\noindent
\begin{minipage}{.49\textwidth}
	 \centering
	 	 \hrulesep
	 	 \includegraphics[width=1\linewidth]{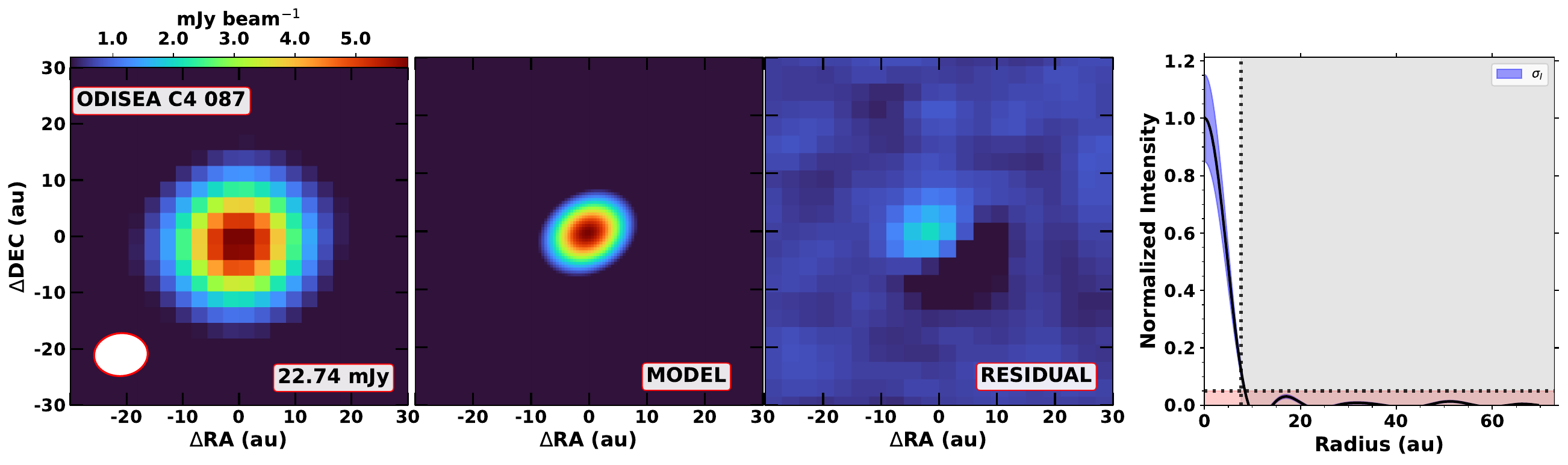}
\end{minipage}%
\vrulesep
\noindent
\begin{minipage}{.49\textwidth}
	 \centering
	 	 \hrulesep
	 	 \includegraphics[width=1\linewidth]{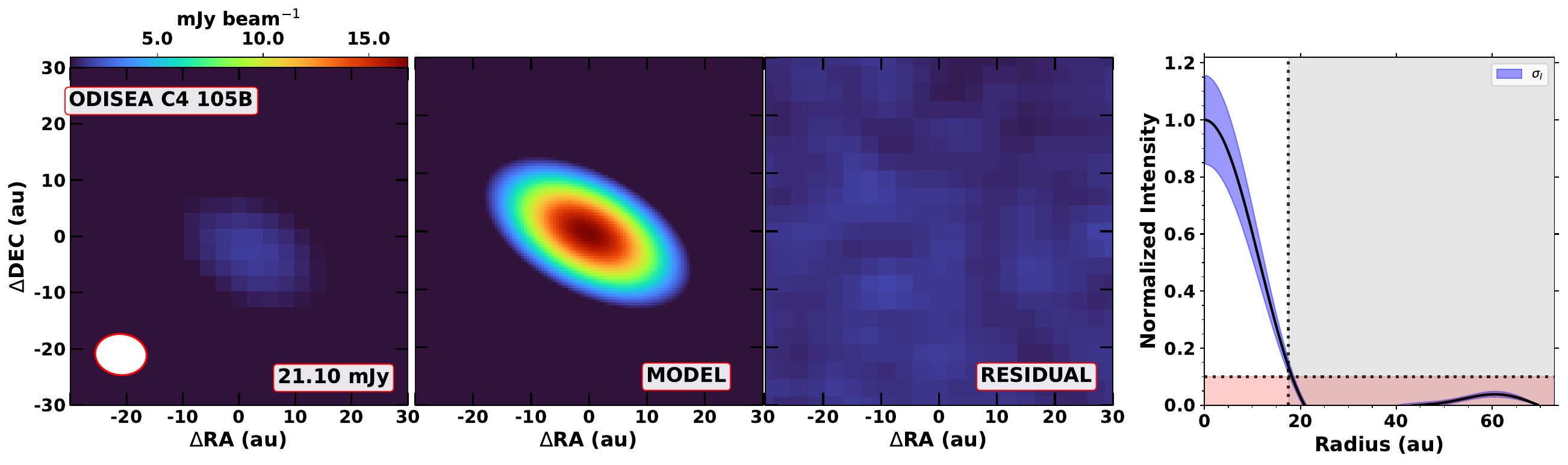}
\end{minipage}%
\vrulesep
\noindent
\begin{minipage}{.49\textwidth}
	 \centering
	 	 \hrulesep
	 	 \includegraphics[width=1\linewidth]{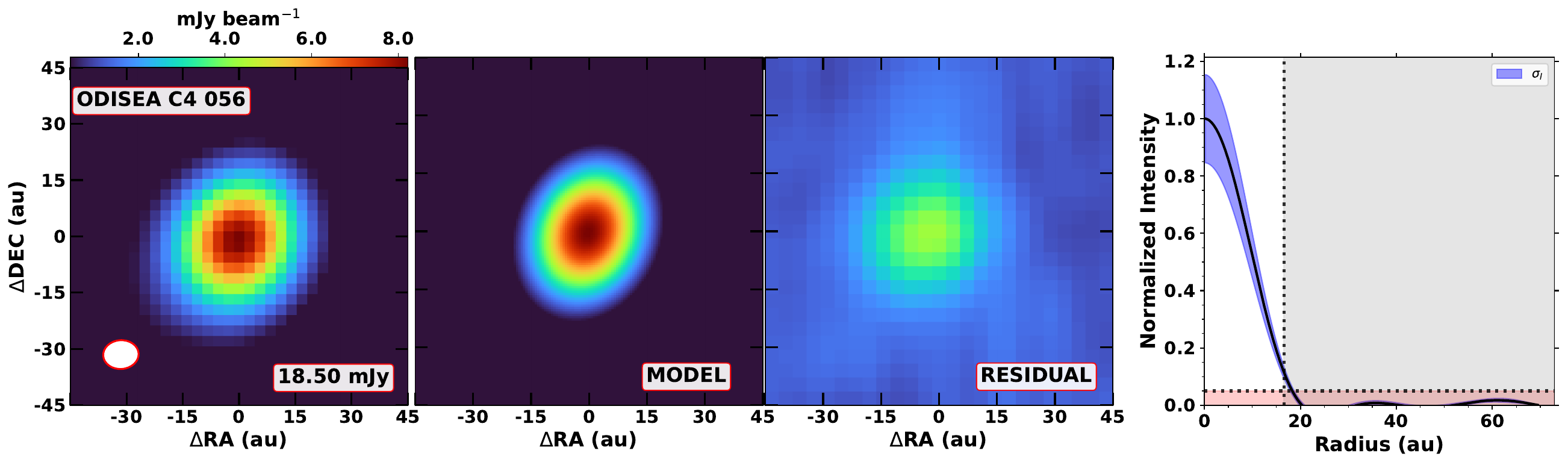}
\end{minipage}%
\vrulesep
\noindent
\begin{minipage}{.49\textwidth}
	 \centering
	 	 \hrulesep
	 	 \includegraphics[width=1\linewidth]{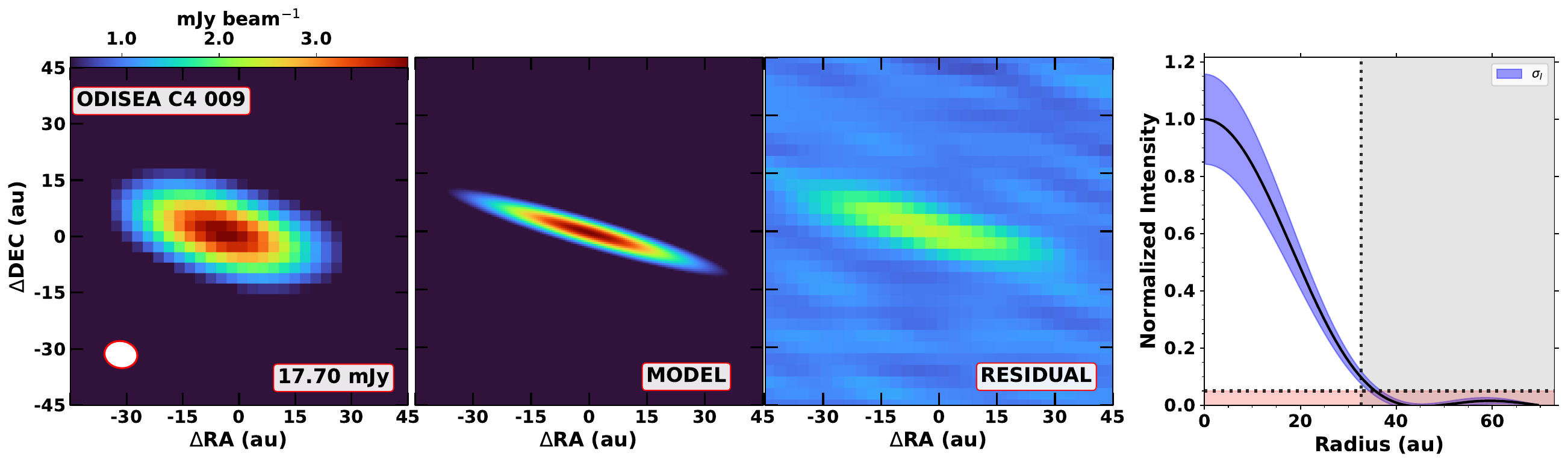}
\end{minipage}%
\vrulesep
\noindent
\begin{minipage}{.49\textwidth}
	 \centering
	 	 \hrulesep
	 	 \includegraphics[width=1\linewidth]{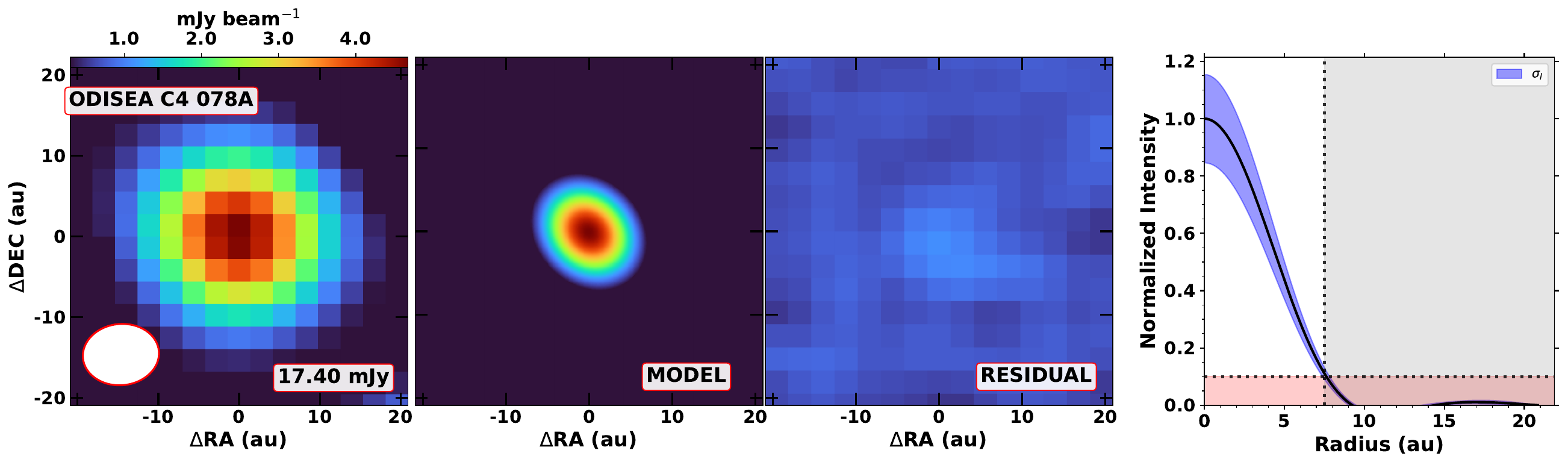}
\end{minipage}%
\vrulesep
\noindent
\begin{minipage}{.49\textwidth}
	 \centering
	 	 \hrulesep
	 	 \includegraphics[width=1\linewidth]{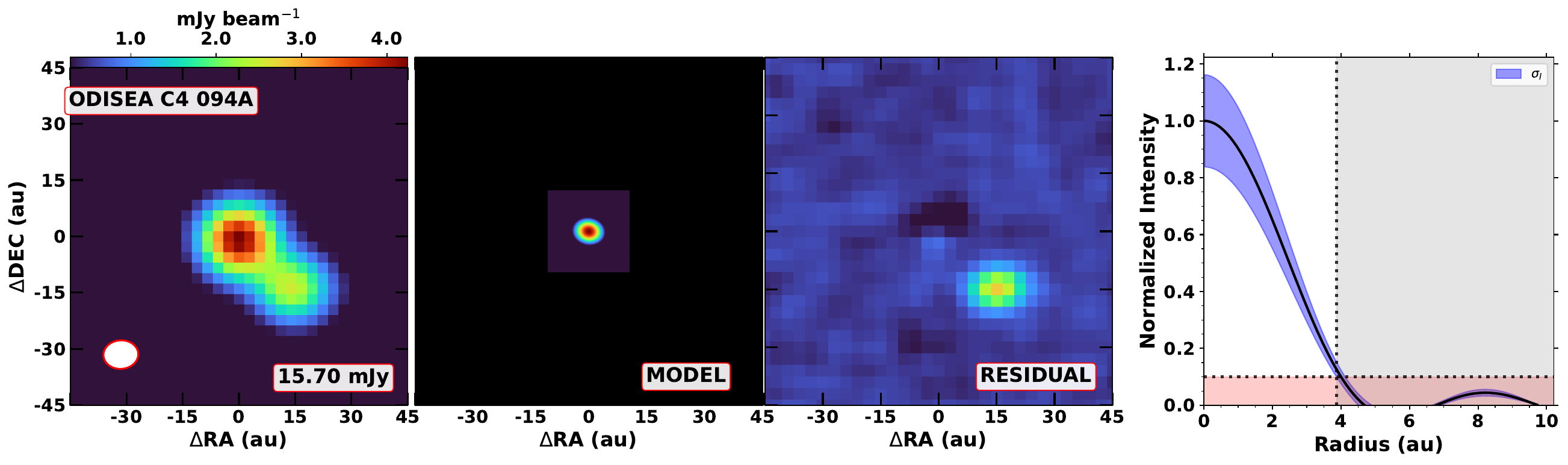}
\end{minipage}%
\vrulesep
\noindent
\begin{minipage}{.49\textwidth}
	 \centering
	 	 \hrulesep
	 	 \includegraphics[width=1\linewidth]{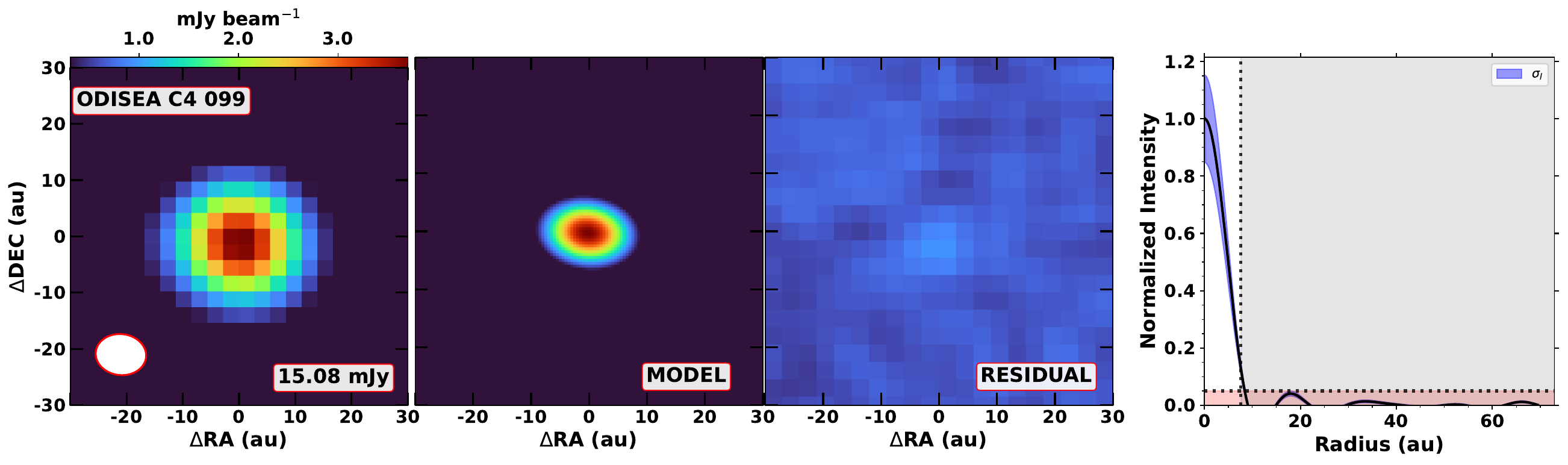}
\end{minipage}%
\vrulesep
\noindent
\begin{minipage}{.49\textwidth}
	 \centering
	 	 \hrulesep
	 	 \includegraphics[width=1\linewidth]{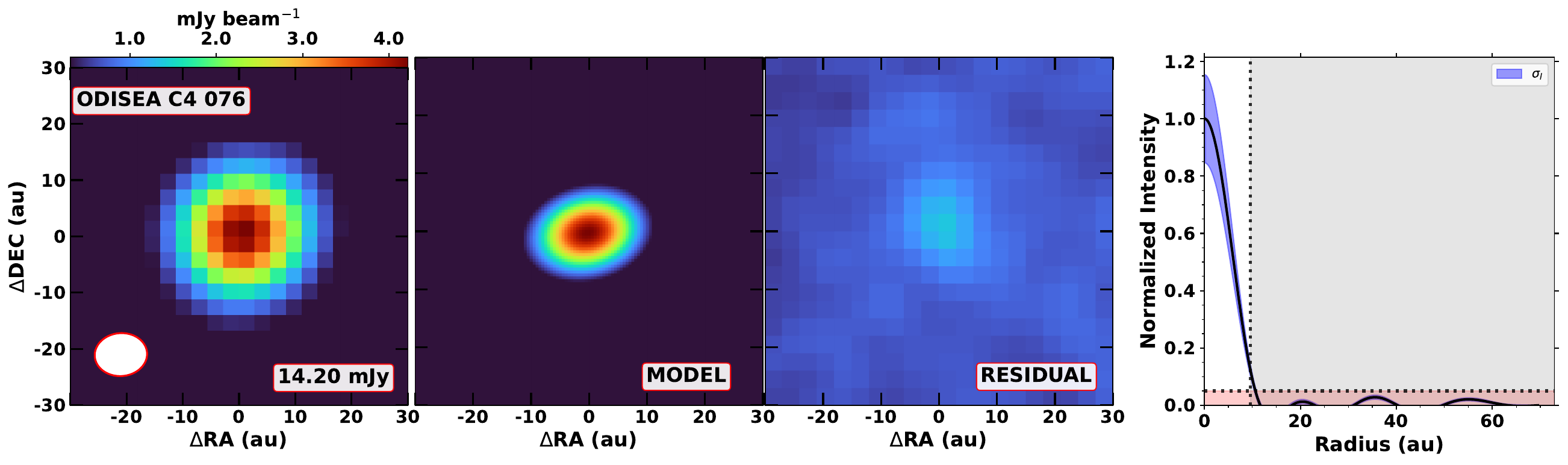}
\end{minipage}%
\vrulesep
\noindent
\begin{minipage}{.49\textwidth}
	 \centering
	 	 \hrulesep
	 	 \includegraphics[width=1\linewidth]{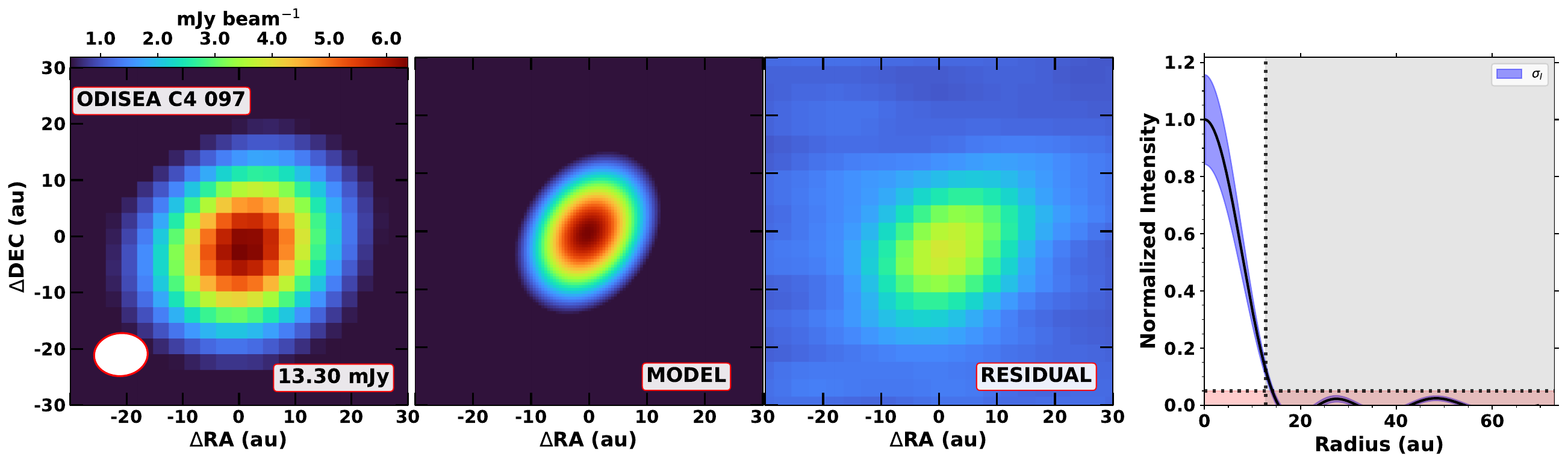}
\end{minipage}%
\vrulesep
\noindent
\begin{minipage}{.49\textwidth}
	 \centering
	 	 \hrulesep
	 	 \includegraphics[width=1\linewidth]{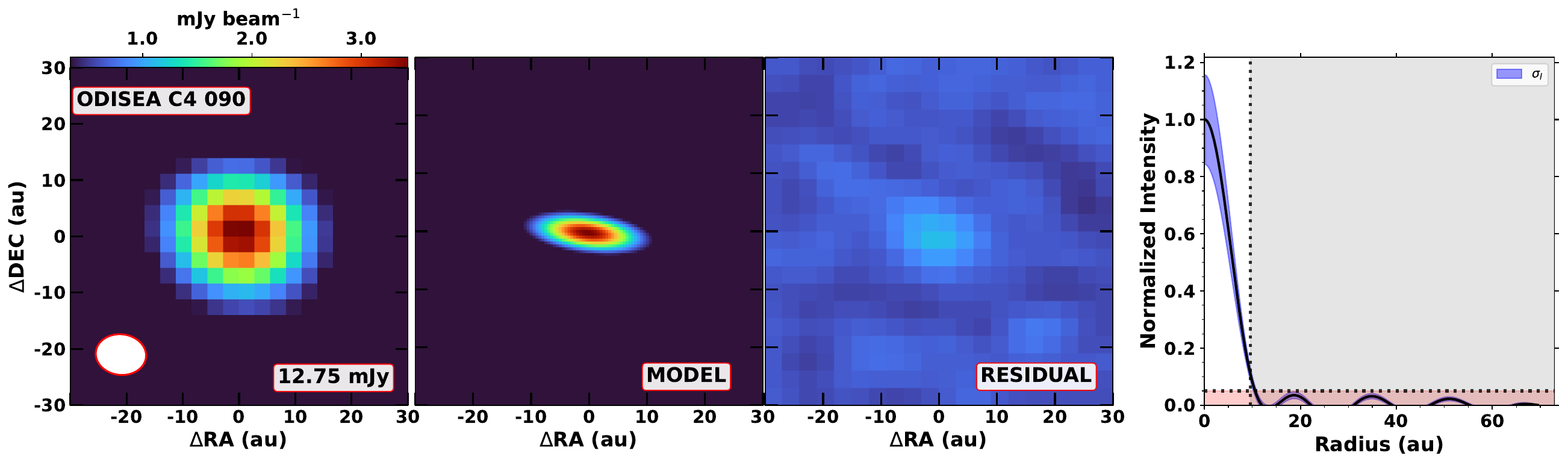}
\end{minipage}%
\vrulesep
\noindent
\begin{minipage}{.49\textwidth}
	 \centering
	 	 \hrulesep
	 	 \includegraphics[width=1\linewidth]{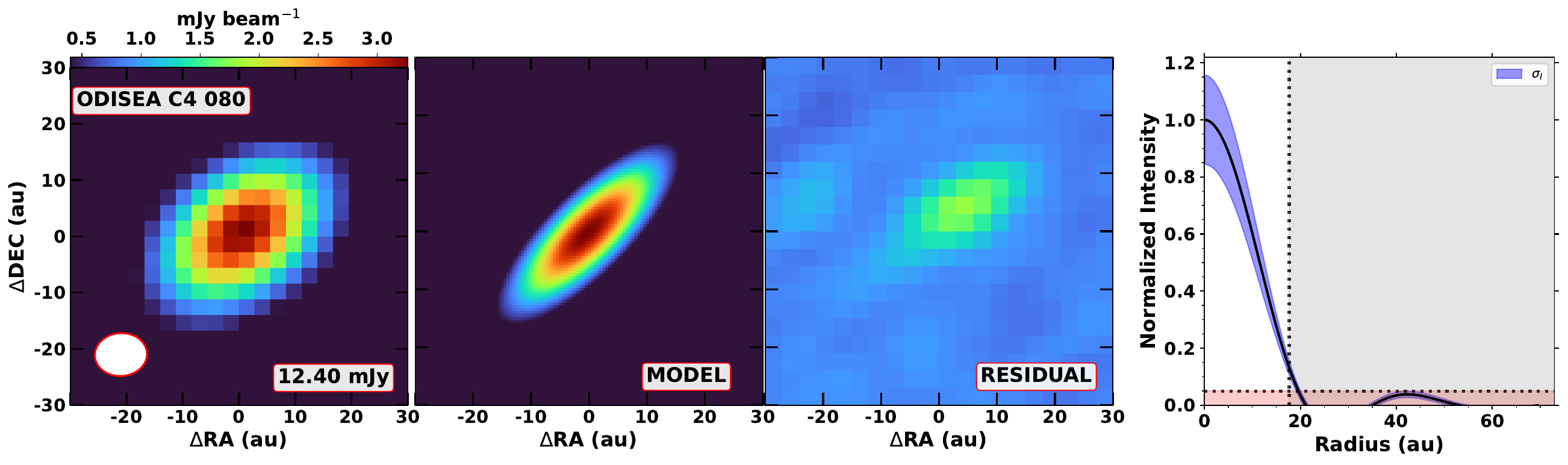}
\end{minipage}%
\vrulesep
\noindent
\begin{minipage}{.49\textwidth}
	 \centering
	 	 \hrulesep
	 	 \includegraphics[width=1\linewidth]{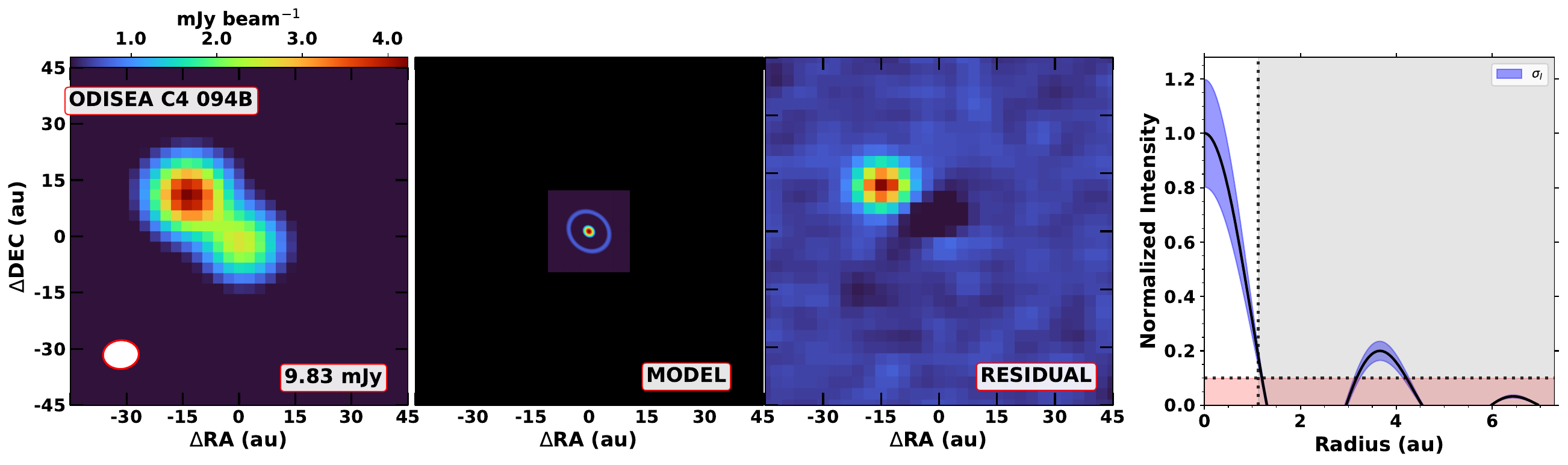}
\end{minipage}%
\vrulesep
\noindent
\begin{minipage}{.49\textwidth}
	 \centering
	 	 \hrulesep
	 	 \includegraphics[width=1\linewidth]{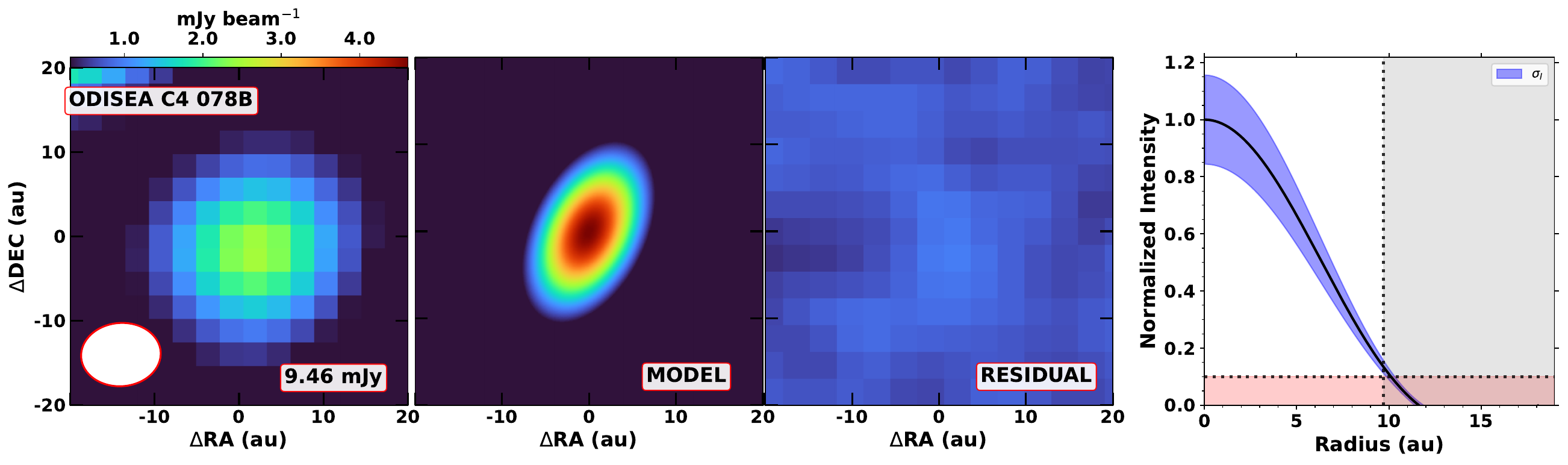}
\end{minipage}%
\vrulesep
\noindent
\begin{minipage}{.49\textwidth}
	 \centering
	 	 \hrulesep
	 	 \includegraphics[width=1\linewidth]{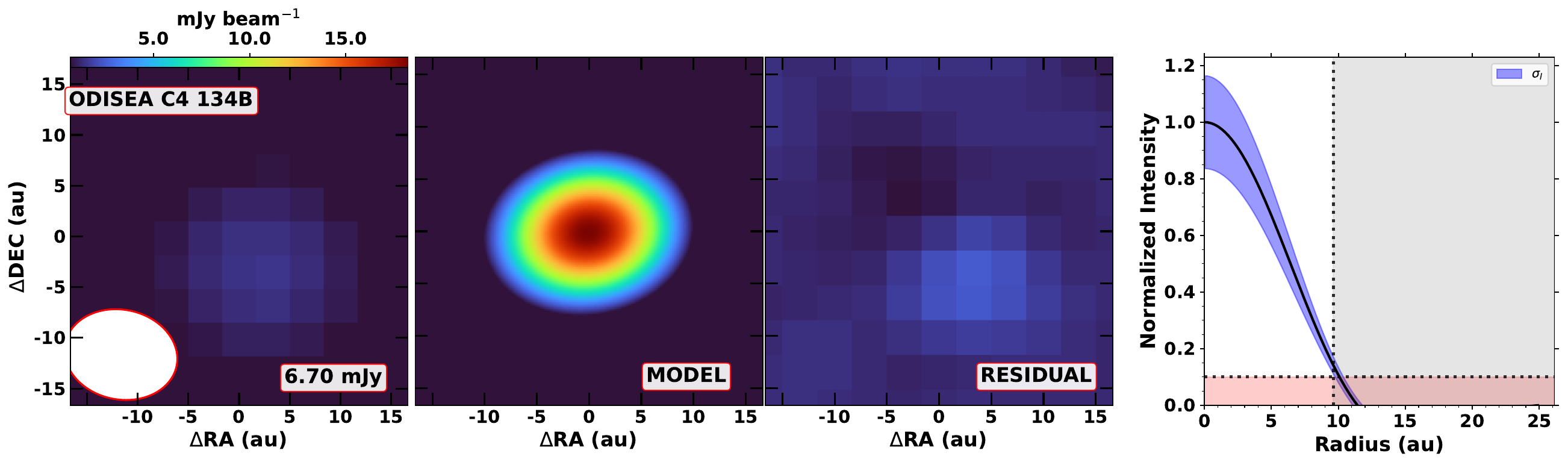}
\end{minipage}%
\vrulesep
\noindent
\begin{minipage}{.49\textwidth}
	 \centering
	 	 \hrulesep
	 	 \includegraphics[width=1\linewidth]{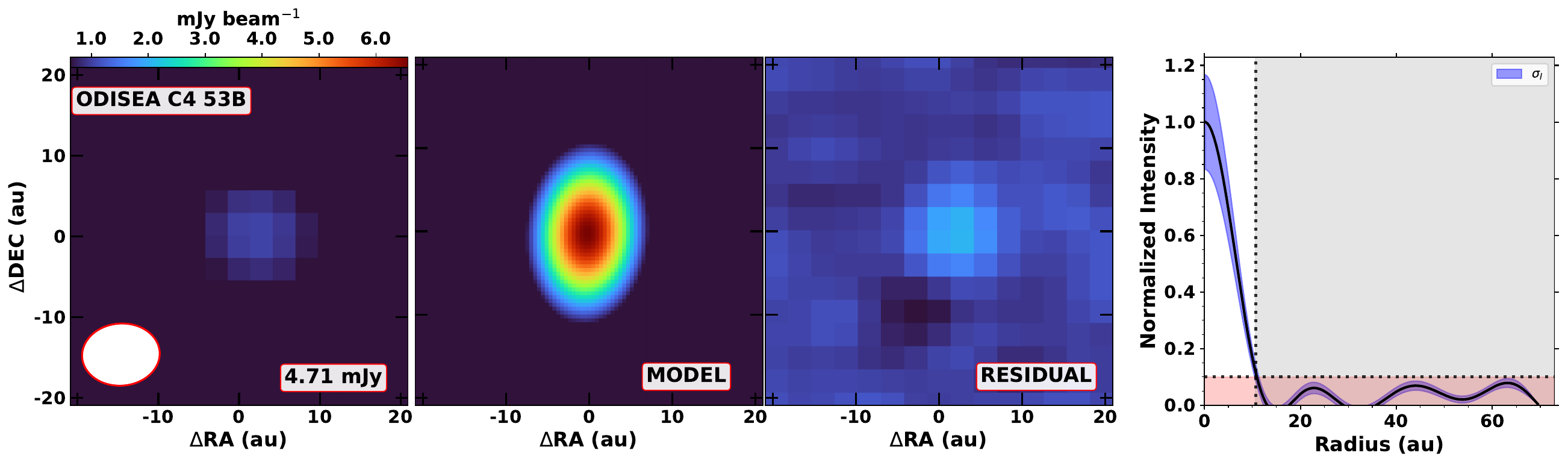}
\end{minipage}%
\vrulesep
\captionof{figure}{Stage 0 disks in embedded sources (Class I/F). 
{Horizontally, the first images in each panel are those created by \texttt{tclean}; the second, third, and fourth images are the models, residuals, and 1D radial profiles created by \texttt{Frank}. The colorbar indicates the flux scale for both \texttt{tclean} image and the residual.} }
\label{fig:0+I_app}
\vspace{0.8cm}

\noindent
\begin{minipage}{.49\textwidth}
	 \centering
	 	 \hrulesep
	 	 \includegraphics[width=1\linewidth]{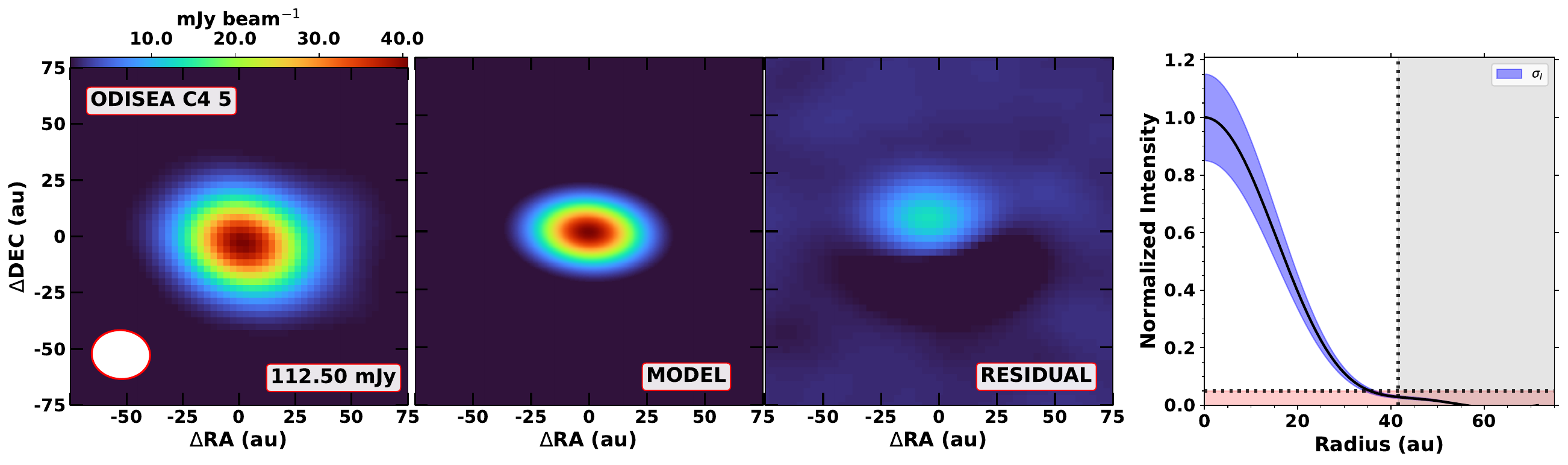}
\end{minipage}%
\vrulesep
\noindent
\begin{minipage}{.49\textwidth}
	 \centering
	 	 \hrulesep
	 	 \includegraphics[width=1\linewidth]{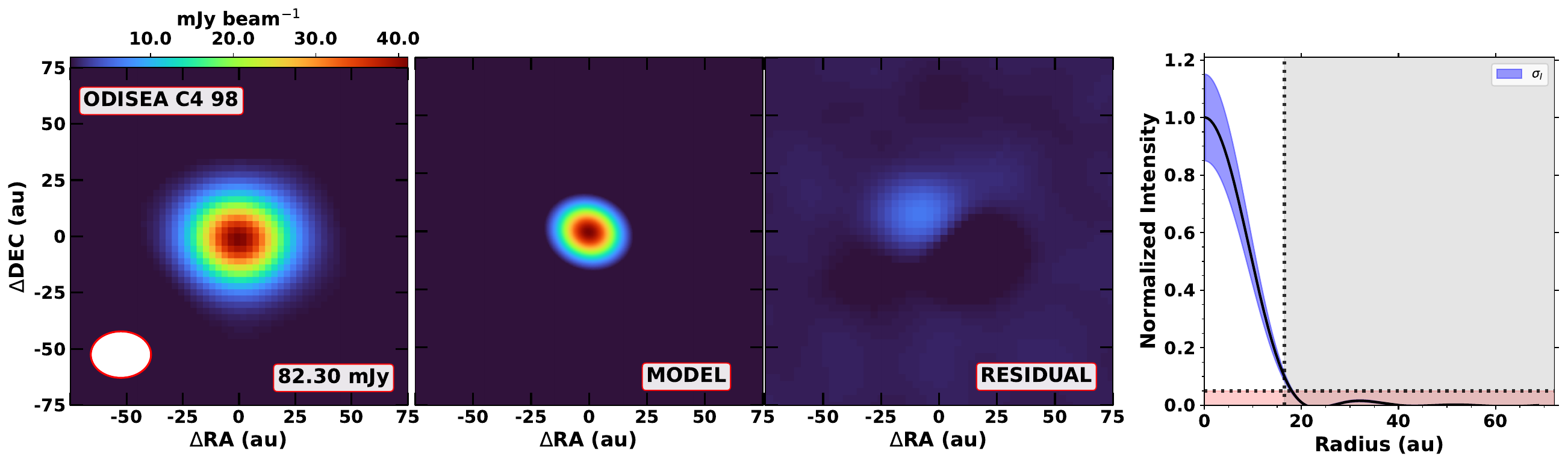}
\end{minipage}%
\vrulesep
\noindent
\begin{minipage}{.49\textwidth}
	 \centering
	 	 \hrulesep
	 	 \includegraphics[width=1\linewidth]{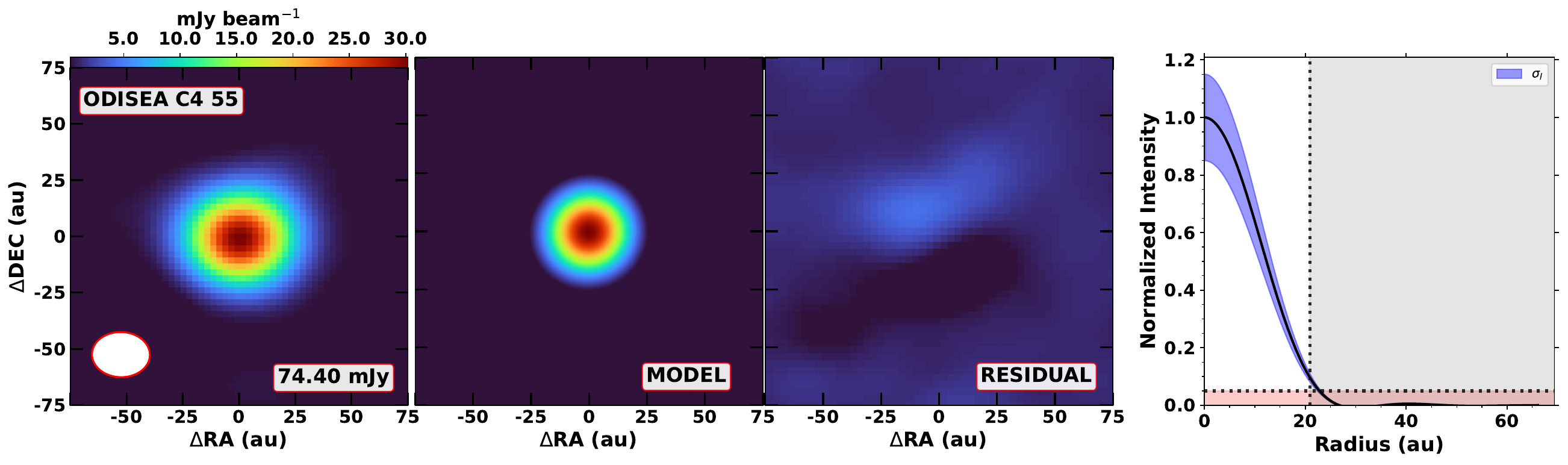}
\end{minipage}%
\vrulesep
\noindent
\begin{minipage}{.49\textwidth}
	 \centering
	 	 \hrulesep
	 	 \includegraphics[width=1\linewidth]{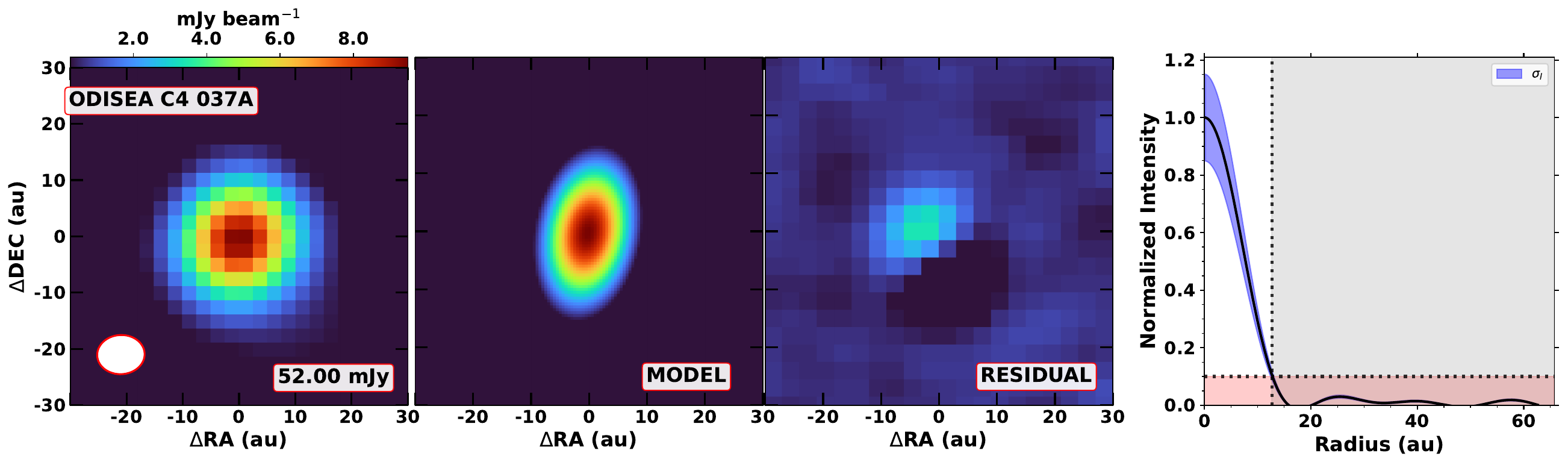}
\end{minipage}%
\vrulesep
\noindent
\begin{minipage}{.49\textwidth}
	 \centering
	 	 \hrulesep
	 	 \includegraphics[width=1\linewidth]{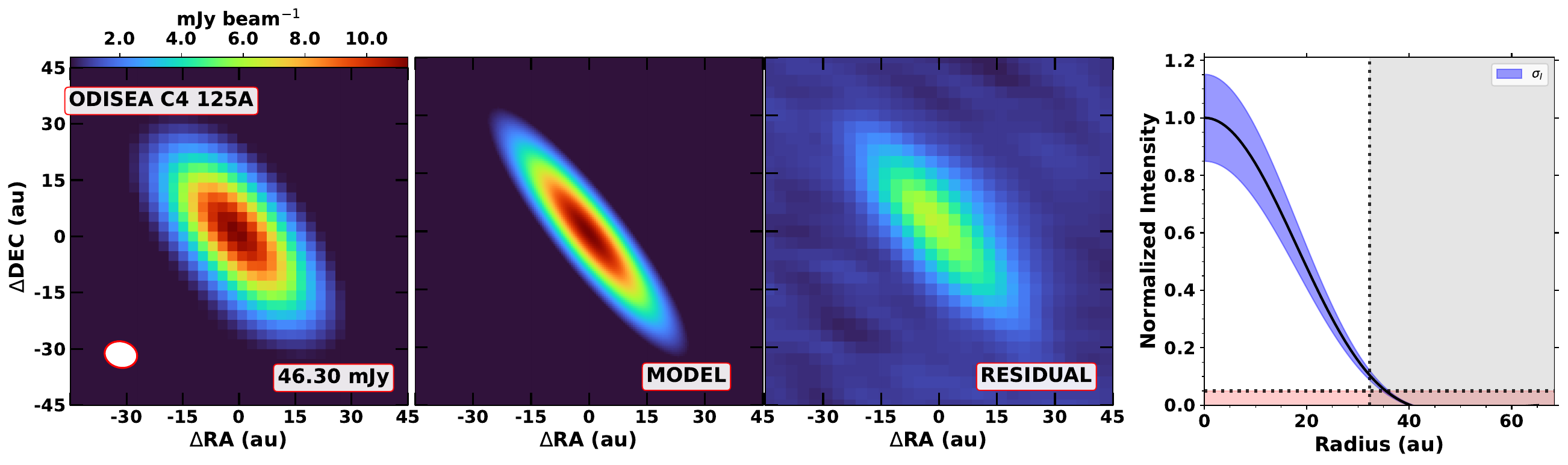}
\end{minipage}%
\vrulesep
\noindent
\begin{minipage}{.49\textwidth}
	 \centering
	 	 \hrulesep
	 	 \includegraphics[width=1\linewidth]{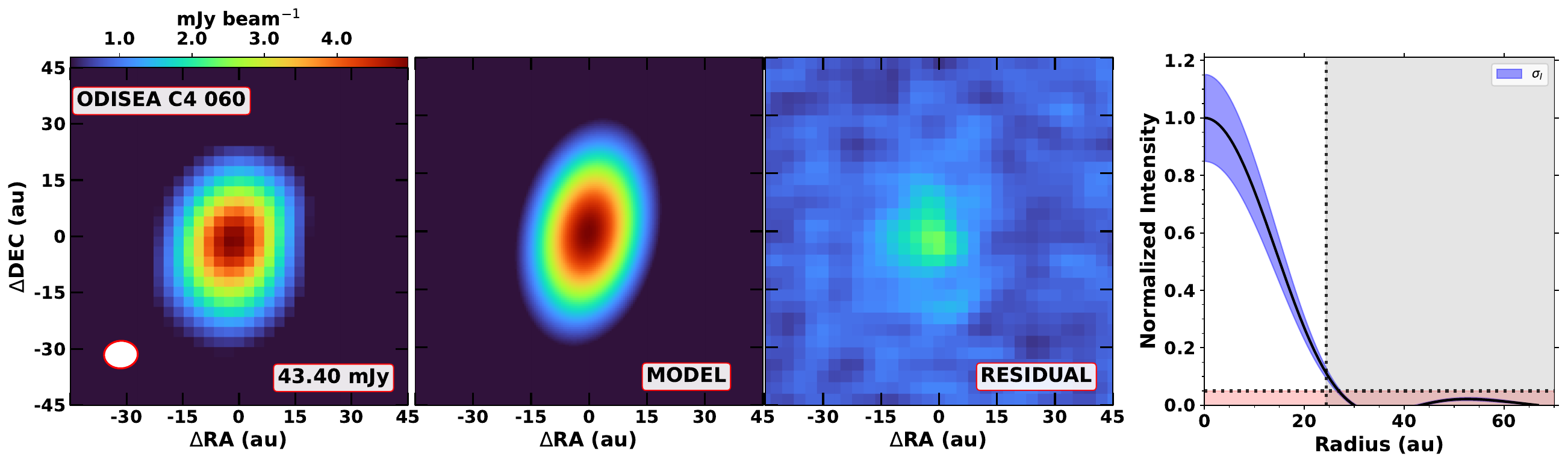}
\end{minipage}%
\vrulesep
\noindent
\begin{minipage}{.49\textwidth}
	 \centering
	 	 \hrulesep
	 	 \includegraphics[width=1\linewidth]{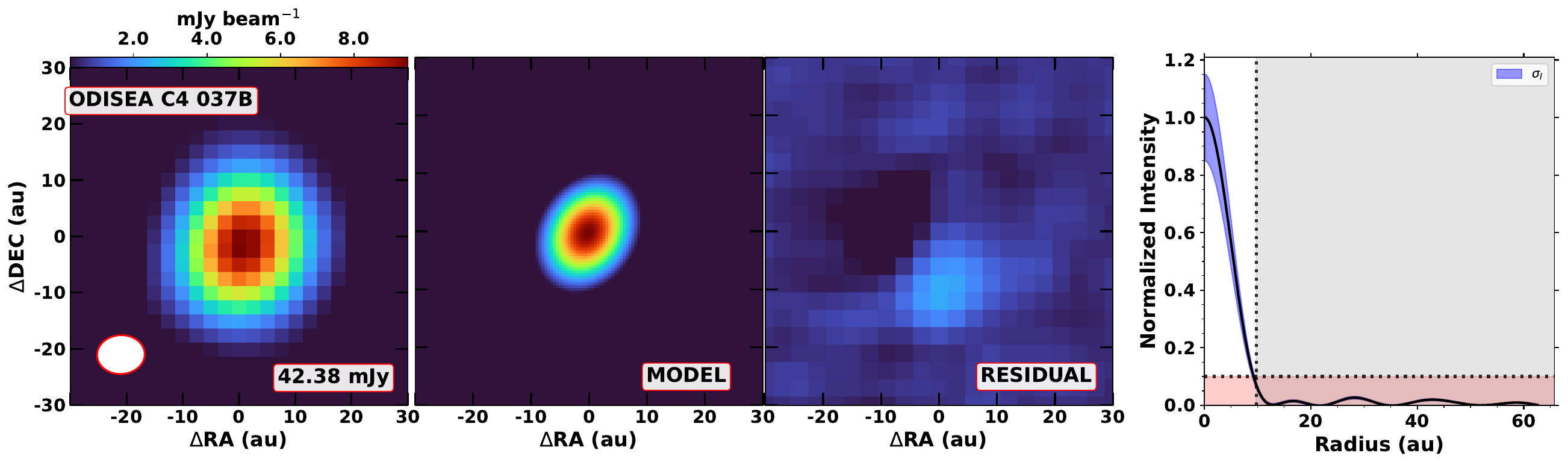}
\end{minipage}%
\vrulesep
\noindent
\begin{minipage}{.49\textwidth}
	 \centering
	 	 \hrulesep
	 	 \includegraphics[width=1\linewidth]{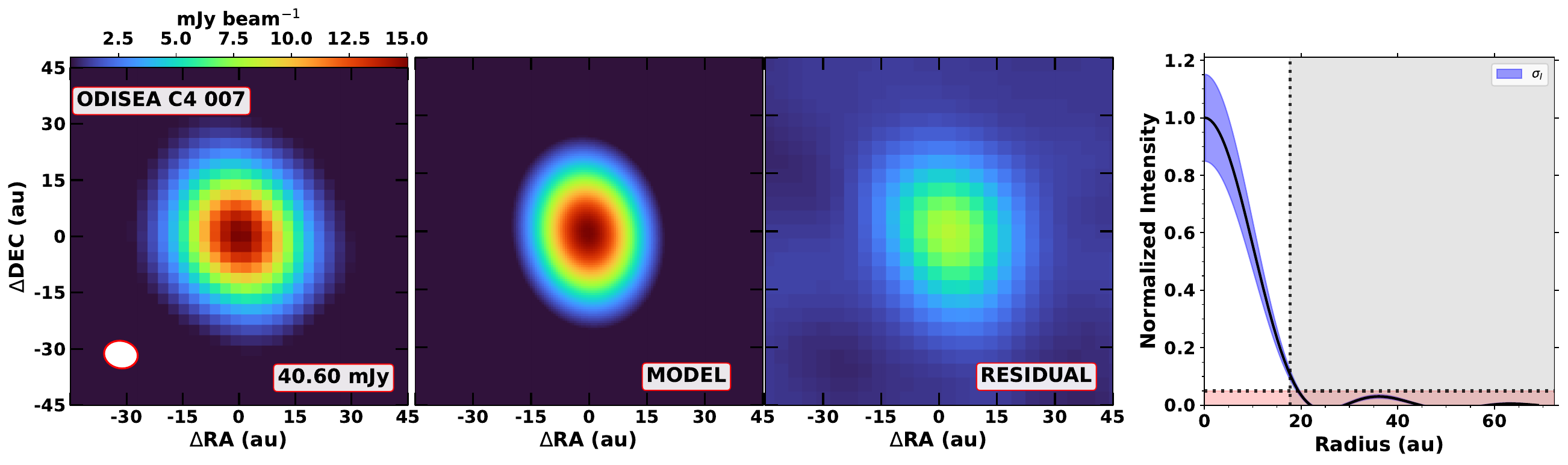}
\end{minipage}%
\vrulesep
\noindent
\begin{minipage}{.49\textwidth}
	 \centering
	 	 \hrulesep
	 	 \includegraphics[width=1\linewidth]{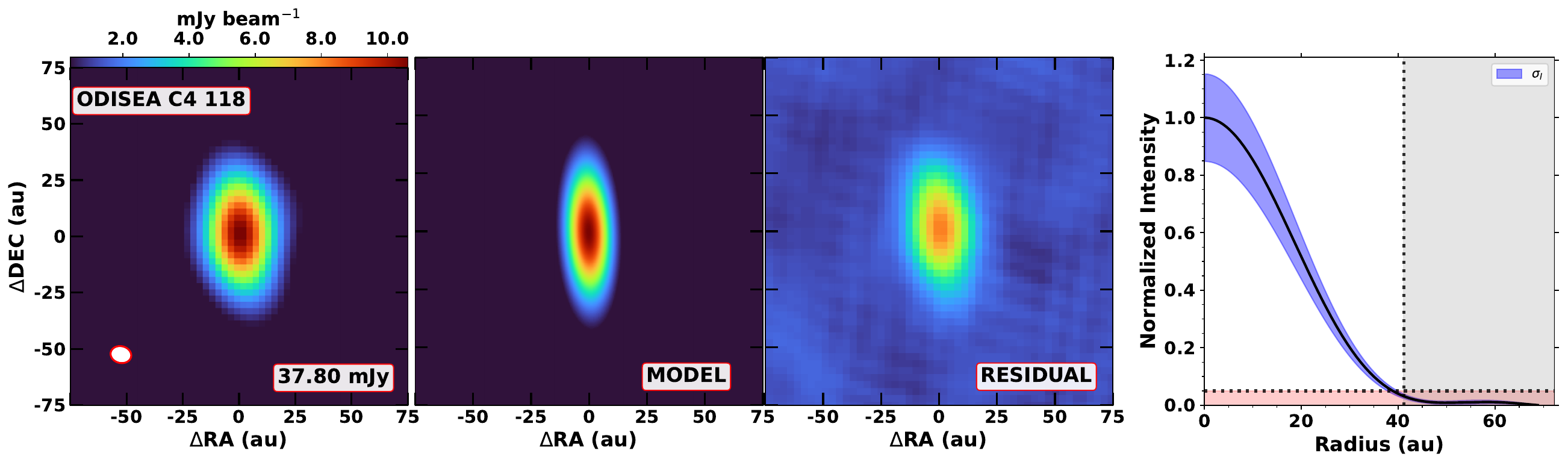}
\end{minipage}%
\vrulesep
\noindent
\begin{minipage}{.49\textwidth}
	 \centering
	 	 \hrulesep
	 	 \includegraphics[width=1\linewidth]{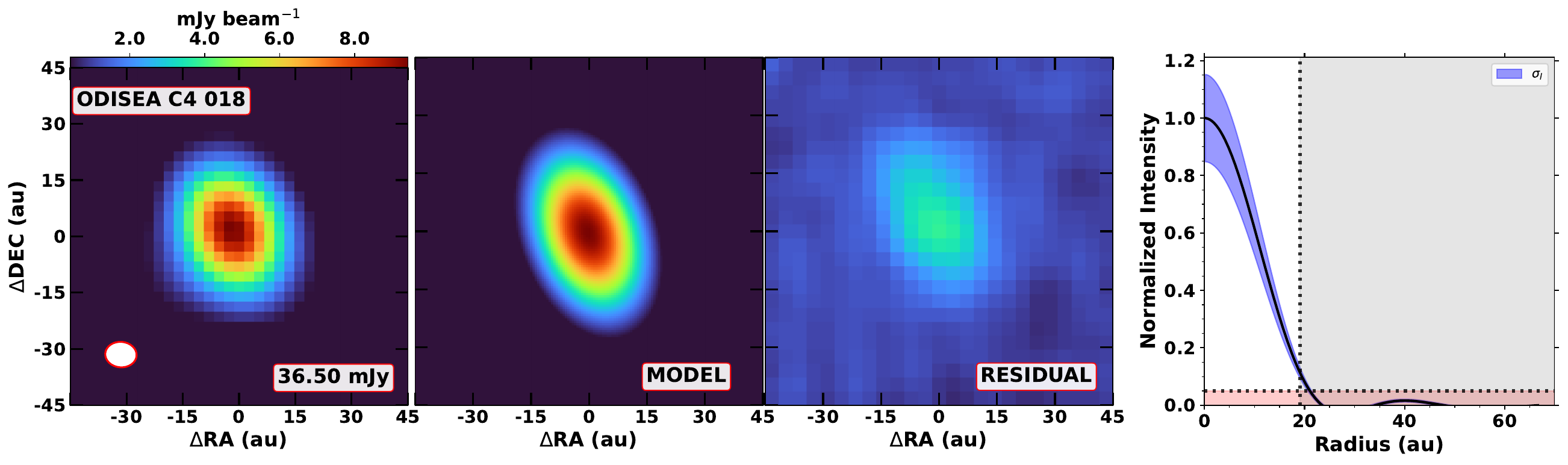}
\end{minipage}%
\vrulesep
\noindent
\begin{minipage}{.49\textwidth}
	 \centering
	 	 \hrulesep
	 	 \includegraphics[width=1\linewidth]{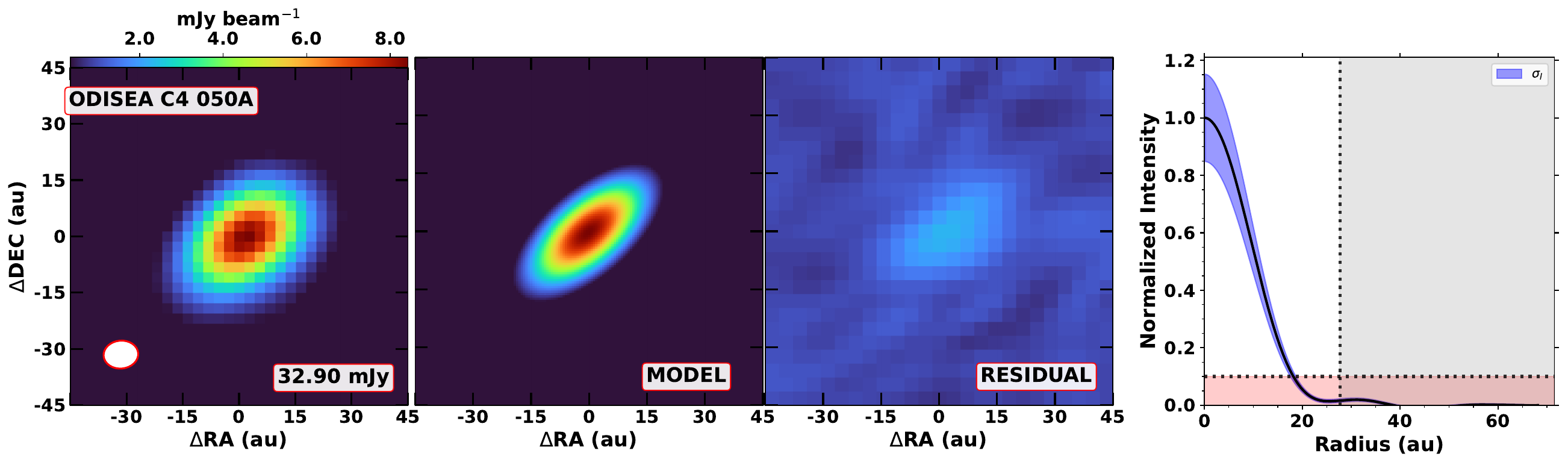}
\end{minipage}%
\vrulesep
\noindent
\begin{minipage}{.49\textwidth}
	 \centering
	 	 \hrulesep
	 	 \includegraphics[width=1\linewidth]{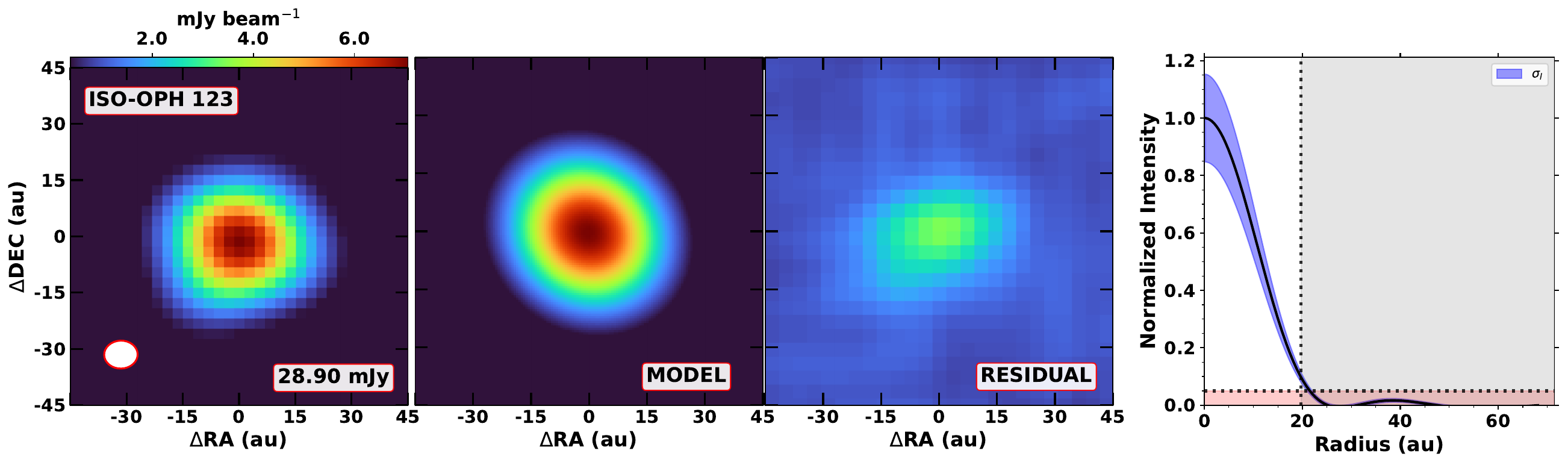}
\end{minipage}%
\vrulesep
\noindent
\begin{minipage}{.49\textwidth}
	 \centering
	 	 \hrulesep
	 	 \includegraphics[width=1\linewidth]{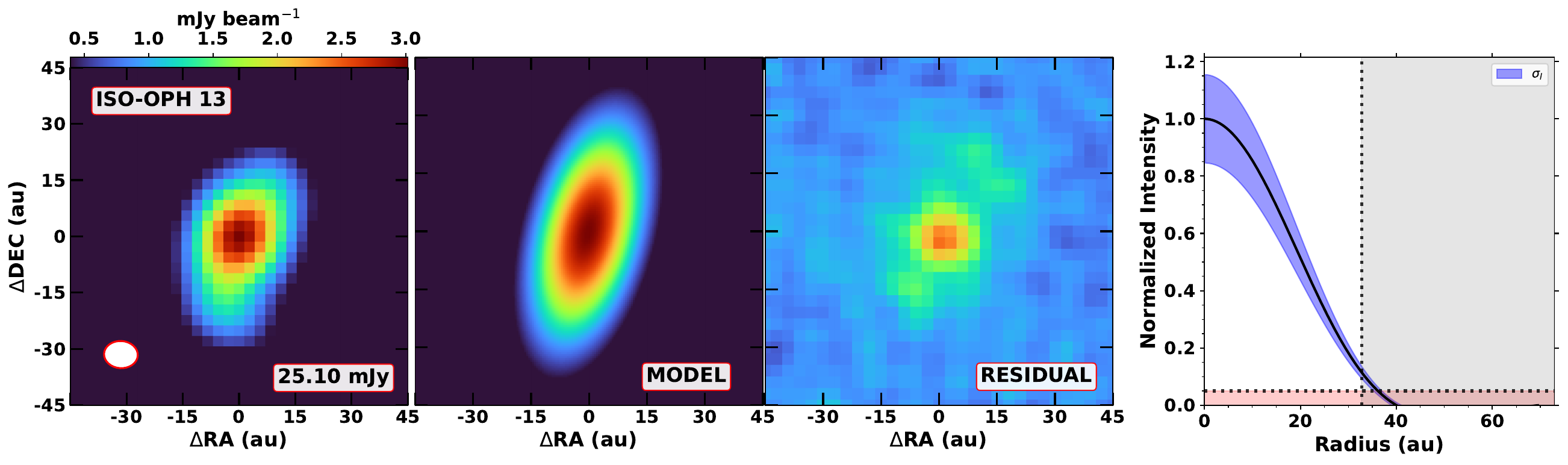}
\end{minipage}%
\vrulesep
\noindent
\begin{minipage}{.49\textwidth}
	 \centering
	 	 \hrulesep
	 	 \includegraphics[width=1\linewidth]{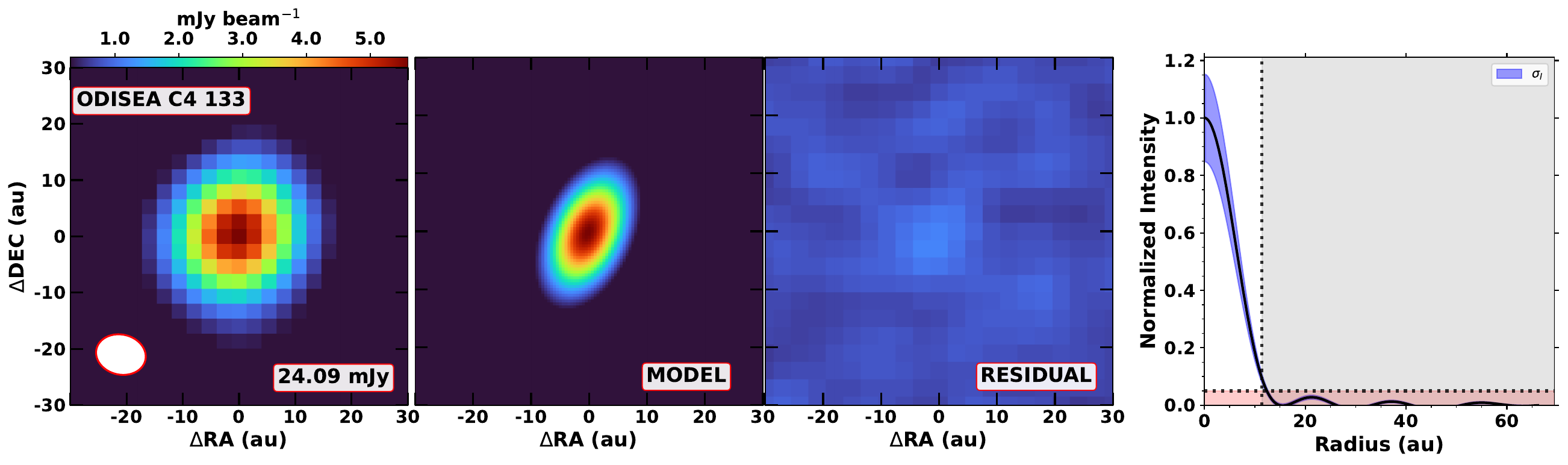}
\end{minipage}%
\vrulesep
\noindent
\begin{minipage}{.49\textwidth}
	 \centering
	 	 \hrulesep
	 	 \includegraphics[width=1\linewidth]{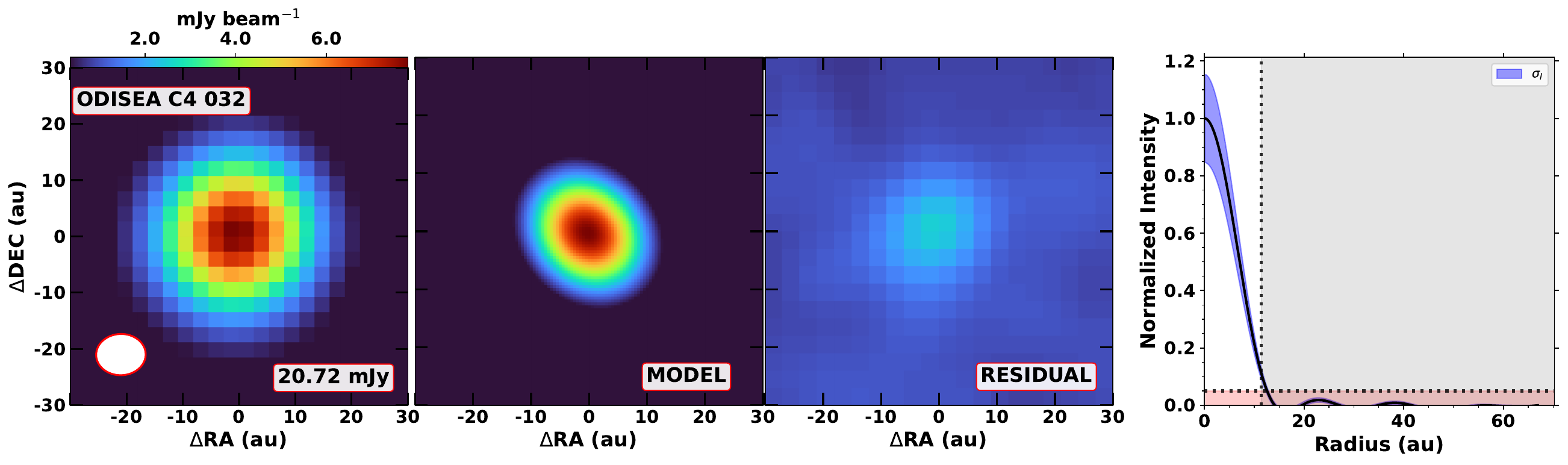}
\end{minipage}%
\vrulesep
\noindent
\begin{minipage}{.49\textwidth}
	 \centering
	 	 \hrulesep
	 	 \includegraphics[width=1\linewidth]{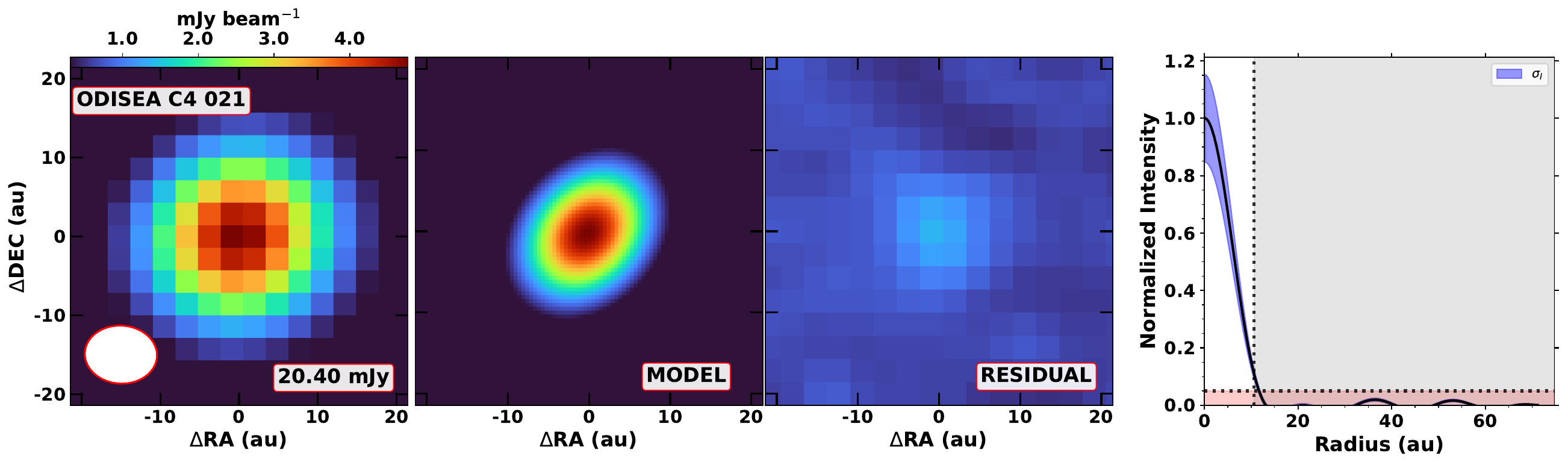}
\end{minipage}%
\vrulesep
\noindent
\begin{minipage}{.49\textwidth}
	 \centering
	 	 \hrulesep
	 	 \includegraphics[width=1\linewidth]{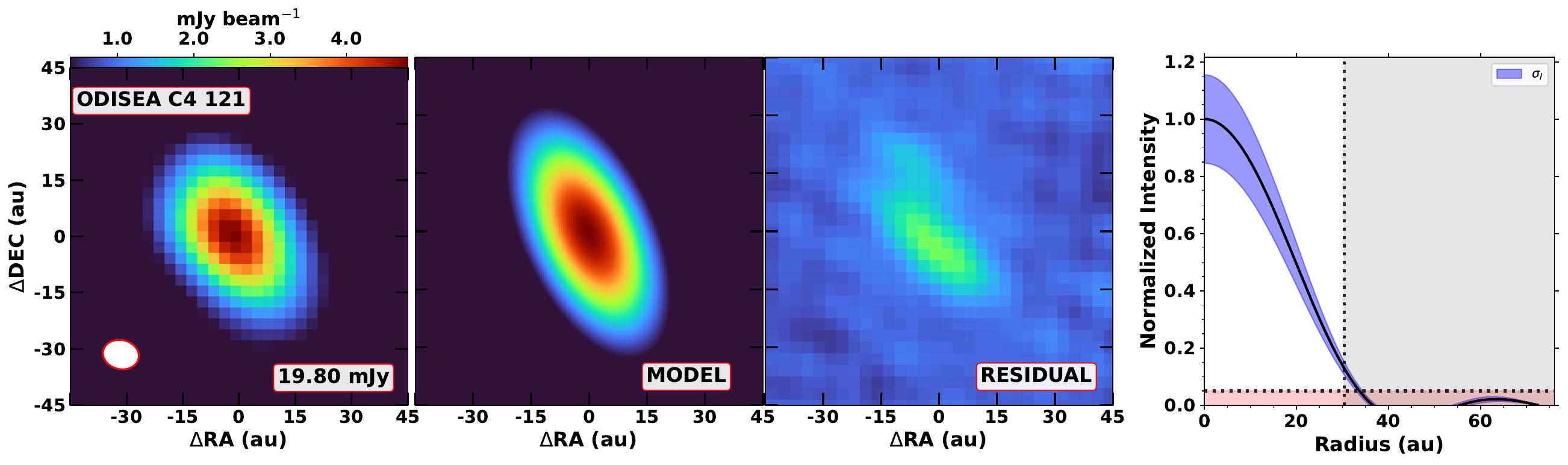}
\end{minipage}%
\vrulesep
\noindent
\begin{minipage}{.49\textwidth}
	 \centering
	 	 \hrulesep
	 	 \includegraphics[width=1\linewidth]{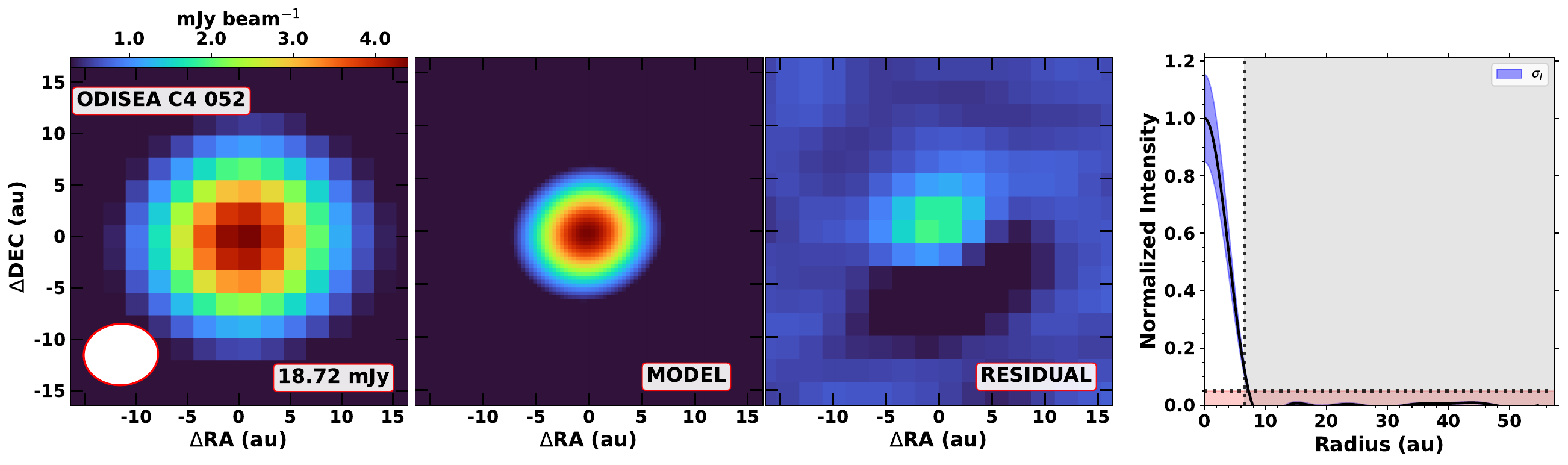}
\end{minipage}%
\vrulesep
\noindent
\begin{minipage}{.49\textwidth}
	 \centering
	 	 \hrulesep
	 	 \includegraphics[width=1\linewidth]{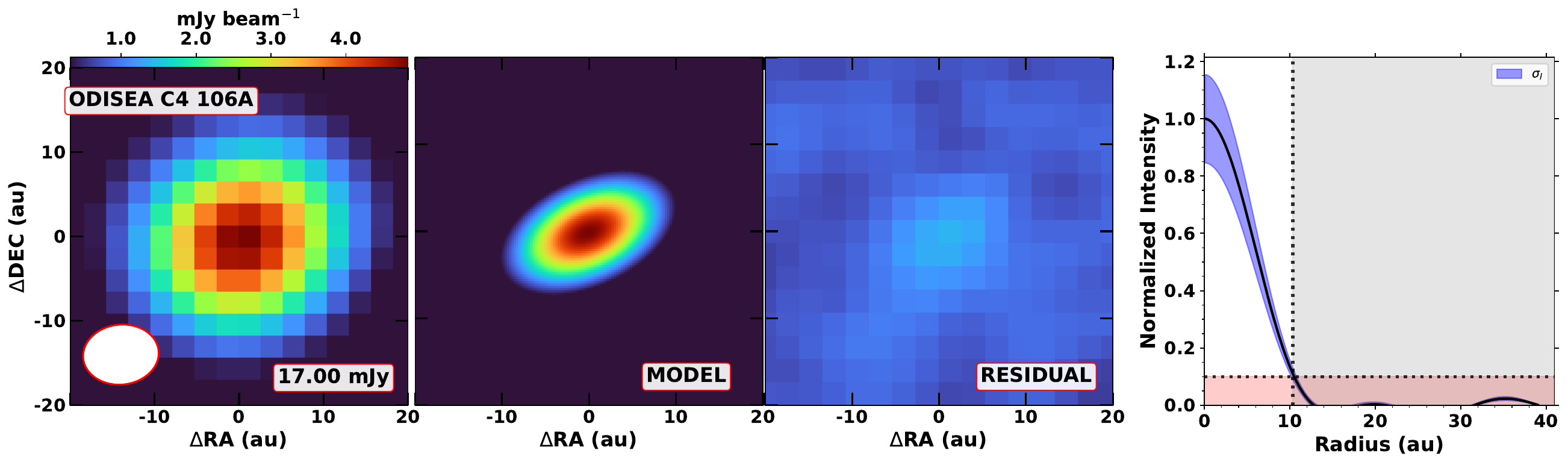}
\end{minipage}%
\vrulesep
\noindent
\begin{minipage}{.49\textwidth}
	 \centering
	 	 \hrulesep
	 	 \includegraphics[width=1\linewidth]{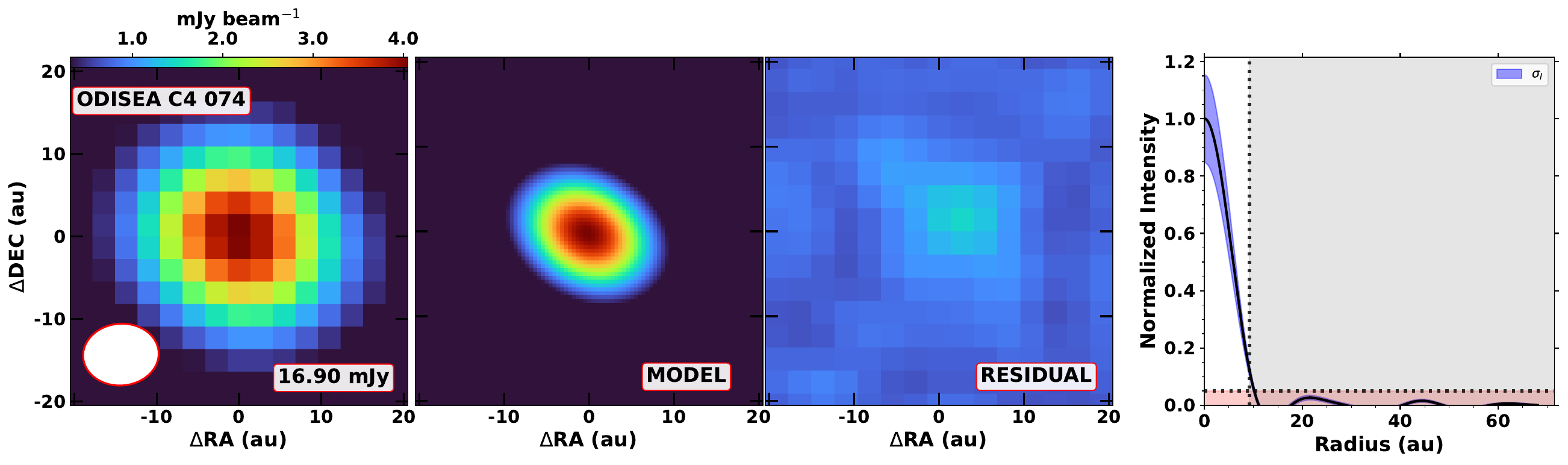}
\end{minipage}%
\vrulesep
\noindent
\begin{minipage}{.49\textwidth}
	 \centering
	 	 \hrulesep
	 	 \includegraphics[width=1\linewidth]{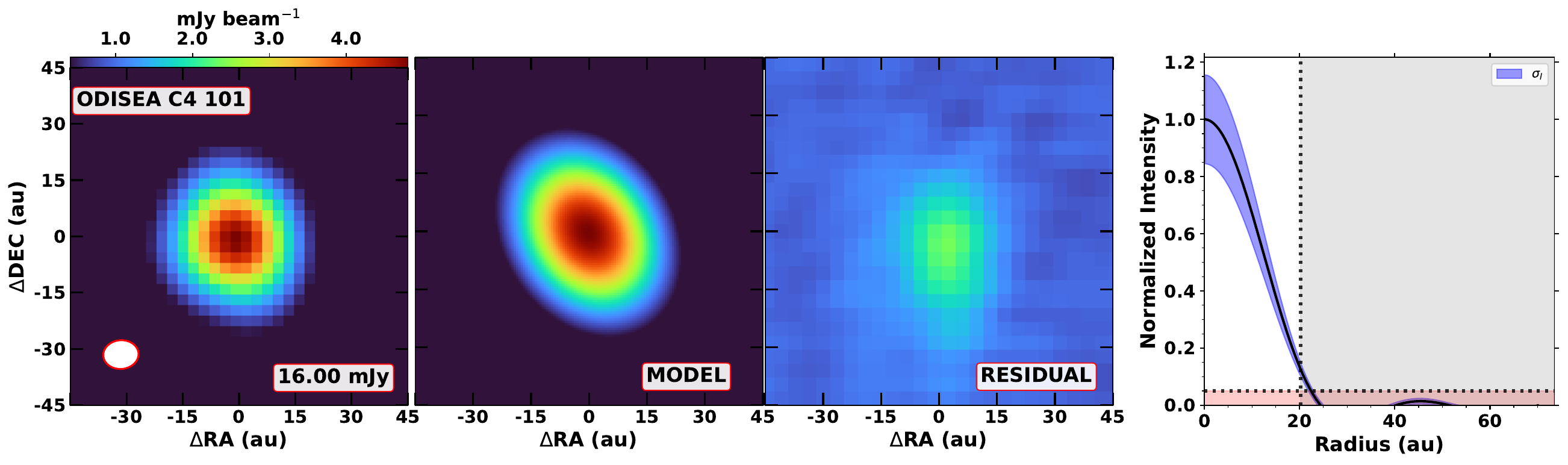}
\end{minipage}%
\vrulesep
\noindent
\begin{minipage}{.49\textwidth}
	 \centering
	 	 \hrulesep
	 	 \includegraphics[width=1\linewidth]{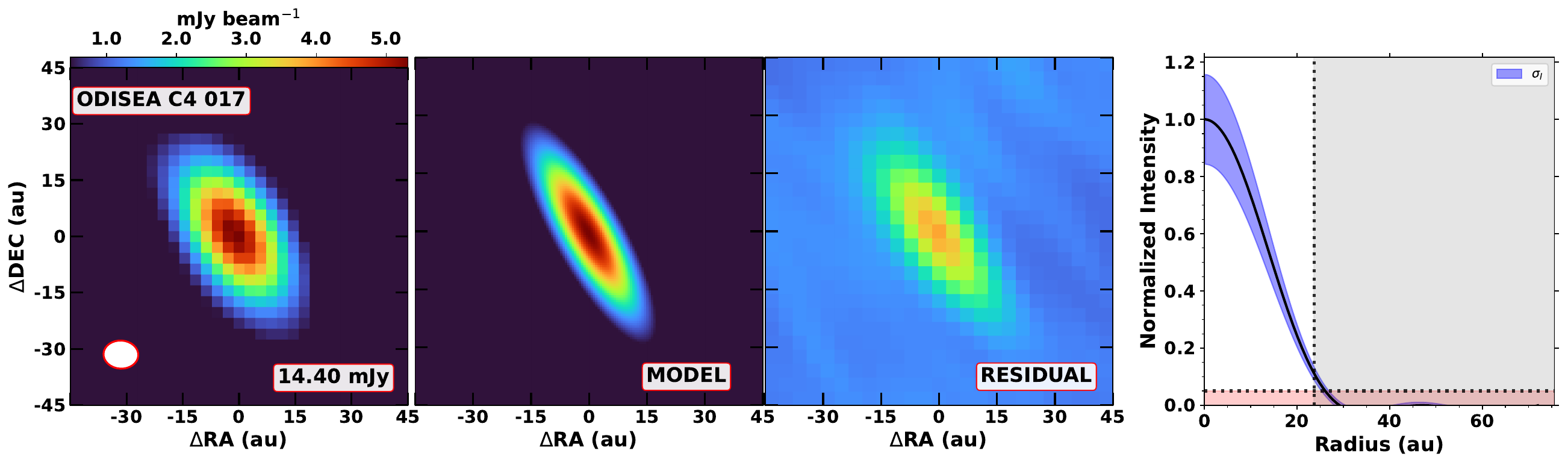}
\end{minipage}%
\vrulesep
\noindent
\begin{minipage}{.49\textwidth}
	 \centering
	 	 \hrulesep
	 	 \includegraphics[width=1\linewidth]{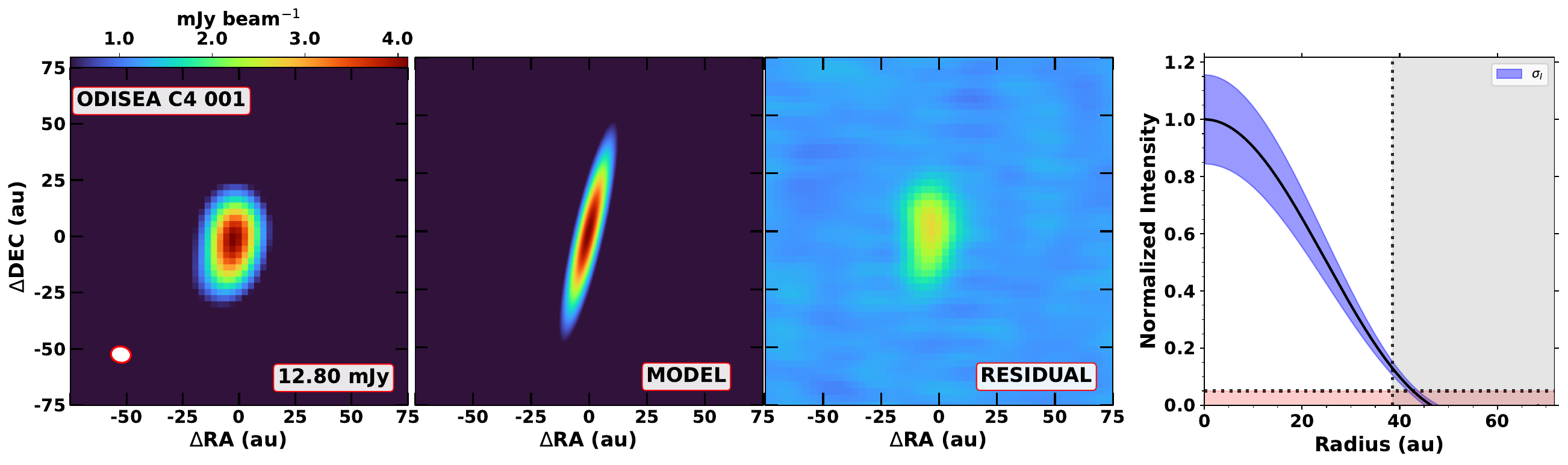}
\end{minipage}%
\vrulesep
\noindent
\begin{minipage}{.49\textwidth}
	 \centering
	 	 \hrulesep
	 	 \includegraphics[width=1\linewidth]{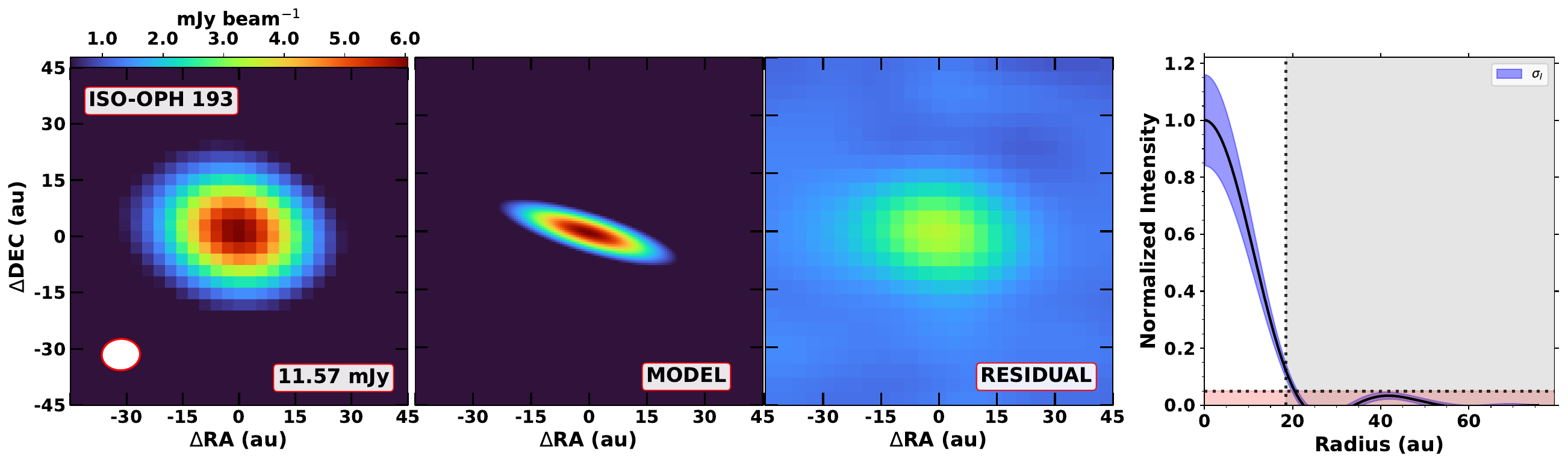}
\end{minipage}%
\vrulesep
\noindent
\begin{minipage}{.49\textwidth}
	 \centering
	 	 \hrulesep
	 	 \includegraphics[width=1\linewidth]{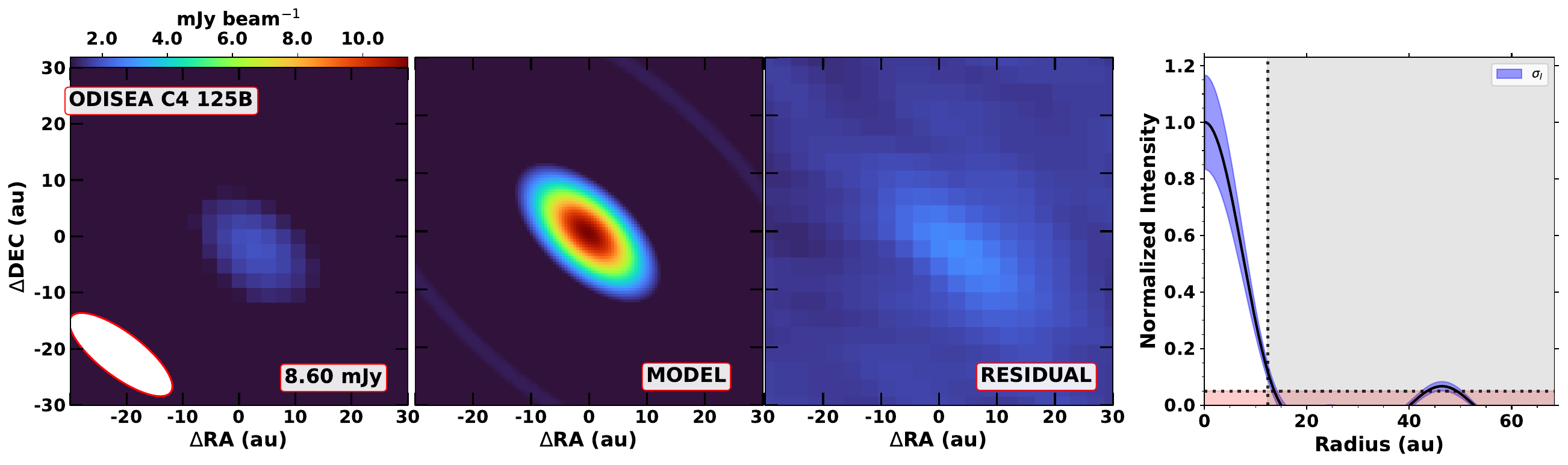}
\end{minipage}%
\vrulesep
\noindent
\begin{minipage}{.49\textwidth}
	 \centering
	 	 \hrulesep
	 	 \includegraphics[width=1\linewidth]{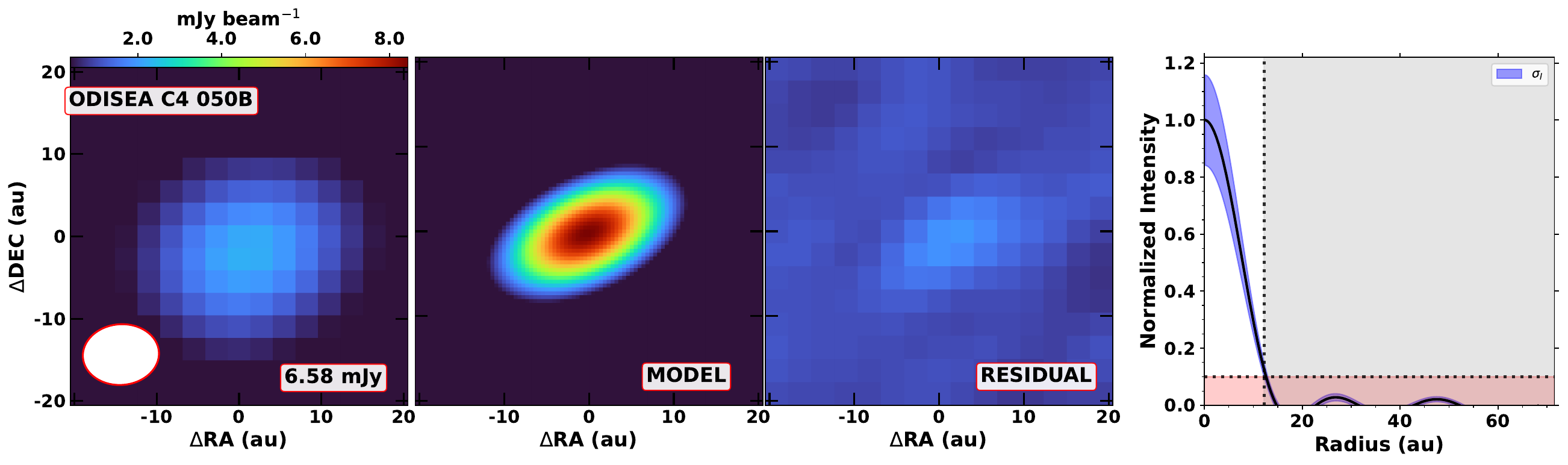}
\end{minipage}%
\vrulesep
\noindent
\begin{minipage}{.49\textwidth}
	 \centering
	 	 \hrulesep
	 	 \includegraphics[width=1\linewidth]{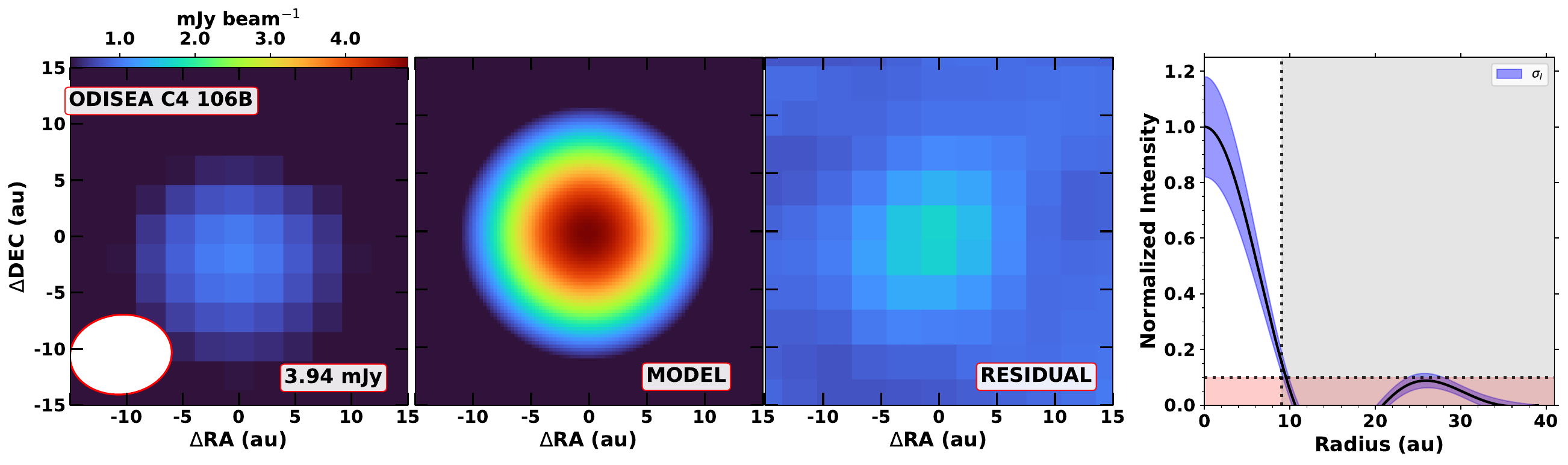}
\end{minipage}%
\vrulesep
\captionof{figure}{Stage 0 disks in  Class II sources. {Horizontally, the first images in each panel are those created by \texttt{tclean}; the second, third, and fourth images are the models, residuals, and 1D radial profiles created by \texttt{Frank}. The colorbar indicates the flux scale for both \texttt{tclean} image and the residual.}}
\label{fig:0+II_app}
\vspace{0.8cm}
\FloatBarrier
\subsection{Stage I}

\noindent
\begin{minipage}{.49\textwidth}
	 \centering
	 	 \hrulesep
	 	 \includegraphics[width=1\linewidth]{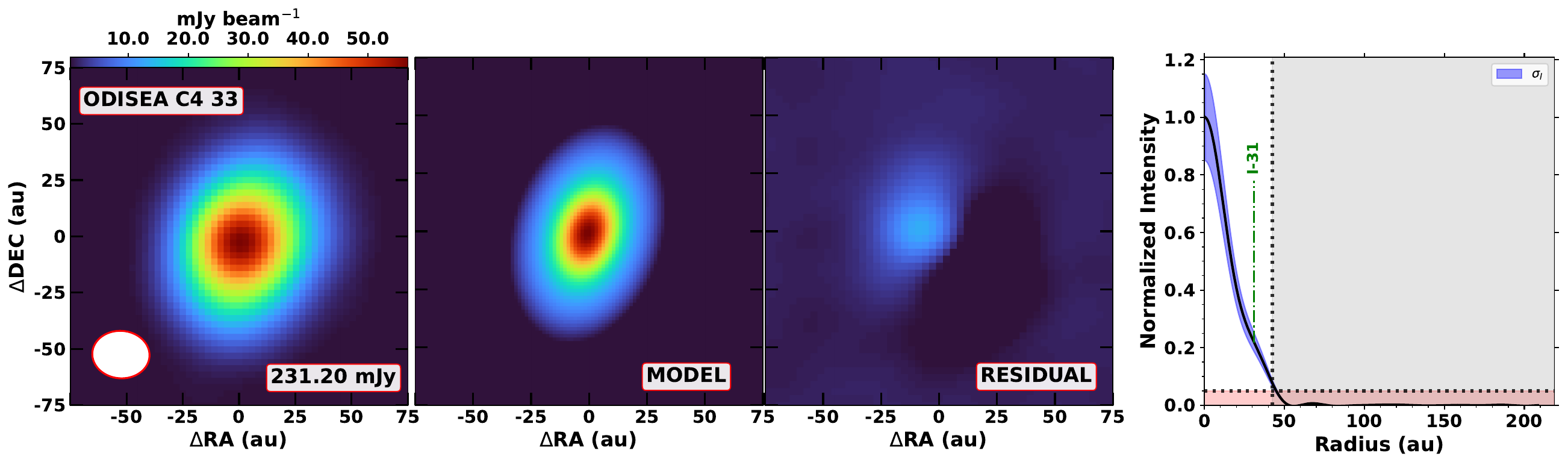}
\end{minipage}%
\vrulesep
\noindent
\begin{minipage}{.49\textwidth}
	 \centering
	 	 \hrulesep
	 	 \includegraphics[width=1\linewidth]{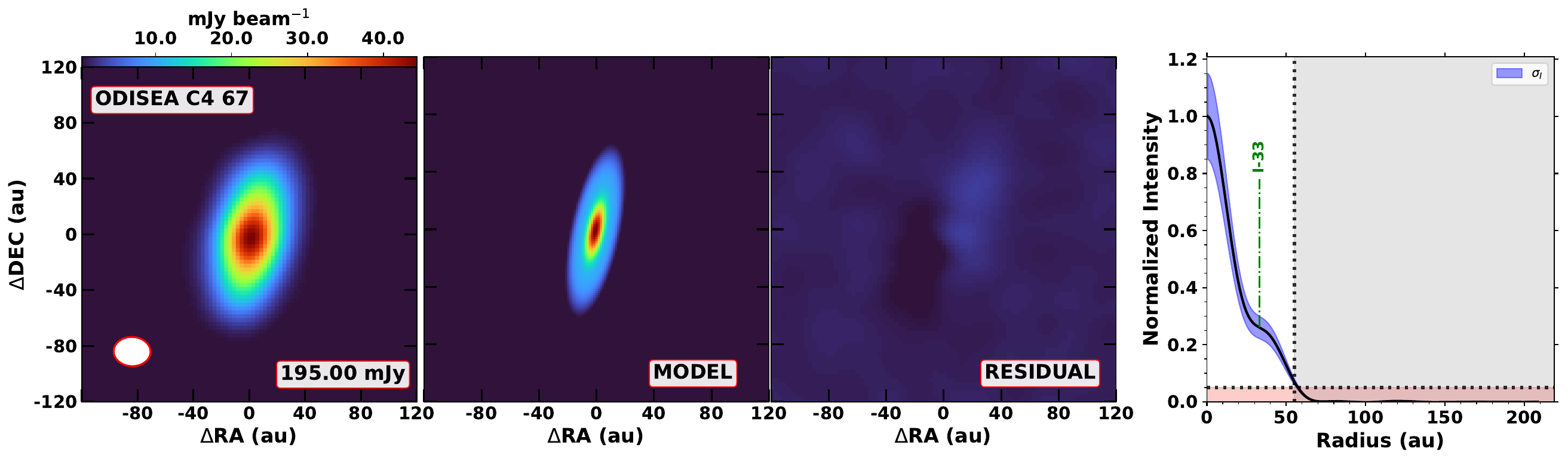}
\end{minipage}%
\vrulesep
\noindent
\begin{minipage}{.49\textwidth}
	 \centering
	 	 \hrulesep
	 	 \includegraphics[width=1\linewidth]{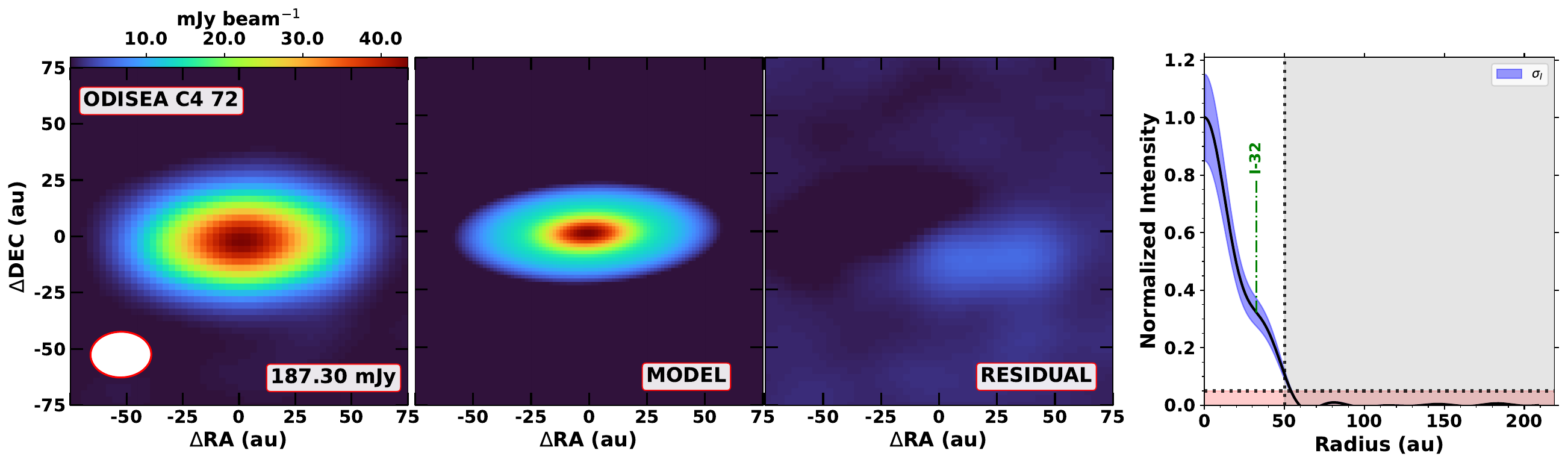}
\end{minipage}%
\vrulesep
\noindent
\begin{minipage}{.49\textwidth}
	 \centering
	 	 \hrulesep
	 	 \includegraphics[width=1\linewidth]{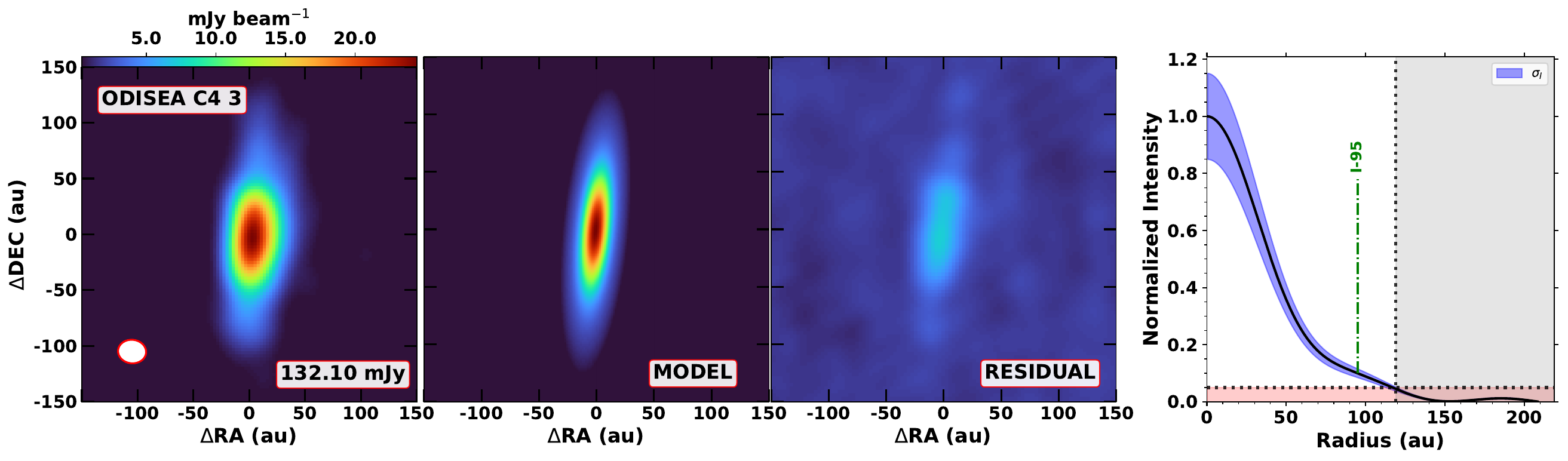}
\end{minipage}%
\vrulesep
\noindent
\begin{minipage}{.49\textwidth}
	 \centering
	 	 \hrulesep
	 	 \includegraphics[width=1\linewidth]{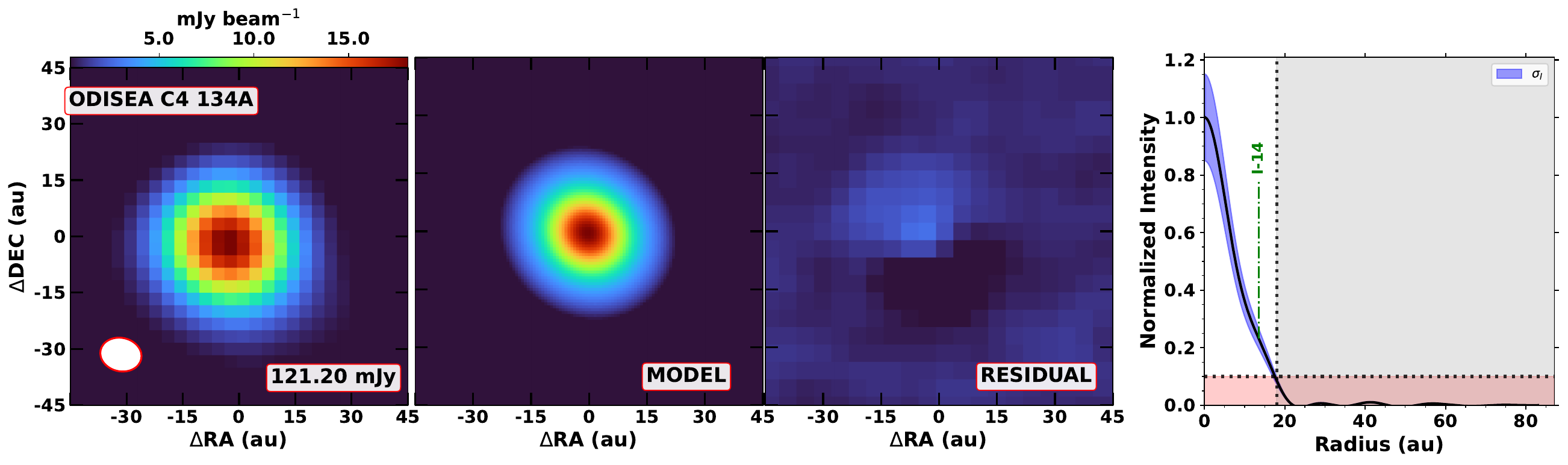}
\end{minipage}%
\vrulesep
\noindent
\begin{minipage}{.49\textwidth}
	 \centering
	 	 \hrulesep
	 	 \includegraphics[width=1\linewidth]{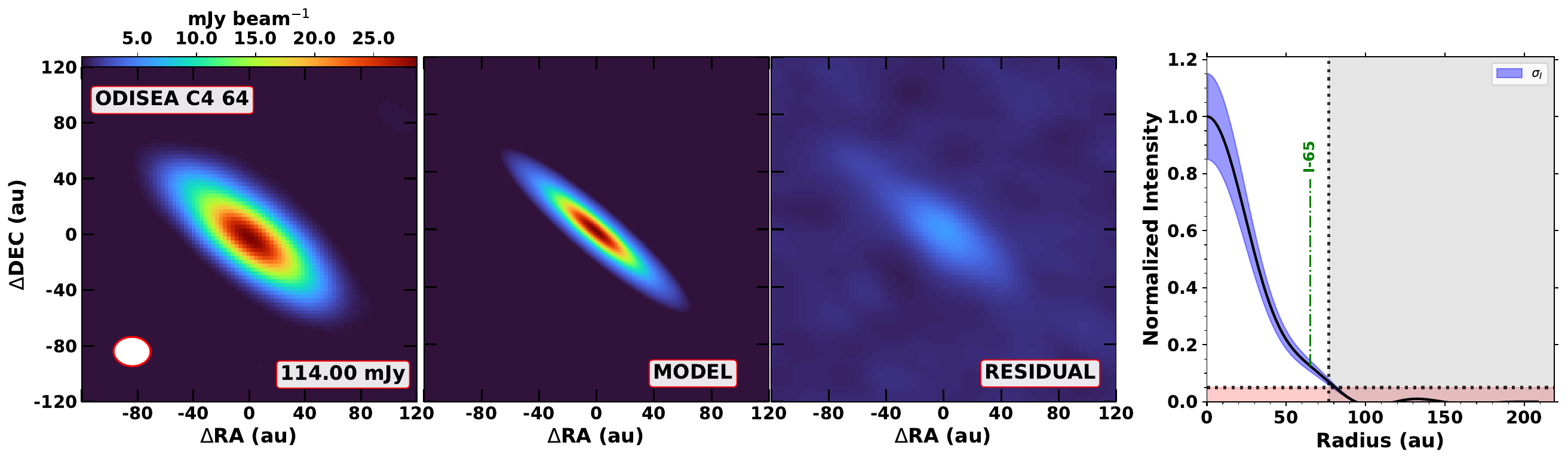}
\end{minipage}%
\vrulesep
\captionof{figure}{Stage I disks in embedded sources (Class I/F). 
{Horizontally, the first images in each panel are those created by \texttt{tclean}; the second, third, and fourth images are the models, residuals, and 1D radial profiles created by \texttt{Frank}. The colorbar indicates the flux scale for both \texttt{tclean} image and the residual.}}
\label{fig:1+I_app}
\vspace{0.8cm}
\FloatBarrier
\subsection{Stage II}
\noindent
\begin{minipage}{.49\textwidth}
	 \centering
	 	 \hrulesep
	 	 \includegraphics[width=1\linewidth]{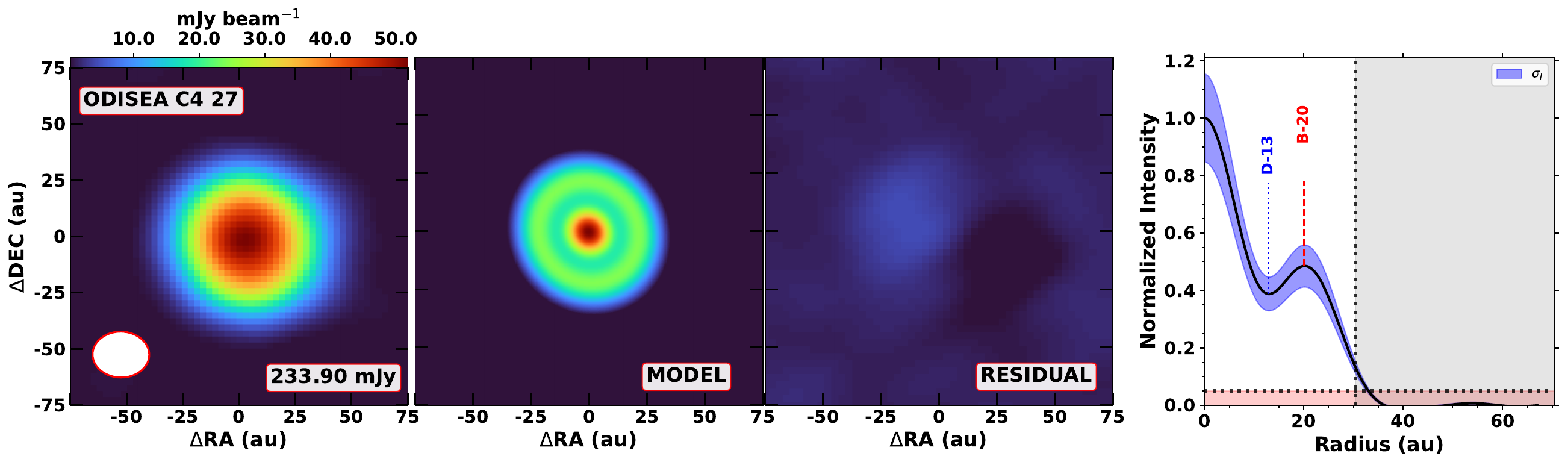}
\end{minipage}%
\vrulesep
\noindent
\begin{minipage}{.49\textwidth}
	 \centering
	 	 \hrulesep
	 	 \includegraphics[width=1\linewidth]{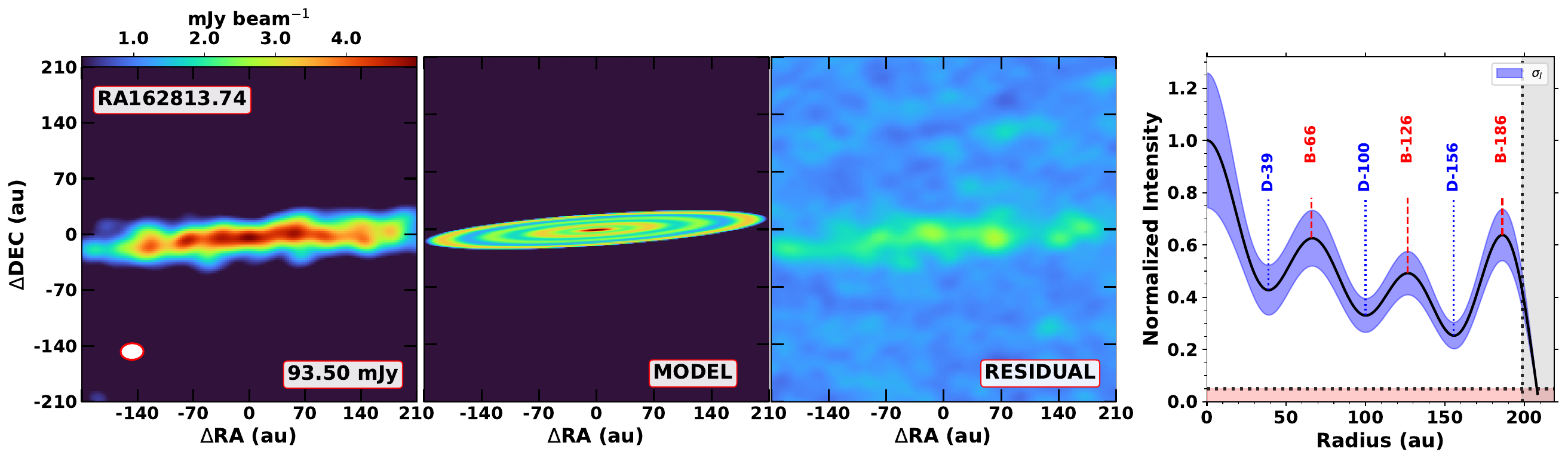}
\end{minipage}%
\vrulesep
\noindent
\begin{minipage}{.49\textwidth}
	 \centering
	 	 \hrulesep
	 	 \includegraphics[width=1\linewidth]{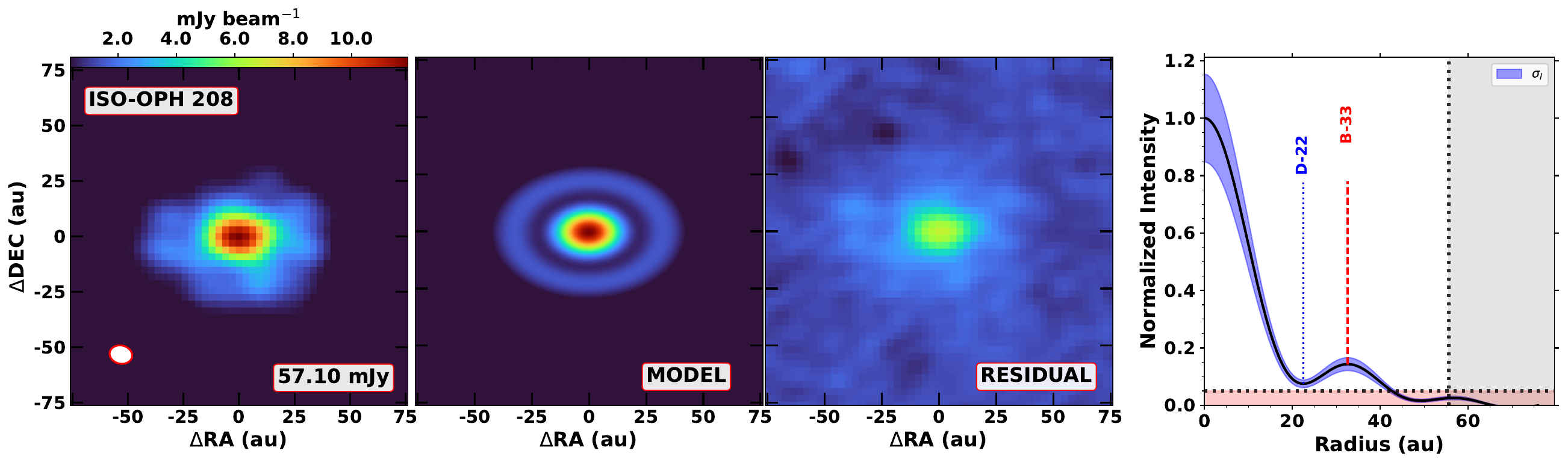}
\end{minipage}%
\vrulesep
\captionof{figure}{Stage II disks in Class II sources.
{Horizontally, the first images in each panel are those created by \texttt{tclean}; the second, third, and fourth images are the models, residuals, and 1D radial profiles created by \texttt{Frank}. The colorbar indicates the flux scale for both \texttt{tclean} image and the residual.}}
\label{fig:2+II_app}
\vspace{0.8cm}

\FloatBarrier
\subsection{Stage V}

\noindent
\begin{minipage}{.49\textwidth}
	 \centering
	 	 \hrulesep
	 	 \includegraphics[width=1\linewidth]{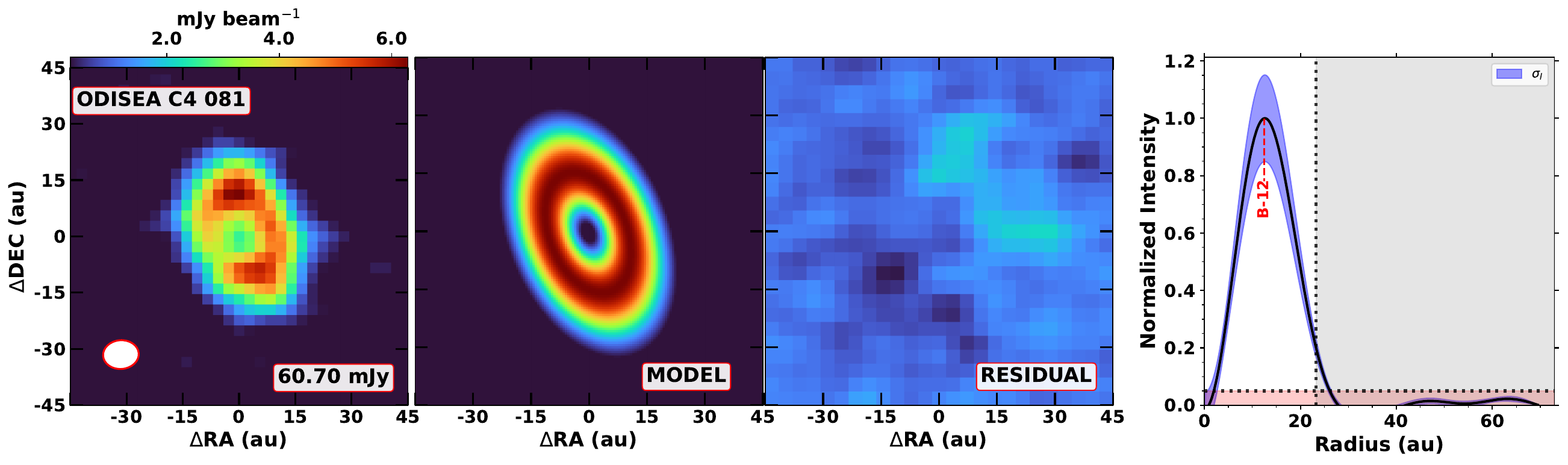}
\end{minipage}%
\vrulesep
\noindent
\begin{minipage}{.49\textwidth}
	 \centering
	 	 \hrulesep
	 	 \includegraphics[width=1\linewidth]{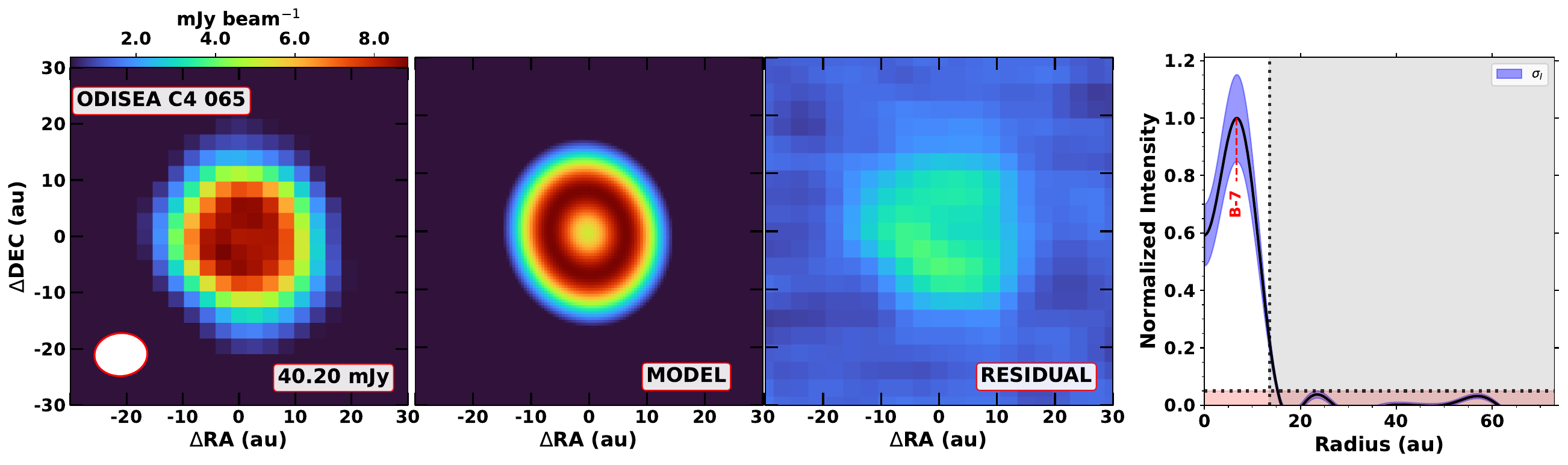}
\end{minipage}%
\vrulesep
\captionof{figure}{Stage V disks in embedded sources (Class I/F). {Horizontally, the first images in each panel are those created by \texttt{tclean}; the second, third, and fourth images are the models, residuals, and 1D radial profiles created by \texttt{Frank}. The colorbar indicates the flux scale for both \texttt{tclean} image and the residual.}}
\label{fig:5+I_app}
\vspace{0.8cm}

\noindent
\begin{minipage}{.49\textwidth}
	 \centering
	 	 \hrulesep
	 	 \includegraphics[width=1\linewidth]{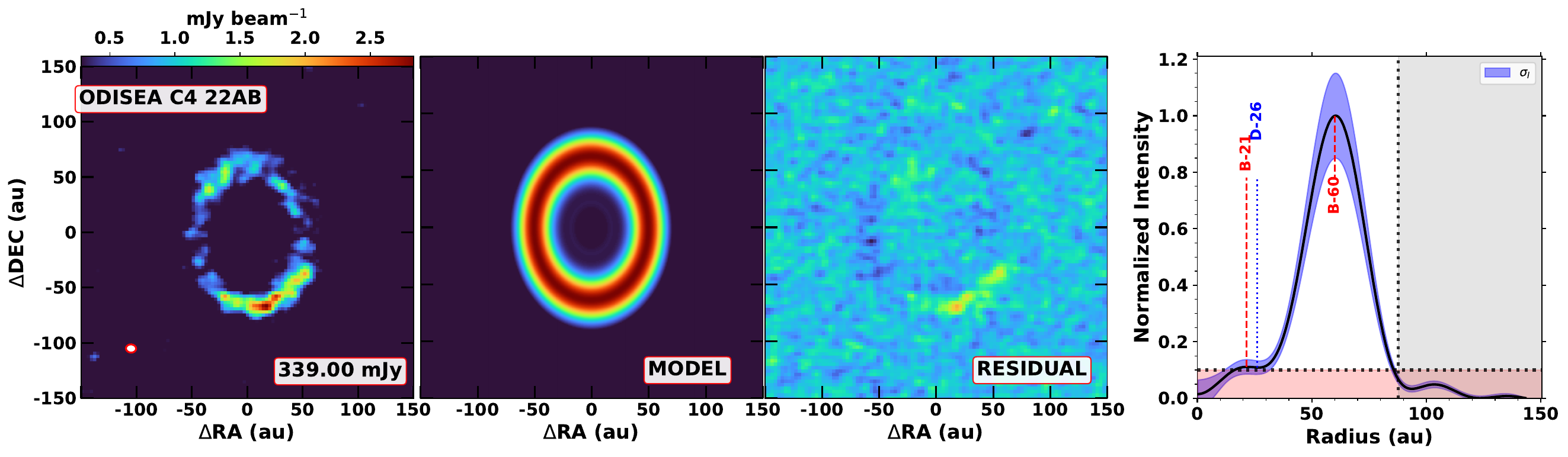}
\end{minipage}%
\vrulesep
\noindent
\begin{minipage}{.49\textwidth}
	 \centering
	 	 \hrulesep
	 	 \includegraphics[width=1\linewidth]{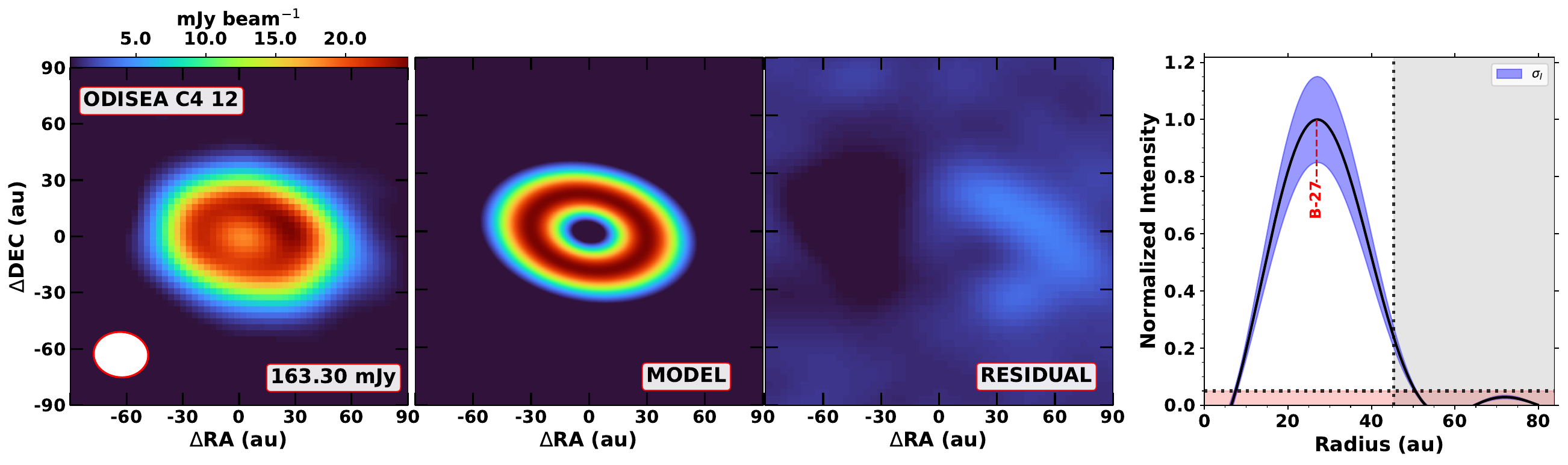}
\end{minipage}%
\vrulesep
\noindent
\begin{minipage}{.49\textwidth}
	 \centering
	 	 \hrulesep
	 	 \includegraphics[width=1\linewidth]{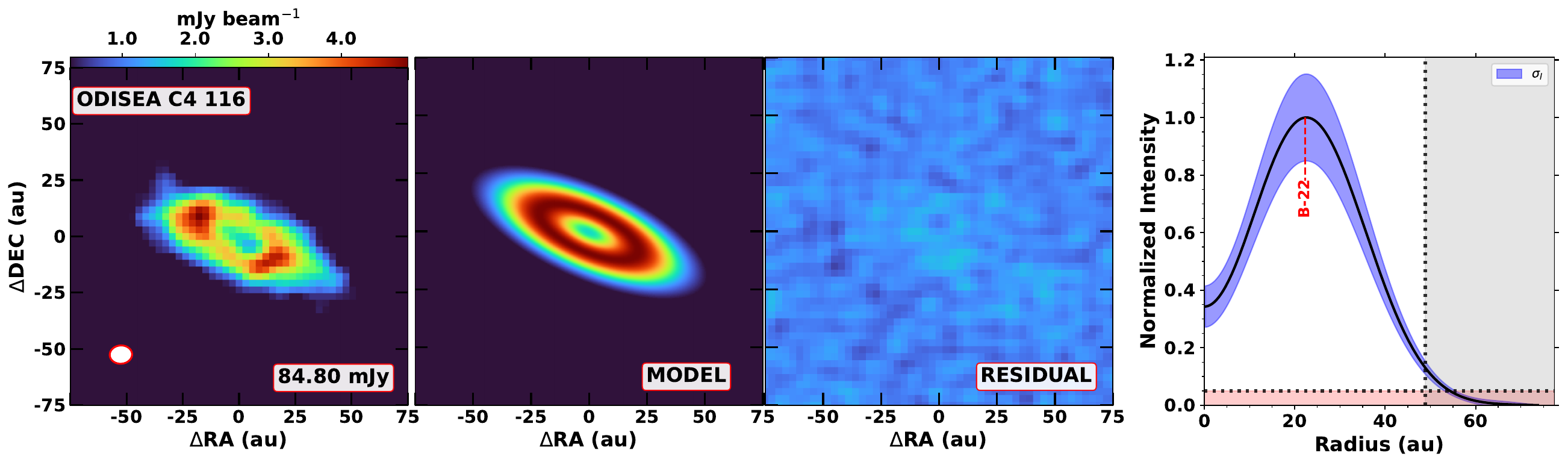}
\end{minipage}%
\vrulesep
\noindent
\begin{minipage}{.49\textwidth}
	 \centering
	 	 \hrulesep
	 	 \includegraphics[width=1\linewidth]{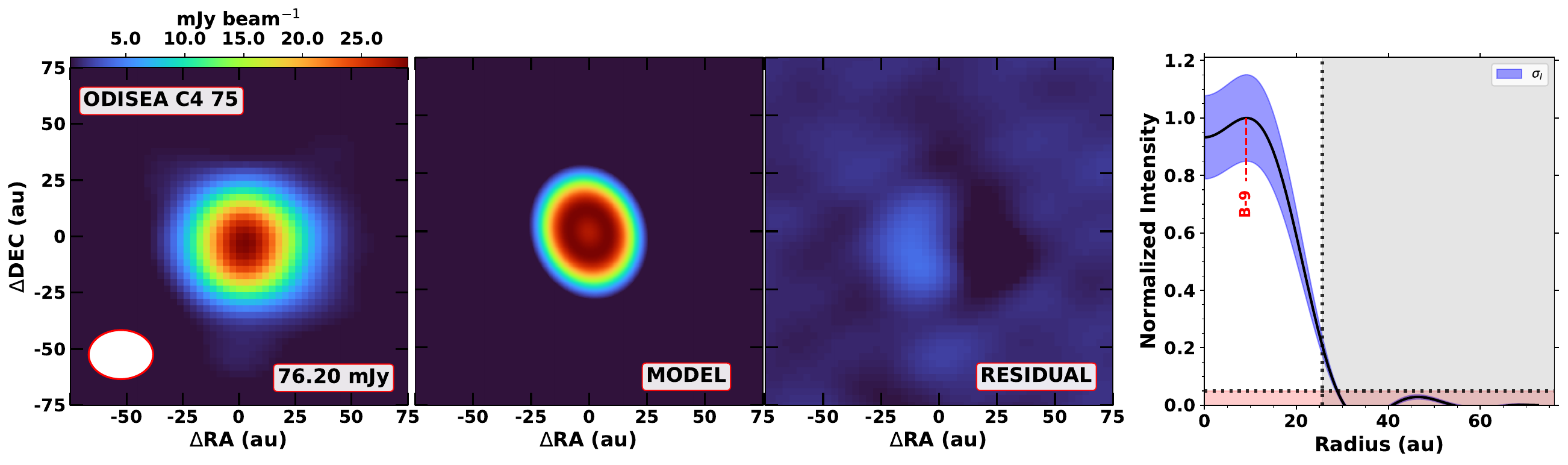}
\end{minipage}%
\vrulesep
\noindent
\begin{minipage}{.49\textwidth}
	 \centering
	 	 \hrulesep
	 	 \includegraphics[width=1\linewidth]{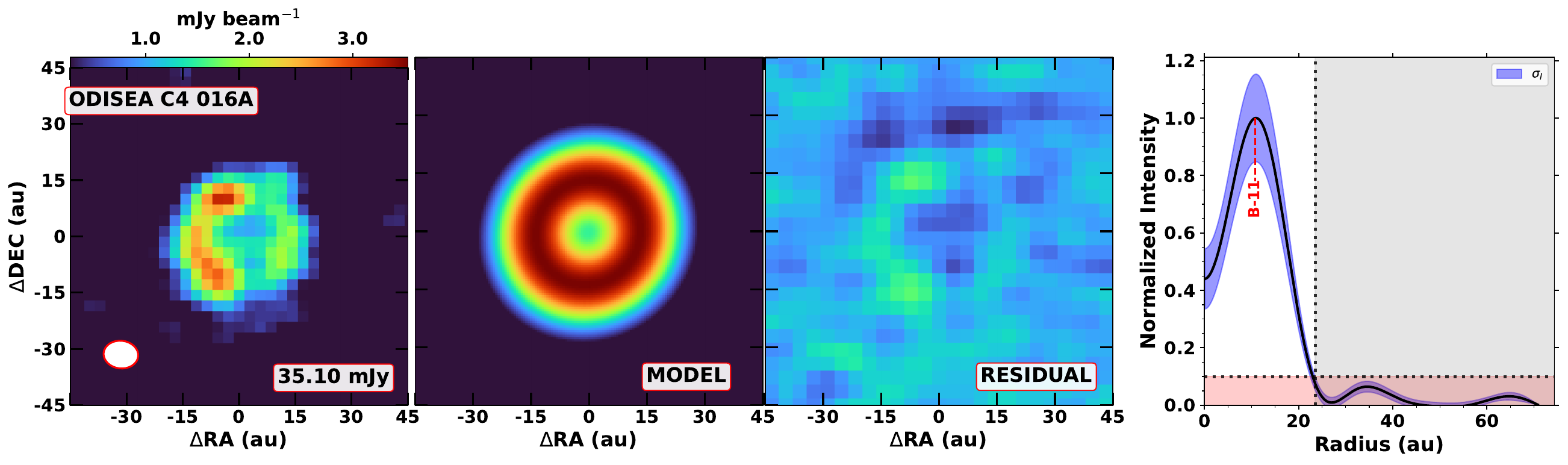}
\end{minipage}%
\vrulesep
\noindent
\begin{minipage}{.49\textwidth}
	 \centering
	 	 \hrulesep
	 	 \includegraphics[width=1\linewidth]{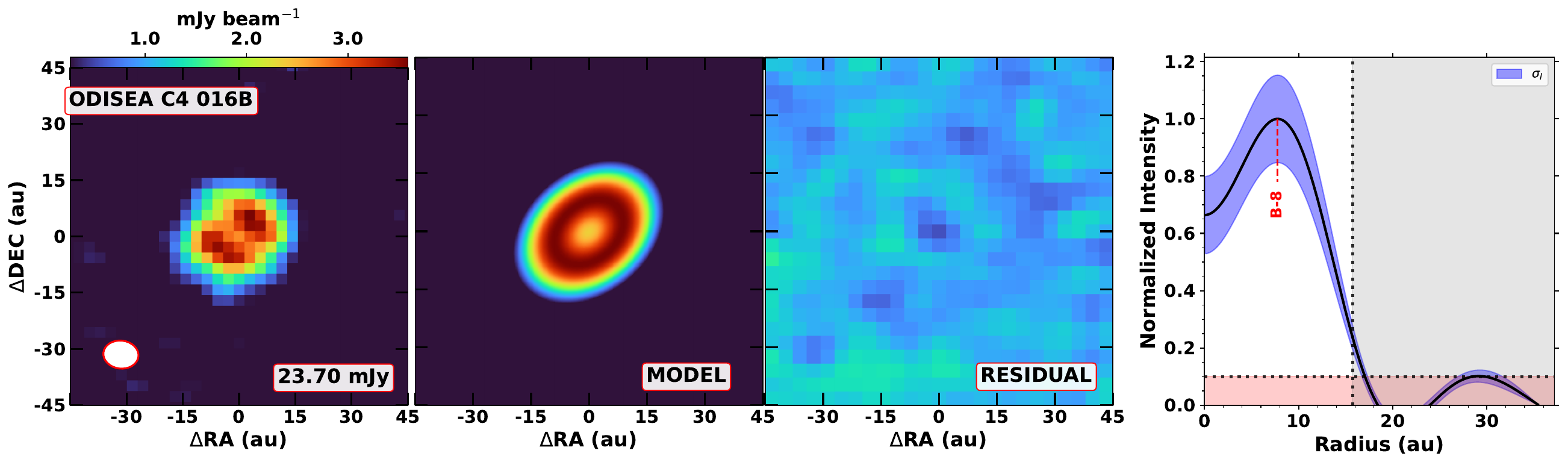}
\end{minipage}%
\vrulesep
\captionof{figure}{Stage V disks in Class II sources.
{Horizontally, the first images in each panel are those created by \texttt{tclean}; the second, third, and fourth images are the models, residuals, and 1D radial profiles created by \texttt{Frank}. The colorbar indicates the flux scale for both \texttt{tclean} image and the residual.}}
\label{fig:5+II_app}
\vspace{0.8cm}

\FloatBarrier
\section{New substructures from the image plane and \texttt{Frank} analysis}

In Stage I, we identify new inflection points in the Class I/F objects ODISEA\_C4\_33 (ISO-Oph 21), ODISEA\_C4\_67 (ISO-Oph 99), ODISEA\_C4\_72 (ISO-Oph 112), and ODISEA\_C4\_64 (ISO-Oph 94, Fig.~\ref{fig:1+I_app}), and in the Class II ODISEA\_C4\_70 disc (WL 10, Fig.~\ref{fig:1+II}). These discs have shown smooth emission profiles in \cite{2019Cieza} and no evidence of features potentially linked to planetary core formation.

Among the binary systems, we detect new inflection points in ODISEA\_C4\_105A (ISO Oph 167) and ODISEA\_C4\_134A (ISO-Oph 204, Fig~\ref{fig:1+I_app}).

In Stage II, we observe faint blobs in the image plane of \ODISEA{043} (ISO-Oph 43, Fig.~\ref{fig:2+I}) and RA162813.74 (Fig.~\ref{fig:2+II_app}), which likely correspond to the edges of rings seen at high inclinations ($i \sim 79\degree$ and $i \sim 85\degree$, respectively).
\texttt{Frank} identifies these same structures as gap–ring pairs.
The disc RA162813.74 was observed at 200 mas resolution; thus, higher-resolution (50 mas) ALMA observations will be required to confirm and resolve these substructures.
We also detect a new gap–ring pair in ISO-Oph 208.

The Stage III object ODISEA\_C4\_38 (ISO-Oph 37) was previously studied by \cite{2021Cieza}, in which a gap-ring pair was identified. 
The increased angular precision achieved by our analysis framework, as described in Section \ref{sec:2.1}, has enabled the identification of an additional inflection point in the source at  $\sim14~\mathrm{au}$.

In Stage V, we detect new cavities in the Class I/F objects ODISEA\_C4\_68 (WL 17) and ODISEA\_C4\_081 (2MASS J16271643-2431145, Fig.~\ref{fig:5+I},~\ref{fig:5+I_app}), and in the Class II objects ODISEA\_C4\_065 (ISO Oph 95), ODISEA\_C4\_116 (SR
20W), and ODISEA\_C4\_75 (ISO Oph 117, Fig.~\ref{fig:5+II_app}).
For ODISEA\_C4\_68 (Fig.~\ref{fig:5+I}) and ODISEA\_C4\_75 (Fig.~\ref{fig:5+II_app}), the cavities are visible only in the \texttt{Frank} model but not in the image plane, indicating that observations at 50 mas or better will be needed to confirm these structures.
\cite{2025Shoshi} classified ODISEA\_C4\_116 as a “Ring” disc, following their morphological classification scheme based on major-axis brightness profiles.
However, they could not clearly identify a central cavity due to the bumpy brightness distribution and the apparent near-edge-on geometry. We confirm that this morphology corresponds to a disc with a clear inner cavity, now robustly detected both in the image plane and in the visibility domain using \texttt{Frank}. Our analysis further indicates a more moderate inclination of $68^\circ$, allowing a more reliable characterization of the disc structure.

In Stage V, the binary system ODISEA\_C4\_16A+B (WSB 19) shows cavities in both discs, resolved here for the first time (Fig~\ref{fig:5+II_app}).
Both components are in the same evolutionary stage and display rings at $9.3 \pm 1.5~\mathrm{au}$, suggesting that the system has co-evolved along a similar evolutionary path.
\texttt{Frank} modeling of the binary systems is based on the 50 mas high-resolution dataset; therefore, confirming these features and resolving them directly in the image plane will require even higher-resolution ALMA observations.

\section{Resolution degradation test}

\begin{figure}[!ht]
    \centering
    \includegraphics[width=0.49\textwidth]{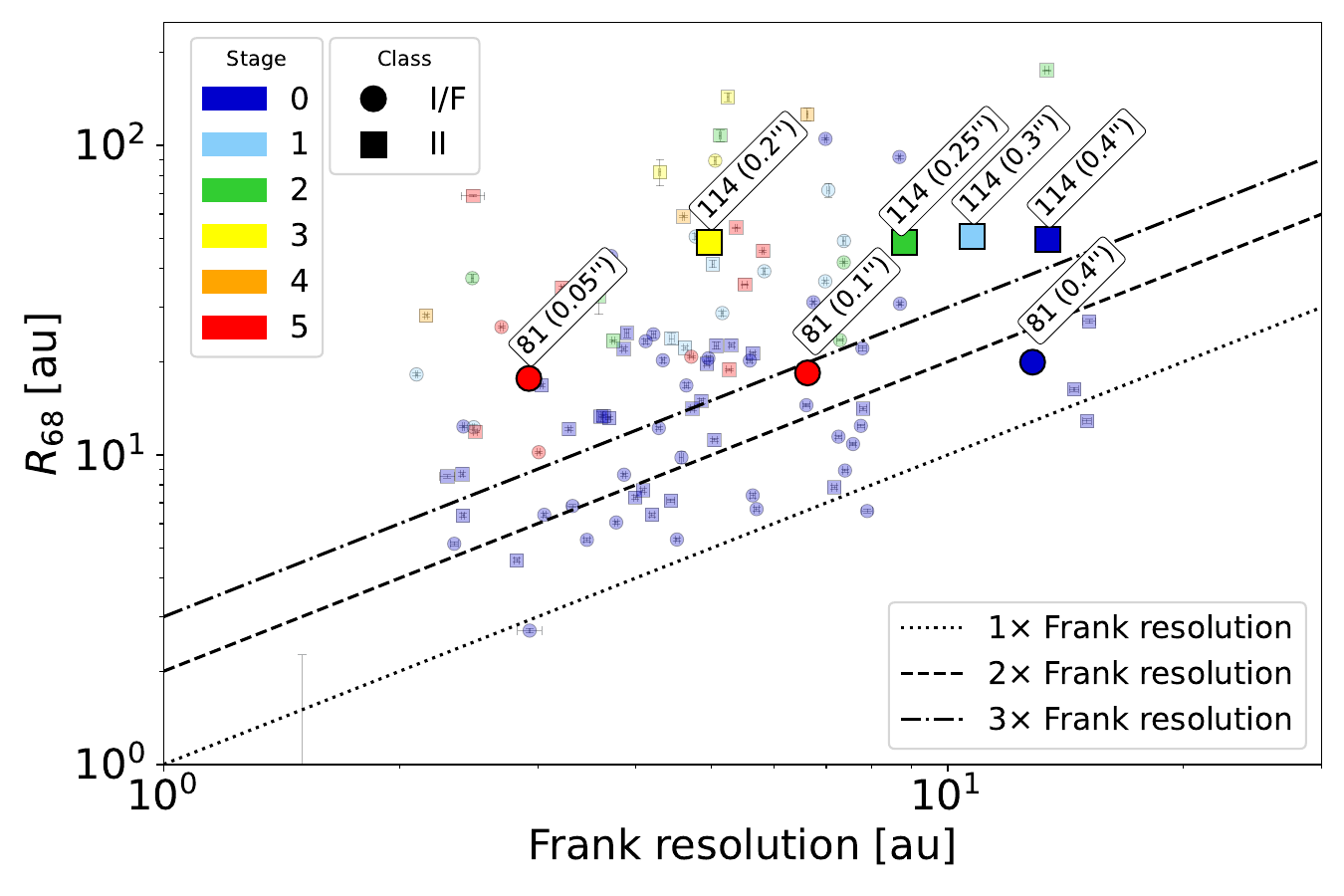}
    \caption{$R_{68}$ as a function of the {\texttt{Frank}} resolution, with diagonal dashed lines indicating the 1$\times$, 2$\times$, and 3$\times$ resolution limits. Color denotes the morphological stage. The dot-dashed line is 3 times the \texttt{Frank} resolution, the dashed line is 2 times the \texttt{Frank} resolution, and the dotted line is 3 times the \texttt{Frank} resolution. The bright points are the two discs \ODISEA{114}and \ODISEA{081} at different resolutions, annotated on the plot. }
    \label{fig:frank_degraded}
\end{figure}

\begin{figure*}[!h]
    \includegraphics[width=0.45\textwidth]{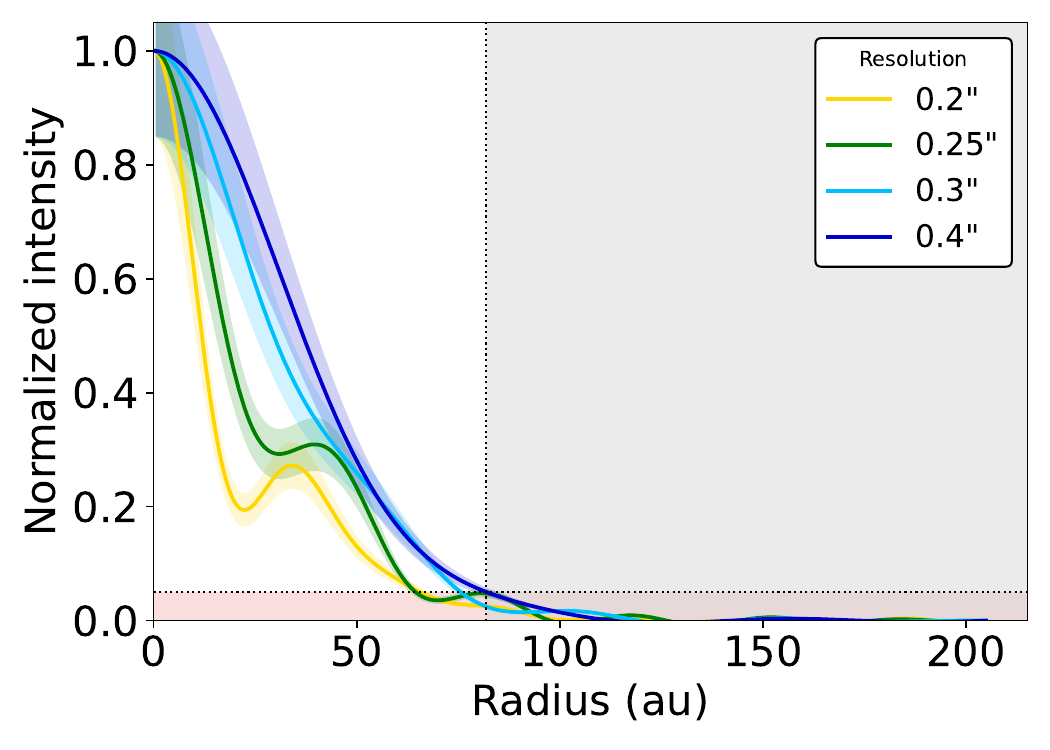}
    \includegraphics[width=0.45\textwidth]{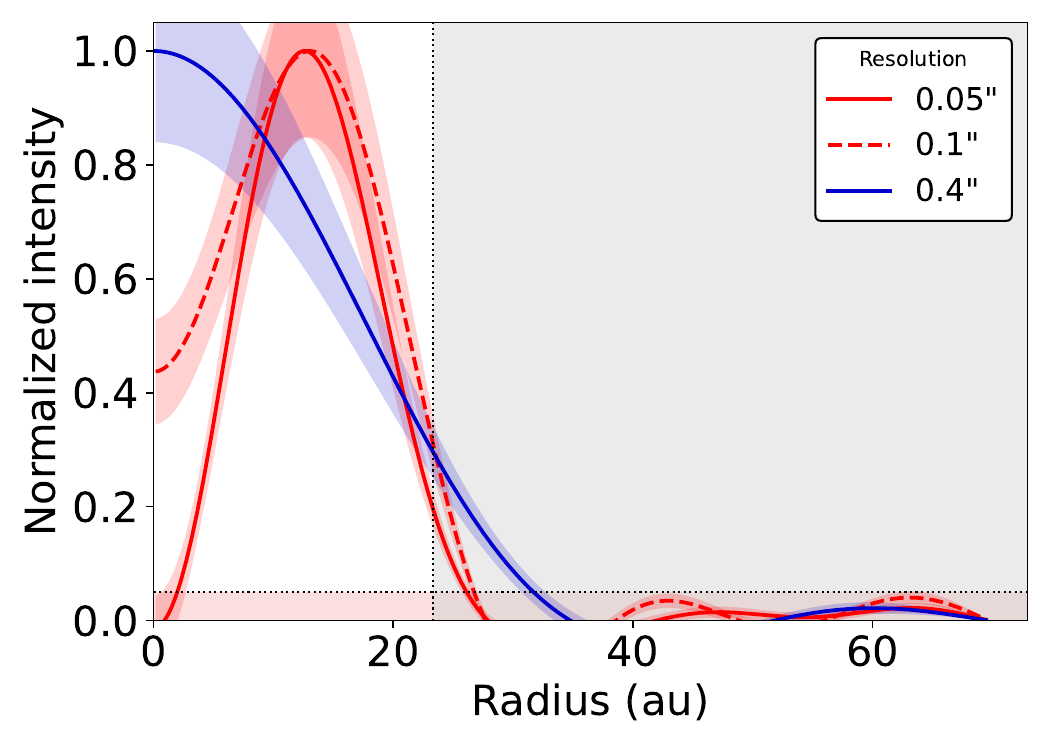}
    \caption{\texttt{Frank} radial profiles for \ODISEA{114} (Left) and \ODISEA{081} (Right) derived from the original visibilities and from data sets progressively degraded in resolution. The colors indicate different morphological stages similar to Fig.\ref{fig:frank_degraded}.}
    \label{fig:frank_degraded_overplot}
\end{figure*}

\begin{figure*}[!t]
\begin{minipage}{\textwidth}
\section{Analysis with bolometric temperature}
\centering
        \includegraphics[ clip,width=0.9\linewidth]{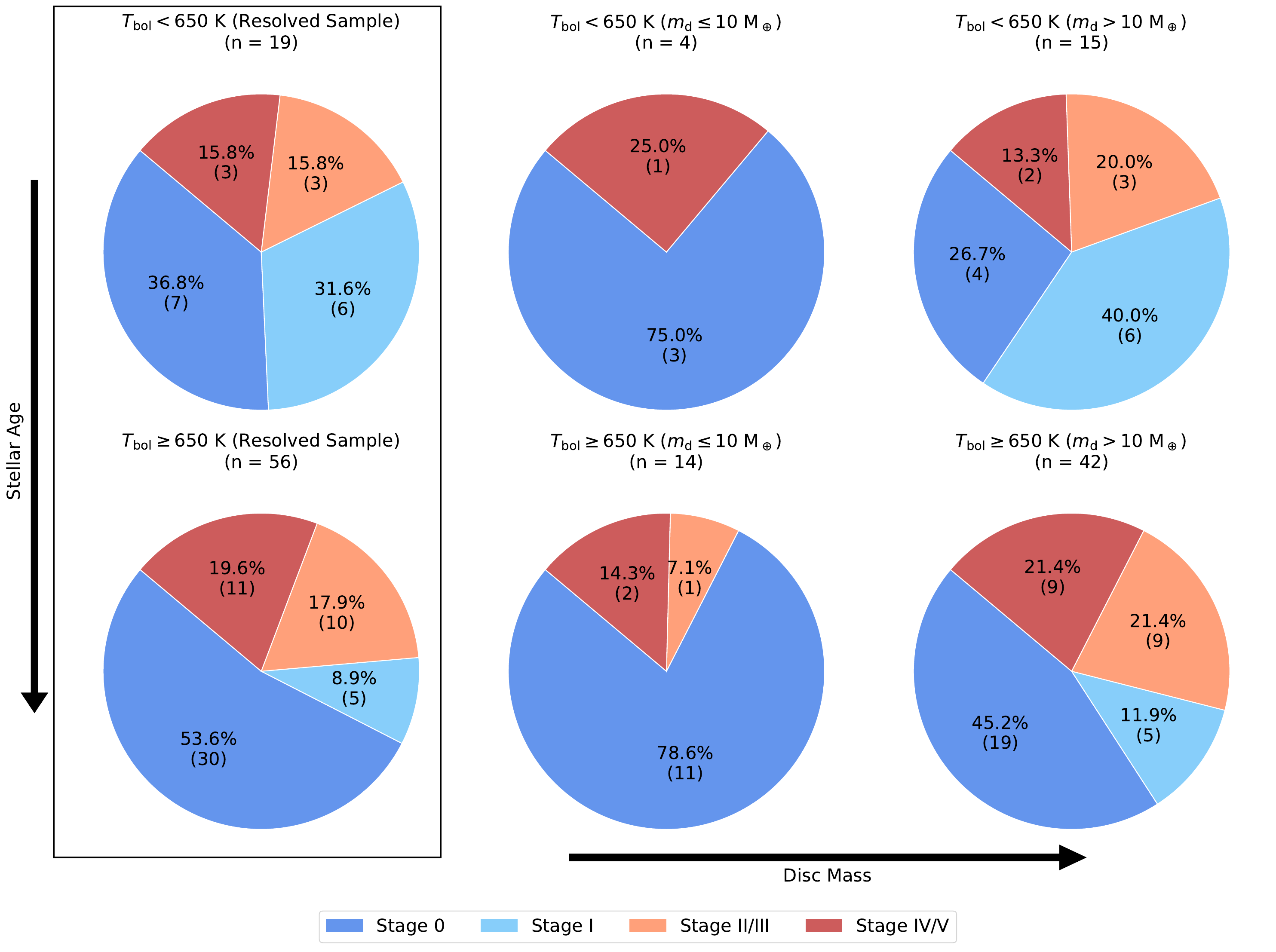}
        \caption{Top and bottom rows show pie charts of the sample divided by two temperature regimes $T_{bol}<650K$ and $T_{bol}\geq650K$. The second and third columns are further divided by disc mass at $M_{\mathrm{d}} = 10~M_{\oplus}$. Each pie chart indicates the fraction of discs in different morphological stages, Stage~0 (featureless), Stage~I (inflection), Stage~II/III (gaps), and Stage~IV/V (cavities) represented in different colors. }
        \label{fig:tbol_pie_chart}
\end{minipage}
\end{figure*}

 \begin{figure*}
    \includegraphics[width=0.49\textwidth]{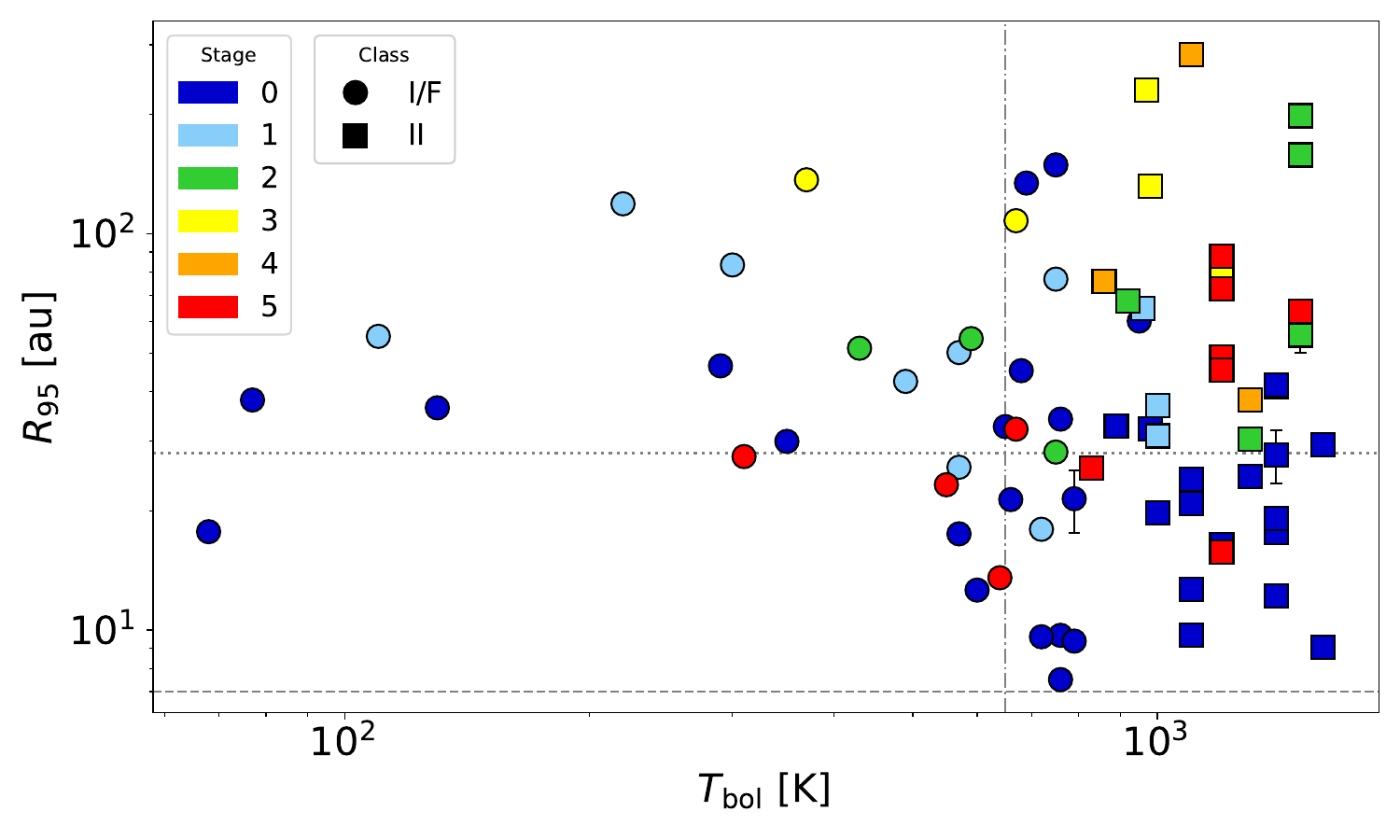}
    \includegraphics[width=0.49\textwidth]{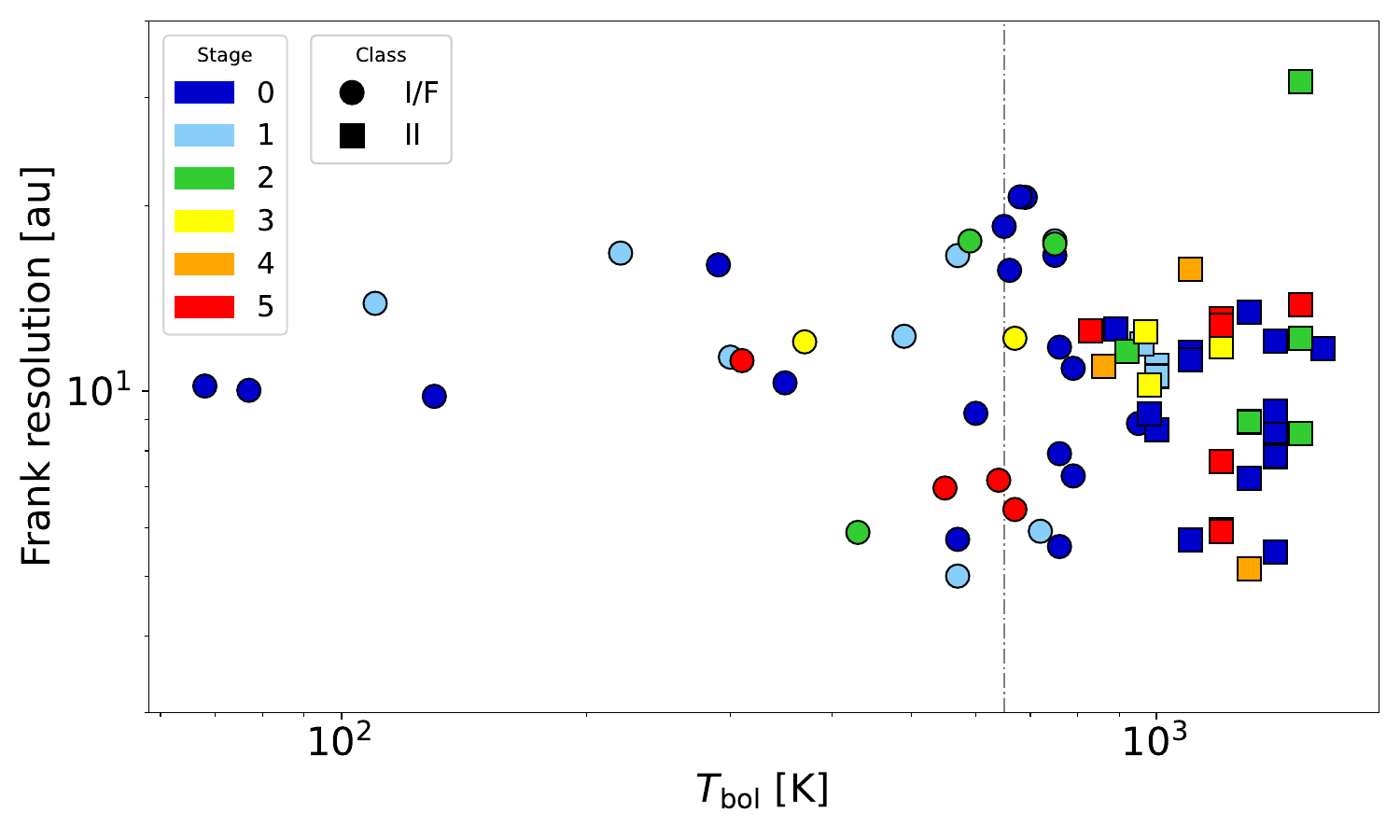}
    \caption{$R_{95}$ (Left) and resolution limits derived from \texttt{Frank} (Right) as a function of bolometric temperature ($T_{\mathrm{bol}}$). Circles and squares represent Class I/F and Class II systems, respectively, with point colors indicating morphological stage. The vertical dashed line marks the $T_{\mathrm{bol}}=650$ K division used to separate cold and warm populations, while the dotted and dashed horizontal lines mark the 22 au (0.2'') and 7 au (0.05'') beam resolution limits of our sample.}
    \label{fig:tbol}
\end{figure*} 

\begin{figure*}
    \includegraphics[width=0.49\textwidth]{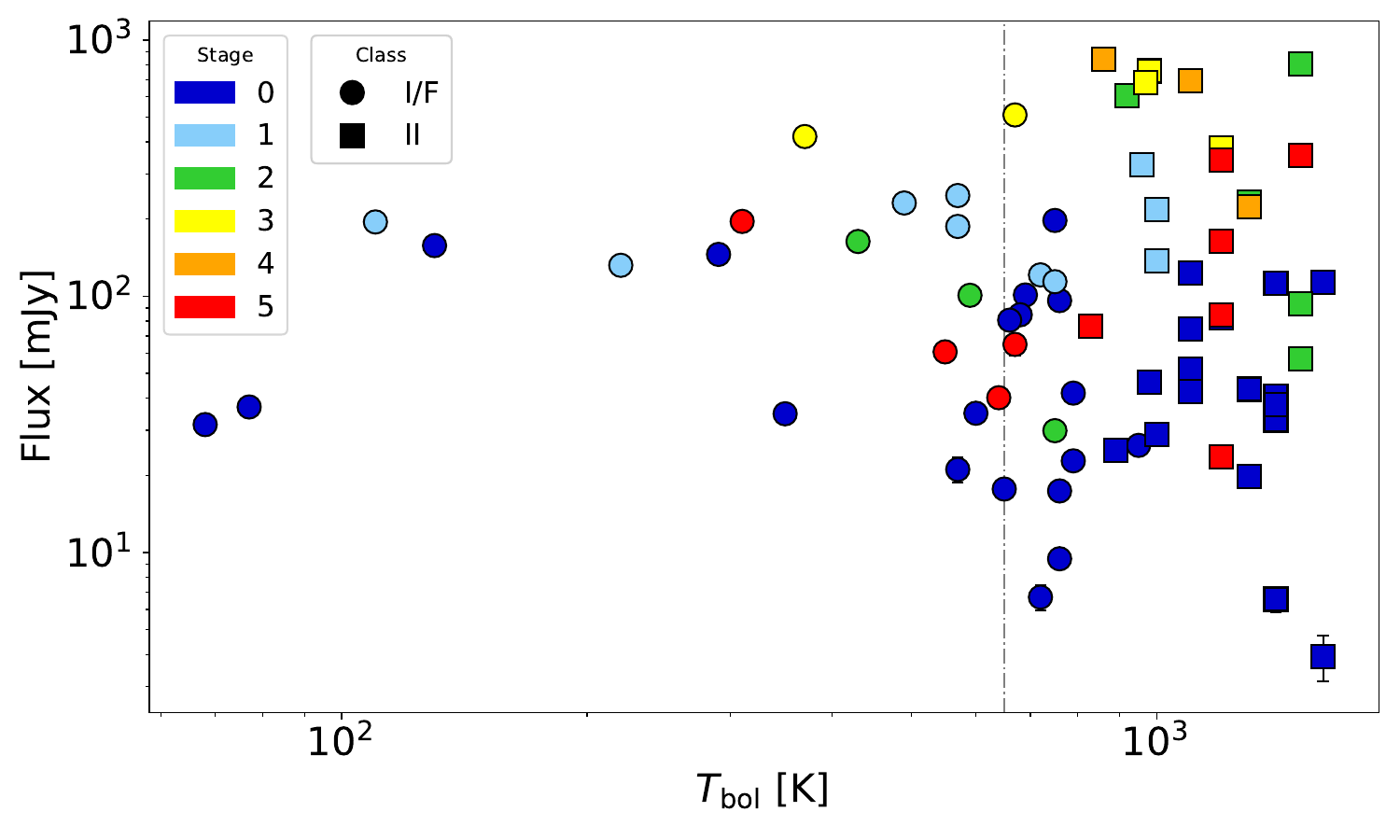}
    \includegraphics[width=0.49\textwidth]{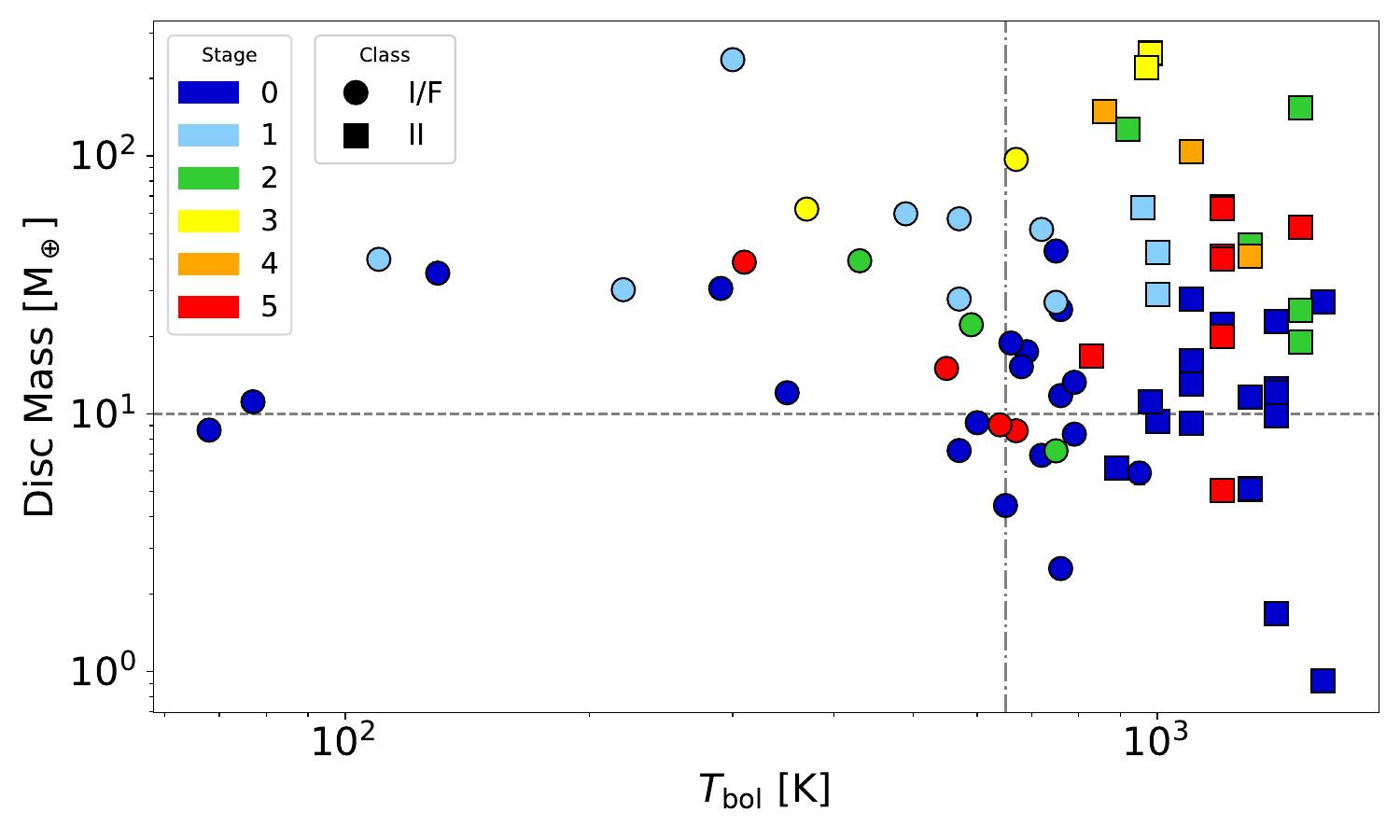}
        \caption{Flux (Left) and dust mass (Right) as a function of bolometric temperature ($T_{\mathrm{bol}}$). Circles and squares represent Class I/F and Class II systems, respectively, with point colors indicating morphological stage. The vertical dashed line marks the $T_{\mathrm{bol}}=650$ K division used to separate cold and warm populations, while the horizontal line marks the $10,M_{\oplus}$ dust-mass threshold.}
    \label{fig:tbol_mass}
\end{figure*}

\end{appendix}

\end{document}